\documentclass[aps,twocolumn]{revtex4-1}
\usepackage{amsmath,euscript,theorem,graphicx,rotating}

\usepackage{chemformula} 
\usepackage[T1]{fontenc} 

\usepackage[latin1]{inputenc}
\usepackage[T1]{fontenc}
\usepackage{txfonts}
\usepackage{url}
\usepackage{xr} 
\usepackage{color}
\usepackage{dcolumn}
\usepackage{bm}
\usepackage{longtable}
\usepackage{xspace}
\usepackage{booktabs}
\usepackage{version}
\usepackage{stackengine}
\usepackage{gensymb}

\bibstyle{achemso}

\usepackage{verbatim}

\usepackage{array}
\newcolumntype{L}{>{$}l<{$}}
\newcolumntype{R}{>{$}r<{$}}
\newcolumntype{C}{>{$}c<{$}}

\newcommand{\abinitio}{{\em ab initio}\xspace}
\newcommand{\Abinitio}{{\em Ab initio}\xspace}
\newcommand{\etal}{\emph{et al.}\xspace}

\newcommand{\rr}{{\mathbf r}}

\newcommand{\RR}{{\mathbf R}}
\newcommand{\rp}{{\mathbf r'}}

\newcommand{\secrff}[1]{{\ \S \ref{#1}}}

\newcommand{\figrff}[1]{Figure~\ref{#1}}
\newcommand{\tabrff}[1]{Table~\ref{#1}}

\newcommand{\eqrff}[1]{eq.~\eqref{#1}}

\newcommand{\Q}[2]{\ensuremath{Q^{#1}_{#2}}\xspace}  

\newcommand{\T}[2]{\ensuremath{T^{#1}_{#2}}\xspace}
\newcommand{\A}[2]{\ensuremath{\alpha^{#1}_{#2}}\xspace}

\newcommand{\kJmol}{\unskip\ensuremath{\,\rm{kJ\,mol}^{-1}}\xspace}

\newcommand{\Cn}[1]{\ensuremath{C_{#1}}\xspace}

\newcommand{\Eelst}[0]{\ensuremath{E^{(1)}_{\rm elst}}\xspace}
\newcommand{\EelstMP}[0]{\ensuremath{E^{(1)}_{\rm elst}[{\rm DM}]}\xspace}
\newcommand{\Eexch}[0]{\ensuremath{E^{(1)}_{\rm exch}}\xspace}
\newcommand{\EINDreg}[1]{\ensuremath{E^{(#1)}_{\rm IND}({\rm Reg})}\xspace}
\newcommand{\EIND}[1]{\ensuremath{E^{(#1)}_{\rm IND}}\xspace}

\newcommand{\EpolCL}[1]{\ensuremath{E^{(#1)}_{\rm pol,cl}}\xspace}
\newcommand{\EPOLCL}[0]{\ensuremath{E_{\rm pol,cl}}\xspace}
\newcommand{\Epol}[1]{\ensuremath{E^{(#1)}_{\rm pol}}\xspace}

\newcommand{\Eindpol}[1]{\ensuremath{E^{(#1)}_{\rm ind,pol}}\xspace}
\newcommand{\Eindpolreg}[1]{\ensuremath{E^{(#1)}_{\rm ind,pol}({\rm Reg})}\xspace}
\newcommand{\Eexind}[1]{\ensuremath{E^{(#1)}_{\rm ind,exch}}\xspace}

\newcommand{\Eexindreg}[1]{\ensuremath{E^{(#1)}_{\rm ind,exch}({\rm Reg})}\xspace}

\newcommand{\Edisp}[1]{\ensuremath{E^{(#1)}_{\rm DISP}}\xspace}

\newcommand{\EDISP}[1]{\ensuremath{E^{(#1)}_{\rm DISP}}\xspace}
\newcommand{\Edisppol}[1]{\ensuremath{E^{(#1)}_{\rm disp,pol}}\xspace}
\newcommand{\Eexdisp}[1]{\ensuremath{E^{(#1)}_{\rm disp,exch}}\xspace}

\newcommand{\EINT}[0]{\ensuremath{E_{\rm int}}\xspace}
\newcommand{\Eint}[1]{\ensuremath{E_{\rm int}^{(#1)}}\xspace}

\newcommand{\ESR}[0]{\ensuremath{E_{\rm sr}}\xspace}

\newcommand{\deltaHF}[0]{\ensuremath{\delta^{\rm HF}_{\rm int}}\xspace}
\newcommand{\PhiA}[1]{\ensuremath{\Phi^{\mathrm{A}}_{#1}}\xspace}

\newcommand{\EA}[1]{\ensuremath{E^{\mathrm{A}}_{#1}}\xspace}

\newcommand{\ECD}[1]{\ensuremath{E_{\rm CD}^{(#1)}}\xspace}
\newcommand{\EPOL}[1]{\ensuremath{E_{\rm POL}^{(#1)}}\xspace}

\newcommand{\regSAPTDFT}[0]{Reg-SAPT(DFT)\xspace}

\newcommand{\betapolIP}[0]{\ensuremath{\beta_{\rm pol}^{\mathrm{IP}}}\xspace}

\newcommand{\BETApol}[1]{\ensuremath{\beta_{\rm pol}^{#1}}\xspace}

\newcommand{\BETAdisp}[1]{\ensuremath{\beta_{\rm disp}^{#1}}\xspace}

\newcommand{\CamCASP}[1]{{\sc CamCASP #1}\xspace}

\newcommand{\DALTON}{{\sc DALTON} 2.0\xspace}
\newcommand{\Dalton}{\DALTON}
\newcommand{\NWChem}{{\sc NWChem}\xspace}
\newcommand{\PsiF}{{\sc Psi4}\xspace}
\newcommand{\Gamess}{{\sc GAMESS(US)} \xspace}

\newcommand{\Molpro}{{\sc Molpro}\xspace}

\newcommand{\R}[1]{\ensuremath{\rho^{#1}}\xspace}

\newcommand{\ISApolLn}[1]{ISA-Pol-L#1\xspace}

\mathchardef\lt="313C \mathchardef\gt="313E
\mathcode`<="4268 \mathcode`>="5269
\newcolumntype{d}[1]{D{.}{.}{#1}}
\newcolumntype{.}{D{.}{.}{-1}}
\newcolumntype{,}{D{,}{,}{2}}

\newcolumntype{L}{>{$}l<{$}}
\newcolumntype{R}{>{$}r<{$}}
\newcolumntype{C}{>{$}c<{$}}

\newcommand{\water}[0]{H$_2$O\xspace}
\newcommand{\waterN}[1]{(H$_2$O)$_{#1}$\xspace}
\newcommand{\EintS}[2]{\ensuremath{E_{\mathrm{int}}\text{[{#1}B-{#2}B]}}\xspace}
\newcommand{\EintB}[1]{\ensuremath{E_{\mathrm{int}}\text{[{#1}B]}}\xspace}

\def\Put(#1,#2)#3{\leavevmode\makebox(0,0){\put(#1,#2){#3}}}

\newcommand{\JCP}[0]{J. Chem. Phys.\ }
\newcommand{\JPCA}[0]{J. Phys. Chem. A\ }

\newcommand{\JCTC}[0]{J. Chem. Theory Comput.\ }
\newcommand{\IJQC}[0]{Int. J. Quantum Chem.\ }
\newcommand{\CPL}[0]{Chem. Phys. Lett.\ }

\newcommand{\PRB}[0]{Phys. Rev. B\ }

\newcommand{\PRL}[0]{Phys. Rev. Lett.\ }
\newcommand{\CR}[0]{Chem. Rev.\ }

\newcommand{\MolP}[0]{Mol. Phys.\ }

\newcommand{\PCCP}[0]{Phys. Chem. Chem. Phys.\ }

\newcommand{\IRPC}[0]{Int. Revs. Phys. Chem.\ }

\newcommand{\ChemComm}[0]{Chem. Commun.\ }

\newcommand{\SurfSciLett}[0]{Surf. Sci. Lett.\ }

\begin{document}

\title[polarization models]
{
First-principles many-body non-additive polarization energies from monomer and 
dimer calculations only : A case study on water
}

\author{Rory A. J. Gilmore}
\affiliation{School of Physics and Astronomy 
and the Thomas Young Centre for Theory and Simulation of Materials
at Queen Mary University of London,
London E1 4NS, U.K.}
\author{Martin T. Dove}
\affiliation{School of Physics and Astronomy 
and the Thomas Young Centre for Theory and Simulation of Materials
at Queen Mary University of London,
London E1 4NS, U.K.}
\author{Alston J. Misquitta}
\affiliation{School of Physics and Astronomy 
and the Thomas Young Centre for Theory and Simulation of Materials
at Queen Mary University of London,
London E1 4NS, U.K.}
\email{a.j.misquitta@qmul.ac.uk}

\begin{abstract}
The many-body polarization energy is the major source of non-additivity
in strongly polar systems such as water. This non-additivity is often 
considerable and must be included, if only in an average manner, 
to correctly describe the physical properties of the system.
Models for the polarization energy are usually parameterized using
experimental data, or theoretical estimates of the many-body effects.
Here we show how many-body polarization models can be developed for water
complexes using data for the monomer and dimer only using ideas recently
developed in the field of intermolecular perturbation theory and 
state-of-the-art approaches for calculating distributed molecular properties based on the 
iterated stockholder atoms (ISA) algorithm. 
We show how these models can be calculated, and validate their
accuracy in describing the many-body non-additive energies of a range
of water clusters. 
We further investigate their sensitivity to the details of the
polarization damping models used.
We show how our very best polarization models yield many-body energies
that agree with those computed with coupled-cluster methods, but at
a fraction of the computational cost.
\end{abstract}

\date{\today}

\maketitle


\section{Introduction}
\label{sec:intro}

\Abinitio intermolecular interaction models have come a long way in recent years. 
We have seen the development both of specific interaction models, for high-accuracy
calculations on specific systems, and of classes of models designed to 
be more generally applicable to a broad range of systems. 
By and large, these models are constructed using well-defined theoretically 
motivated functional forms, with parameters either extracted from 
molecular properties, or fitted to a range of \abinitio interaction energy
calculations, or a combination of these. 
The successes of these approaches have now been well documented and include the fields of 
molecular organic crystal structure prediction, \cite{WelchKMSP08,CooperHJD08,PriceLWHPKD10,MisquittaWSP08}
simulation of liquids such as water, \cite{BukowskiSGA07}
high-accuracy spectroscopy 
(see for example \cite{JankowskiS98a,GroenenboomEtAl00a,vanderAvoirdEtAl10,VissersHJWvanderA05}),
and other bulk properties.

In principle, creating a model from first principles may seem reasonably 
straightforward: a functional form with a fixed number of parameters which 
are determined using as large a data set of \abinitio interaction energies 
as is needed. 
This is of course a simplistic view of what is a complex fitting problem in
a high dimensional space with parameters that are often heavily correlated. 
However, for the two-body interaction (here and elsewhere in this paper by `body' we mean
an interacting unit, that is, a molecule), particularly with the rigid-body 
approximation, this problem has been largely solved using advanced, hierachical
fitting techniques in which the bulk of the parameters are extracted from the charge-density
and static and frequency-dependent density linear-response functions or 
are given prior values (in the Bayesian sense) via a density-overlap model
\cite{MisquittaS16,StoneM07,MisquittaWSP08,VanVleetMSS15,VanVleetMS18,MetzPS16}. 
Additionally, machine learning algorithms are being used, particularly for high-accuracy 
modelling of small systems \cite{uteva_interpolation_2017},
though these techniques are being used on larger systems too 
\cite{li_machine_2017,popelier_qctff_2015}. 

Many-body non-additive effects complicate matters considerably.
Even with the rigid-body approximation, the dimension of the fitting space is 
much larger when many-body effects are included. 
Here, more than ever, we need recourse to physical models rather than 
brute-force fitting to computed data. 

The importance of many-body effects cannot be disputed. The non-trivial consequences
of many-body dispersion energies have been extensively documented in the recent
literature in the condensed phase, or in low-dimensional systems, these effects
can lead to qualitative deviations from the two-body additive description.
Likewise, many-body polarization effects have been shown to play an important role in
molecular organic crystals, \cite{WelchKMSP08,cooper_molecular_2008}
 interfaces, \cite{harder_many-body_2009,misra_insights_2017}
ion solvation,
the anomalous properties of water,
and more generally, in any system with strong permanent multipoles and high polarizabilities.
Besides these there are many-body effects arising from the exchange energy,
charge-delocalization energy (often termed ``charge-transfer'', but see the discussion below),
and the various couplings between these and the dispersion and polarization. 
In this paper we will be concerned with many-body polarization models only. 

In some cases, it is possible to mimic the average effect of the many-body interactions 
by adjusting the parameters in the two-body model. For example, in bulk water the 
many-body polarization leads to a charge movement in the water molecules that can 
be used to approximate the effects of polarization without the need for an effective
polarization model. 
This is of course only approximate and leads to models such as TIP4P that 
use a fixed enhanced dipole moment but are limited
in applicability to situations where the water is `bulk-like', and fail where the
water molecules are in a manifestly different state, for example in clusters
\cite{cisneros_modeling_2016} and at interfaces \cite{harder_many-body_2009,misra_insights_2017}.
If we seek to develop models that are applicable to an ever wider range of applications, then
we much describe these complex effects correctly, preferably from first principles. 

In principle the development of a many-body polarization model is straightforward: the system is 
assigned permanent multipoles and static (zero-frequency) polarizabilities. These
are coupled together and the polarization energy is determined through the self-consistent 
multipole-moment changes in a manner described by Stone \cite{Stone:book:13}. 
In some models Drude oscillators are used in place of point polarizabilities, but these
do not differ in a fundamental way, except that we understand better how to increase the
ranks of the point polarizabilities.
The multipoles and polarizabilities need to be distributed for all but the smallest of 
molecules and they need to include terms of high enough rank to ensure convergence.
The difficulty with the polarization model lies in the choice of damping: all polarization models
need some form of damping to ensure that the many-body polarization energies are meaningful.
While the choice of damping may not appear to have a significant effect on the two-body energies,
the consequences for the many-body energies can be enormous, particularly as we increase the 
ranks of the polarizabilitity tensors. 

Thus far there have been two main approaches to developing many-body models for water:
the first approach combines a two-body (2B) interaction model with an $N$-body (NB) polarization model, 
and the second uses both a two-body and three-body (3B) model with non-additive contributions
involving four and more bodies described using a polarization model. 
We will refer to these approaches as the ``2B+NBpol'' and ``2B+3B+NBpol'' models, respectively.
Note that here and elsewhere in this paper we refer to a (water) molecule as a ``body''.
The advantage of the first approach is simplicity: such a model will be more efficiently 
evaluated in a simulation, but the latter approach is expected to be more accurate as
it offers the possibility of including three-body non-additivity from the exchange and
dispersion effects in addition to that from the induction contribution to the interaction energy.
The disadvantage of the second approach is one of dimensionality: for rigid water molecules
the 2B model is 6-dimensional, but the 3B model is 12-dimensional. If the water molecules
are considered flexible then the number of dimensions increases to 12 and 21 respectively. 
This has meant that recent models based on the second approach, 
in particular the MB-pol model \cite{babin2013development,babin2014development}, which includes 
molecular flexibility, and the CCpol23+ model, which does not, 
have been parameterised using an enormous amount of data.
The three-body part of MB-pol used 12,000 water trimer calculations and CCpol23+ used
more than 71,000 trimers. Additionally CCpol23+ used tetramers of water extracted from the known 
water hexamers. 

Any model based on the 2B+NBpol approach will need far less data for its
parameterization. However as this class of models relies entirely on
the classical polarizable model for the many-body non-additivity, some, carefully
chosen clusters of water are needed in practice to help tune the parameters in the polarization model 
so as to result in a reasonable description of the many-body effects. 
This is the approach taken by the DPP and DPP2 models. \cite{kumar2010second}
Interestingly, the one exception to this is are the ASP-W potentials \cite{MillotSCHS98} 
which are constructed without recourse to cluster data.

While extensive data sets of clusters of more than two molecules is a viable approach for
the water molecule, for larger systems both the 2B+3B+NBpol and 2B+NBpol approaches can prove both
computationally formidable and also cumbersome. 
As we seek to develop many-body interaction models for larger systems we need to find ways of 
reducing our dependence on data and also reduce fitting to a minimum. 
Indeed, with these considerations in mind, over the last few years we have developed techniques 
for constructing many-body polarization models from accurate distributed multipoles
\cite{MisquittaSF14} and distributed polarizabilities \cite{misquitta_isa-pol_2018}, based both 
on a basis-space version \cite{MisquittaSF14} of the iterated stockholder atoms (ISA) algorithm
\cite{LillestolenW08} and 
a detailed understanding of the link between the polarization, charge-delocalization and
induction energies made through \regSAPTDFT \cite{Misquitta13a}.
While we have used some of these methods in developing the many-body model for the pyridine complex
and have successfully used this model to find the two known forms as well as a new, third form of the
pyridine crystal, \cite{aina_dimers_2017} we have never put these models to detailed tests
of their capacity to describe the many-body energies of complexes. 
This is what we do now, using some of the extensive datasets
available for the water system. Additionally, as we wish to explore the limits of
this approach --- that uses only monomer and dimer data --- in the prediction of the 
many-body non-additive energies, we pose the following questions:
\begin{itemize}
  \item Q1: Can many-body polarization models be developed from the 
    properties of the monomer and dimer energy calculations only?
  \item Q2: Can we develop a systematic hierarchy of polarizability models of increasing 
    rank, and how do the accuracies of these models vary with rank?
  \item Q3: How accurate are the damping models obtained from \regSAPTDFT and 
    how sensitive are the many-body energies and cluster geometries to deviations from them?
\end{itemize}
The first question is central to this paper: we will show that under some assumptions, the
many-body polarization energy can indeed be determined from monomer properties and dimer energies
only; no more information is needed. 
The second question is posed as this question has never been addressed in a systematic study: 
most polarization models make use of dipole-dipole polarizabilities only, and only 
ASP-W4 \cite{MillotSCHS98} and SCME \cite{wikfeldt_transferable_2013} make use of quadrupolar
polarizabilities, but neither fare well on reproducing the energies of the water 
hexamers \cite{cisneros_modeling_2016}. 
As we shall show, the dipole-dipole polarizability model is a
sweet spot for water, but for higher accuracy, particularly for systems with heavier atoms,
higher ranking terms may be needed. 
Finally we pose the third question as we have previously argued that the true polarization energy
(at second order) is well-defined through \regSAPTDFT \cite{Misquitta13a} and 
this allows us to determine the polarization damping needed for a specified polarization model.
Here we test just how close to optimum is this approach and what happens to the errors in the
many-body energies if the damping parameters are altered from the proposed values.

\section{Computational details}
\label{sec:computational_details}

The molecular properties used in this work were computed using the modified basis-space
iterated stockholder atoms (BS-ISA) and the ISA-Pol algorithms implemented in \CamCASP{6.0}
\cite{CamCASP}.
The Kohn--Sham orbitals and orbital energies needed for these calculations were calculated
using the DALTON2006 \cite{DALTON2} code with a patch from the SAPT \cite{SAPT2008} code.
We used the asymptotically corrected PBE0 \cite{AdamoB99a} functional with the d-aug-cc-pVTZ
basis set. For the asymptotic correction we used the Fermi--Amaldi \cite{FermiA34} long-range form 
with the Tozer \& Handy splicing \cite{TozerH98} and a shift computed using the vertical 
ionization energy of $0.4638$ Hartree \cite{Lias00}.
The linear-response density functional calculations were performed using the 
hybrid ALDA+CHF kernel in \CamCASP{6.0} \cite{CamCASP,MisquittaS16}. 

The SAPT(DFT) interaction energy calculations used to determine the induction energies and
regularized induction energies for the polarization models were computed using \CamCASP{6.0}
\cite{CamCASP} with the above numerical details except that we used the aug-cc-pVTZ 
basis in the MC+BS format \cite{WilliamsMSJ95} using a 3s2p1d1f mid-bond basis.
Note that in \CamCASP{6.0} the second-order exchange-induction energy is computed without
the single-exchange ($S^2$) approximation using a closed-shell implementation of the 
general energy expression derived by Schaffer \& Jansen \cite{SchafferJ12}. 

To develop the non-polarization terms in the two-body potential we used 2048 dimers selected
at random using an algorithm based on Shoemake's uniform random rotations scheme \cite{Shoemake92}
which we have described in previous publications \cite{StoneM07,MisquittaWSP08,MisquittaS16}
and have implemented in the \CamCASP{} program. SAPT(DFT) interaction energies were computed
using the above basis but with \CamCASP{5.9}, that is, the $S^2$ approximation was used
with \Eexind{2}. However as observed by Schaffer \& Jansen, the error is largely cancelled
with a corresponding error of the opposite sign in the \deltaHF term.

\section{Notation \& Nomenclature}
\label{sec:notation}

In this paper we adopt the notation for the SAPT(DFT) interaction energy components
introduced in a recent paper \cite{MisquittaS16}: 
\Eelst and \Eexch are the first-order electrostatic and exchange-repulsion
energies, $\EIND{2} = \Eindpol{2} + \Eexind{2}$ is the total second-order
induction energy, 
$\EINDreg{2} = \Eindpolreg{2} + \Eexindreg{2}$ is the total {\em regularized}
second-order induction energy \cite{Misquitta13a},
$\EDISP{2} = \Edisppol{2} + \Eexdisp{2}$ is the total 
dispersion energy, and $\deltaHF$ is the estimate of effects
of third and higher order, primarily induction \cite{JeziorskaJC87,MoszynskiHJ96}.

We will partition the total second-order induction energy, \EIND{2}, into a 
charge-delocalization (CD) and true polarization (POL) energy at second-order: 
\ECD{2} and \EPOL{2}. Higher-order terms will be defined below.
Note that in previous works we have termed the charge-delocalization energy as the 
``charge-transfer'' energy, which is the term used by much of the literature.
Also note that what we will call the polarization energy is often, and confusingly, termed
the ``induction'' energy in texts \cite{Stone:book:13} and our past papers 
\cite{MisquittaS08a,MisquittaSP08}.

Note that the SAPT energy components without exchange effects included are also termed
``polarization'' \cite{JeziorskiMS94}, but these are not the same as the 
polarization energies that arise from the conventional electromagnetic polarization that
will be the focus of this paper.

\section{The models: Derived Intermolecular Force-Fields (DIFF)}
\label{sec:DIFF_models}

We have described the procedure used to develop the derived intermolecular force-fields,
or DIFF models, in previous works \cite{MisquittaS16}. 
In brief, at long-range,
distributed multipole expansions for the electrostatics, polarization
and dispersion terms are computed from the unperturbed molecular properties,
and at short-range the polarization and dispersion expansions are damped
and electrostatic penetration, exchange-repulsion and charge-delocalization energies
are described using an anisotropic Born--Mayer functional form, with the 
anisotropy described through atomic shape-functions.
Here we list some of the important numerical choices made that are relevant to this work;
the full potential specifications and parameters are provided in the Supplementary Information.
\begin{itemize}
  \item {\em Electrostatic multipoles} were calculated using the ISA-A functional from the
    modified version \cite{misquitta_isa-pol_2018} of the basis-space iterated stockholder atoms
    (BS-ISA) algorithm \cite{MisquittaSF14}. 
    Terms of maximum rank 4 were kept on all atoms, and the electrostatic expansion is not damped. 
  \item {\em Polarizability models} were calculated using the ISA-Pol algorithm 
    \cite{misquitta_isa-pol_2018} in a manner consistent with the electrostatic moments.
    The non-local polarizabilites from the ISA-Pol algorithm include terms from ranks 0 to 4
    and are cumbersome to be used directly, consequently we have localized the polarizabilities
    \cite{misquitta_isa-pol_2018} to result in three models of maximum ranks 1, 2 and 3. 
    These polarizability models are local but anisotropic and include all {\em intramolecular}
    couplings. 
    The polarization damping models are described below in detail.
  \item {\em Dispersion models} for the water dimer are identical to those reported
    in an earlier work \cite{misquitta_isa-pol_2018}. 
    They were calculated in a similar manner from the frequency-dependent
    localized and isotropic ISA-Pol polarizabilities, with all site pairs including terms
    from \Cn{6} to \Cn{12}. We have used the Tang \& Toennies damping functions \cite{TangT92} 
    with damping parameters determined using the scaled-ISA-exponents algorithm 
    \cite{misquitta_isa-pol_2018} which gives damping parameters:
    $\BETAdisp{\mathrm{OO}} = 1.7794$, $\BETAdisp{\mathrm{OH}} = 1.9011$, 
    and $\BETAdisp{\mathrm{HH}} = 2.0227$ atomic units.
  \item {\em Short-range terms} were modelled using an anisotropic Born--Mayer functional
    form in site--site form \cite{MisquittaS16}. The parameters in this part of the model
    were determined by fitting to reference SAPT(DFT) interaction energies which fully account
    for the electrostatic penetration, exchange-repulsion and charge-delocalization energies.
\end{itemize}
The full specifications of the model parameters are included in the supporting information.

Note that in previous works we have fitted the short-range interaction model in a multi-step
procedure that used the distributed density-overlap model as a means to obtain prior
values to the model parameters \cite{MisquittaS16}. We have not done this here as the water 
molecule is small enough that all parameters are well determined by the computed data.

\subsection{The polarization models}
\label{sec:pol_model_theory}

The damped classical polarization model is described comprehensively in texts such as
that by Stone \cite{Stone:book:13} so we outline only the points that relate
to this paper. The classical polarization energy of a molecule $A$ in a cluster is defined as:
\begin{align}
  \EPOLCL{}(A) &= \frac{1}{2} 
               \sum_{a \in A} \sum_{B\ne A} \sum_{b \in B} \sum_{tu}
                \Delta \Q{a}{t} f_{n(tu)}(\BETApol{ab} R_{ab}) \T{ab}{tu} \Q{b}{u},
                \label{eq:pol_classical}
\end{align}
where the ranks of the moments are given in the compact form 
$t \equiv l\kappa$ where $l=0,1,2,\dots$ is the angular momentum quantum number and 
$\kappa=0,1c,1s,\dots,lc,ls$ labels the real components of the spherical harmonics of rank $l$
(see Appendix B in Stone \cite{Stone:book:13}).
$\Q{a}{t}$ is the multipole moment operator for moment $t$ at site $a$,
and $\T{ab}{tu}$ is the interaction tensor \cite{Stone:book:13} which
describes the interaction between a multipole $\Q{b}{u}$ at site $b$ and a multipole
$\Q{a}{t}$ at site $a$.
$f_{n(tu)}(\BETApol{ab} R_{ab})$ is a damping function of order $n$. 
Here $n$ is a function of the tensor ranks $t$ and $u$, and if $t=l_1\kappa_1$
and $u=l_2\kappa_2$, then $n=l_1 + l_2 + 1$.
We assume that the damping depends only on the distance
$R_{ab}$ between sites $a$ and $b$ and not on their relative orientation.
This is an approximation that needs assessment, but we do not address this issue in 
this paper.
The strength of the damping is governed by the damping parameter \BETApol{ab}.
In the above expression, $\Delta \Q{a}{t}$ is the change in multipole moment $t$
at $a$ due to the self-consistent polarization of site $a$ in the 
field of all sites on {\em other} molecules, and is given by 
\begin{align}
  \Delta \Q{a}{t} &= - \sum_{a' \in A} \sum_{B\ne A} \sum_{b \in B} \sum_{t'v}
                   \A{aa'}{tt'} f_{n(t'v)}(\BETApol{a'b} R_{a'b})
                   \T{a'b}{t'v} (\Q{b}{v} + \Delta \Q{b}{v}),
                \label{eq:deltaQ}
\end{align}
where $\A{aa'}{tt'}$ is the distributed polarizability for sites
$(a,a')$ which describes the response of the multipole moment component
$\Q{a}{t}$ at site $a$ to the $t'$-component of the field at site $a'$.
To find $\Delta \Q{a}{t}$ we need to solve \eqrff{eq:deltaQ} iteratively
and this leads to the significant computational cost of the classical
polarization model, though there are now methods to reduce this cost
\cite{albaugh_new_2017,lagardere_tinker-hp_2017}.
If $\Delta \Q{b}{v}$ is dropped from the right-hand-side of this equation then 
the resulting $\Delta \Q{a}{t}$, when inserted in \eqrff{eq:pol_classical} leads to the 
second-order polarization energy, \EpolCL{2}.

In the polarization models used in this paper we assume the Tang--Toennies form for the
damping functions \cite{TangT84}. 
Further we use the localized form \cite{MisquittaS08a,MisquittaSP08} of the 
distributed polarizability tensor, that is, the non-local polarizability $\A{aa'}{tu}$ 
in \eqrff{eq:deltaQ} is replaced by $\A{a}{tu} \delta_{aa'}$, where $\delta_{aa'}$ 
is the Kronecker-delta and \A{a}{tu} is the localized polarizability tensor of the same
rank.

Without the damping functions, the classical polarization model can be used to determine
the polarization energy at long-range only, where orbital overlap effects are negligible.
At short-range, damping needs to be included to suppress the $1/R$ divergences that set
in as $R \rightarrow 0$.
Besides the elimination of the mathematical singularity, the damping functions are also required
to enable the {\em damped} classical polarization energies to match the reference, 
non-expanded polarization energies. 
This poses a problem as neither SAPT(DFT) nor SAPT have an explicit polarization energy term, 
instead in both theories the induction energy can be considered the sum of the 
polarization (POL) and charge-delocalization (CD) energies \cite{Misquitta13a}.
To complicate matters, just about any partitioning of the induction energy into CD and POL
is consistent with {\em a particular} damped classical polarization model as long as 
the CD component is exponentially decaying (and thereby lacks a multipole expansion).
That is, there are infinitely many damped classical polarization models that are consistent
with the SAPT(DFT) induction energies. 
This would not be a problem if we were interested in dimer energies alone, but, as was already
indicated by Misquitta \cite{Misquitta13a}, and as we demonstrate here with extensive datasets,
the choice of damping can lead to vastly different predictions for the
{\em many-body} polarization energies, and the system geometries.
Consequently the choice of charge-delocalization energy is crucial.

\subsubsection{The charge-delocalization (CD) energy}
\label{sec:charge-delocalization}

The charge-delocalization may be thought of as a quantum delocalization process 
in which there is a (typically small) probability of the electronic charge
density of a molecule to tunnel onto the atomic sites of a neighbouring molecule
\cite{Misquitta13a}.
The energy of stabilization of this process is the charge-delocalization energy.
The physical origin of the CD energy is distinct from that of the classical polarization
energy which originates from local responses in the charge denstiy.
In fact, no polarization model with only rank 1 (dipole-dipole) polarizabilities
and those of higher ranks can describe the CD process. 
Instead the charge movement in the CD process probably needs to be described by the 
rank 0 by 0 polarizabilities, also called charge-flow polarizabilities 
(see Chapter 9 in Stone \cite{Stone:book:13}).
These are non-local and are thus able to describe charge movement across a 
system \cite{MisquittaS06,MisquittaSSA10,LiuAD11} but are rarely applied in polarization 
models.
While the non-local polarizabilities, including the charge-flow terms, are computed as
part of the ISA-Pol calculation \cite{misquitta_isa-pol_2018}, they are subsequently 
transformed away into higher-ranking polarizabilities using localization 
algorithms \cite{LeSueurS94,LillestolenW07}.
Even if these terms were retained, it is not clear how to use them directly in 
calculations of the CD energy.
Consequently when developing classical polarization models with the localized polarizability
models such as those we will use here, we need to ensure that the models are constructed
to model only that part of the induction energy associated with local responses. 
That is, the charge-delocalization energy, which involves longer ranged charge 
movement, must be separated out from the induction.

{\bf Regularized SAPT(DFT) in brief:}
We use the definition of the CD energy based on regularised SAPT(DFT) \cite{Misquitta13a},
or \regSAPTDFT. 
In this approach, the second-order induction energy is split into (second-order)
polarization and (second-order) charge-delocalization contributions using a modified 
electron-nuclear interaction operator. 
Conceptually this involves interpreting the charge-delocalization energy as arising
from a tunneling of the charge of one monomer into the attractive potential well 
arising from an electron-deficient site in a partner monomer. 
More specifically, we begin with the second-order induction energy of a molecule (A):
this is the energy of stabilization in response to the total electrostatic potential
arising from the unperturbed partner (B). 
In \regSAPTDFT we write the {\em regularized} electrostatic potential 
of monomer B as $\omega^{\rm B}_{\rm Reg}$ given by:
\begin{align}
  \omega^{\rm B}_{\rm Reg}(\rr) &= 
       - \sum_{\beta \in \mathrm{B}} Z_{\beta} \frac{1}{|\rr-\RR_{\beta}|}
         \left( 1 - e^{-\eta |\rr-\RR_{\beta}|^2} \right) 
       + \int \frac{\rho^{\rm B}(\rp)}{|\rr-\rp|} d\rp,
          \label{eq:VB-reg}
\end{align}
where $\beta$ labels the nuclei of monomer B, $Z_{\beta}$ is the
nuclear charge located at position $\RR_{\beta}$, $\rho^{\rm B}(\rr)$ is the unperturbed
electronic density of B, and $\eta$ is the regularization parameter.
The regularized second-order ``polarization'' (not to be confused with the classical 
polarization energy) component of the induction energy is then defined as:
\begin{align}
  \Eindpolreg{2}\mathrm{[A]} &= 
      {\sum_{r \ne 0}}
      \frac{|< \PhiA{r} | \hat{\Omega}^{\rm B}_{\rm Reg} | \PhiA{0} >|^2}
           { \EA{0} - \EA{r} },
\end{align}
where 
$\hat{\Omega}^{\rm B}_{\rm Reg} = \sum_{i} \omega^{\rm B}_{\rm Reg}(\rr_i)$ 
is the many-body form of the electrostatic potential of monomer B, 
and $\PhiA{r}$ and $\EA{r}$ are the excited states and energies of monomer A. 
A similar expression applies for $\Eindpolreg{2}\mathrm{[B]}$. 
$\Eindpolreg{2}$ is then the sum of the contributions from monomers A and B.
This expression is readily evaluated using linear-response theory within the
SAPT(DFT) framework, as is the accompanying second-order regularized exchange-induction 
contribution \cite{Misquitta13a} without the single-exchange approximation
\cite{SchafferJ12}.
These techniques are available in the \CamCASP{6.0} code and have also been recently implemented
in the \Molpro code \cite{Hesselmann_regInd_2019}.

The extent of the regularization is controlled by $\eta$: as $\eta \rightarrow \infty$ the 
regularization is switched off and we recover \EIND{2}. 
For any finite and positive value of $\eta$ we suppress some fraction of the second-order
induction energy, and with $\eta=3.0$ a.u., Misquitta has demonstrated
that is possible to suppress the CD contribution to \EIND{2} thereby allowing us to define
the second-order polarization energy as
\begin{align}
    \EPOL{2} &= \EINDreg{2} \equiv \Eindpolreg{2} + \Eexindreg{2}.
\end{align}
And the second-order charge-delocalization energy is defined as the difference:
\begin{align}
    \ECD{2} &= \EIND{2} - \EINDreg{2}  \\
            &\equiv \left( \Eindpol{2} - \Eindpolreg{2} \right)
                  + \left( \Eexind{2} - \Eexindreg{2}  \right).
\end{align}
Notice that both \EPOL{2} and \ECD{2} contain contributions from the exchange-induction energy.
This fundamentally differentiates \EPOL{2} from the second-order induction ``polarization'' 
energy, \Eindpol{2}, defined in SAPT(DFT).

The strength of this approach is that both \EPOL{2} and \ECD{2} are well-defined in the 
complete basis set limit and \ECD{2} is exponentially decaying as would be expected on 
physical grounds, even for very strongly bound complexes \cite{Misquitta13a}.
However the downsides of this approach are that there is as yet no rigorous way of determining
which value of $\eta$ exactly suppresses the charge-delocalization states in all cases, and
in its present implementation the technique is applicable at second-order only.
As has been demonstrated by Misquitta \& Stone \cite{MisquittaS16}, the second difficulty 
can be overcome using the classical polarization model. We describe how this is done in 
\secrff{sec:infinite_order_CD} below. 
The first issue is the more challenging one as, while Misquitta has argued 
\cite{Misquitta13a} that $\eta=3.0$ a.u.\ is an appropriate value for the regularization 
parameter, a more general algorithm to determine $\eta$ is needed. 

{\bf Fitting the polarization model damping to \EPOL{2}:}
Once we have determined the \EPOL{2} energies for a representative set of dimers, 
the polarization model damping in \eqrff{eq:pol_classical} and \eqrff{eq:deltaQ}
can be determined by fitting to these energies. 
Since \EPOL{2} is equivalent to the classical polarization energy evaluated at the first 
step of the self-consistent process, no iterations are performed when determining the damping.
That is, at this stage, the polarization energy is computed in response to the permanent fields only.

It may seem paradoxical that the many-body polarization energies can depend strongly
on the definition of the charge-delocalization energy; after all the former arises from 
long-range interactions while the latter is a short-range phenomenon. 
The key issue here is that the polarization model damping (a short-range effect) depends on the 
definition of the CD energy, and so the induced moments also depend on the CD energy. 
These induced moments then interact with other bodies in the system through the long-range
electrostatic interaction, and thereby can have major consequences for the energy of clusters
or the condensed phase.

\subsubsection{Polarization in the DIFF models for water}
\label{sec:DIFF_pol_models}

We have created three DIFF polarization models for water with maximum polarizability 
ranks 1, 2 and 3, respectively. 
As already stated, the distributed polarizability models were computed using
the ISA-Pol algorithm \cite{misquitta_isa-pol_2018} and these include anisotropy, and consequently
also include couplings between the ranks. These localized models are fully coupled in the sense 
that polarization effects {\em within} the water molecule are fully accounted for. 
However, like the ISA-DMA multipoles, they have been computed for a fixed water geometry and
will therefore be incorrect when applied to water molecules in different conformations.
We note here that in previous works we have often limited the polarizability ranks 
\cite{MisquittaS08a,MisquittaSP08} on the hydrogen atoms to rank 1 (dipole-dipole),
but here we have used the same maximum rank on all atoms. 

\begin{figure}
	\begin{center}
		\stackunder[-5pt]{\includegraphics[scale=0.17]{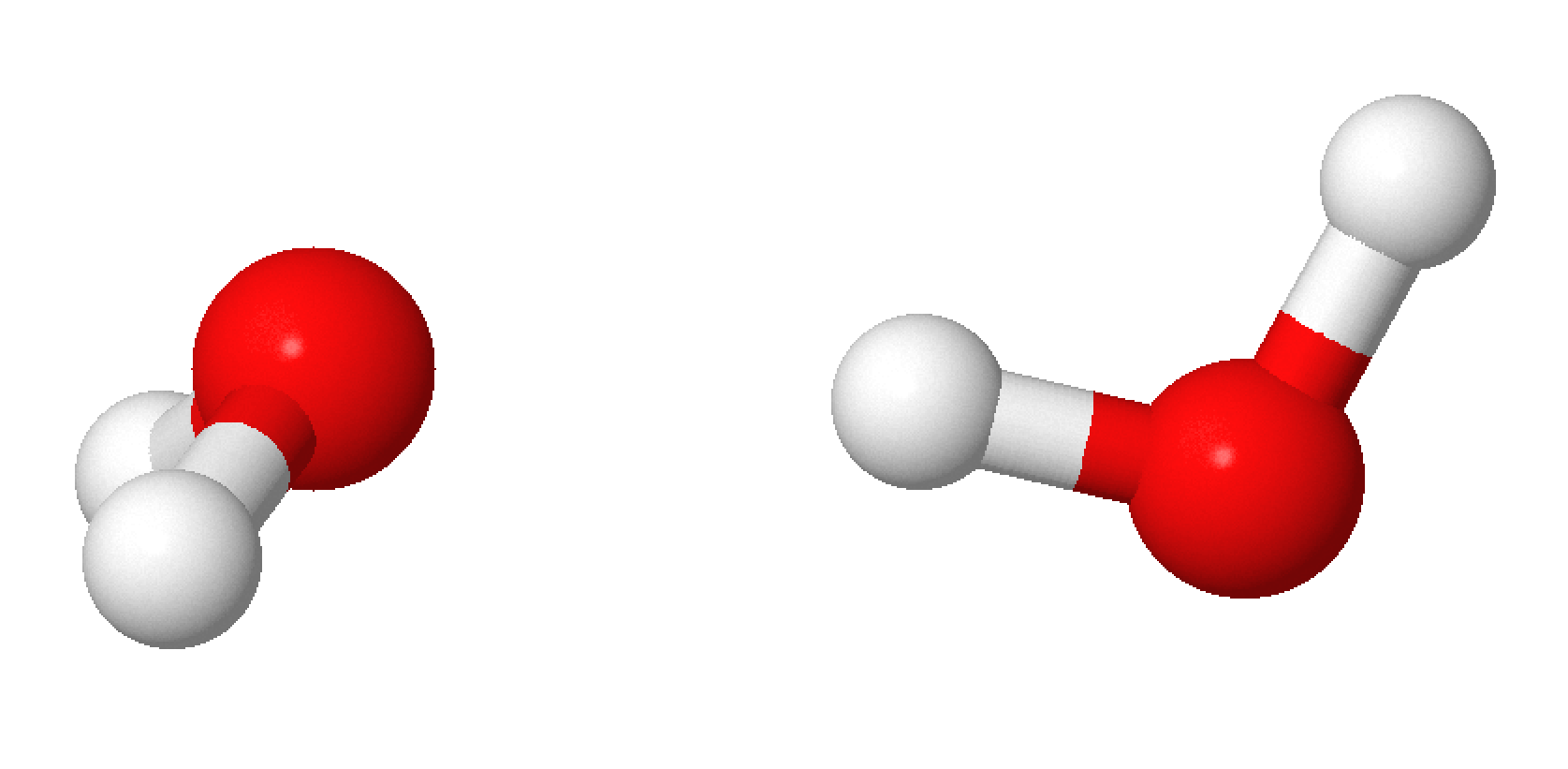}}{(a) O..H contact}
		\stackunder[-5pt]{\includegraphics[scale=0.20]{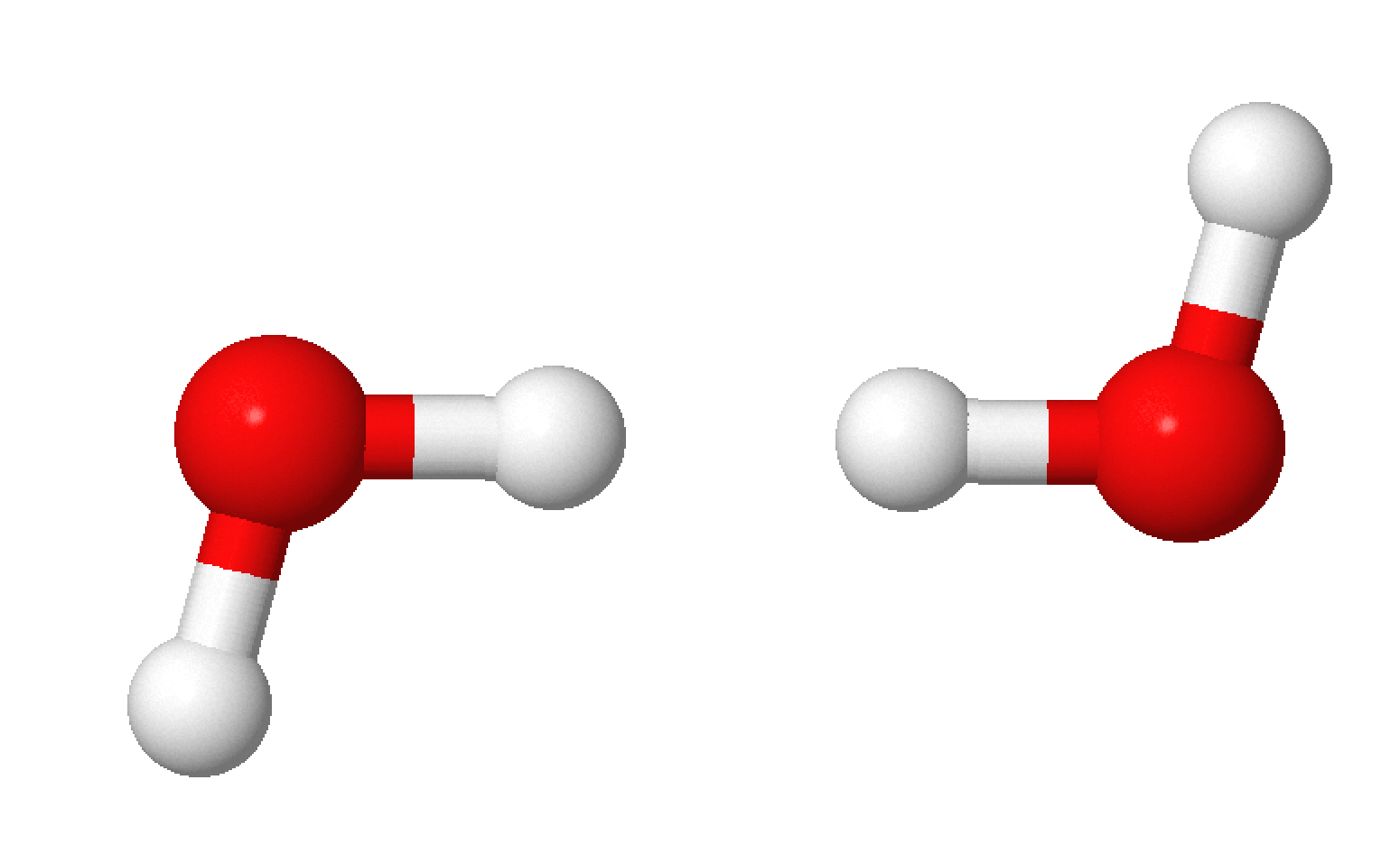}}{(b) H..H contact}
		\stackunder[-5pt]{\includegraphics[scale=0.20]{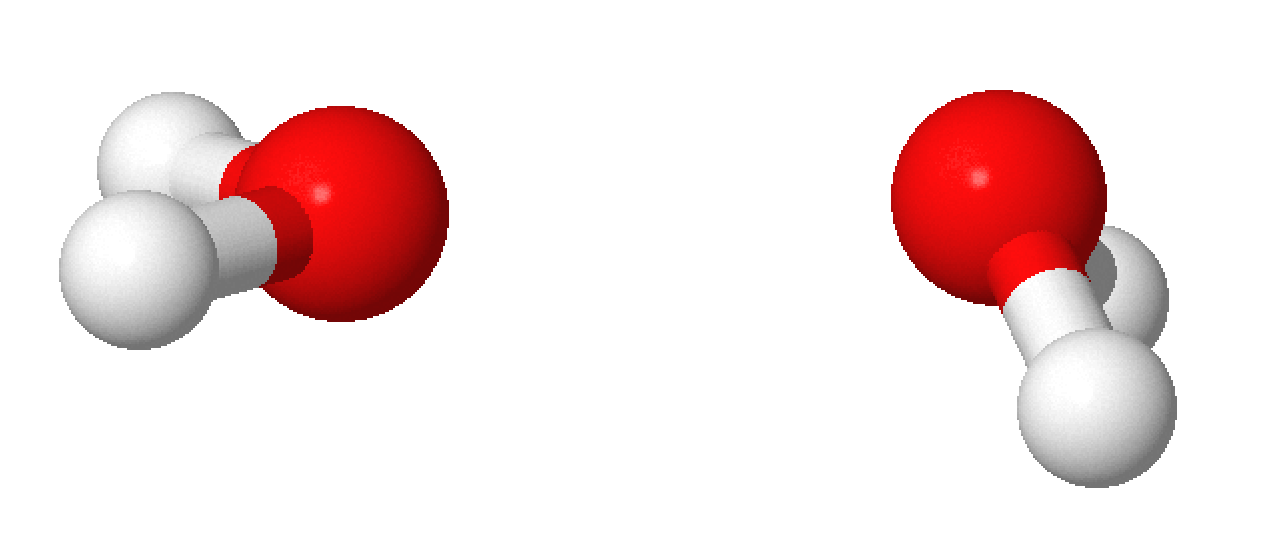}}{(c) O..O contact}
	\end{center}
	\caption{
      A representative sample of the water dimer configurations used in determining the
      polarization model damping parameters. 
      In case (c) the angles of the planes of the two water molecules were sampled in 
      30\textdegree\ intervals. 
      In all cases the dimer separations were sampled along the O..O separation vector.
    }
	\label{fig:dimers_for_damping}
\end{figure}

\begin{figure}
    \begin{center}
        \includegraphics[width=0.48\textwidth]{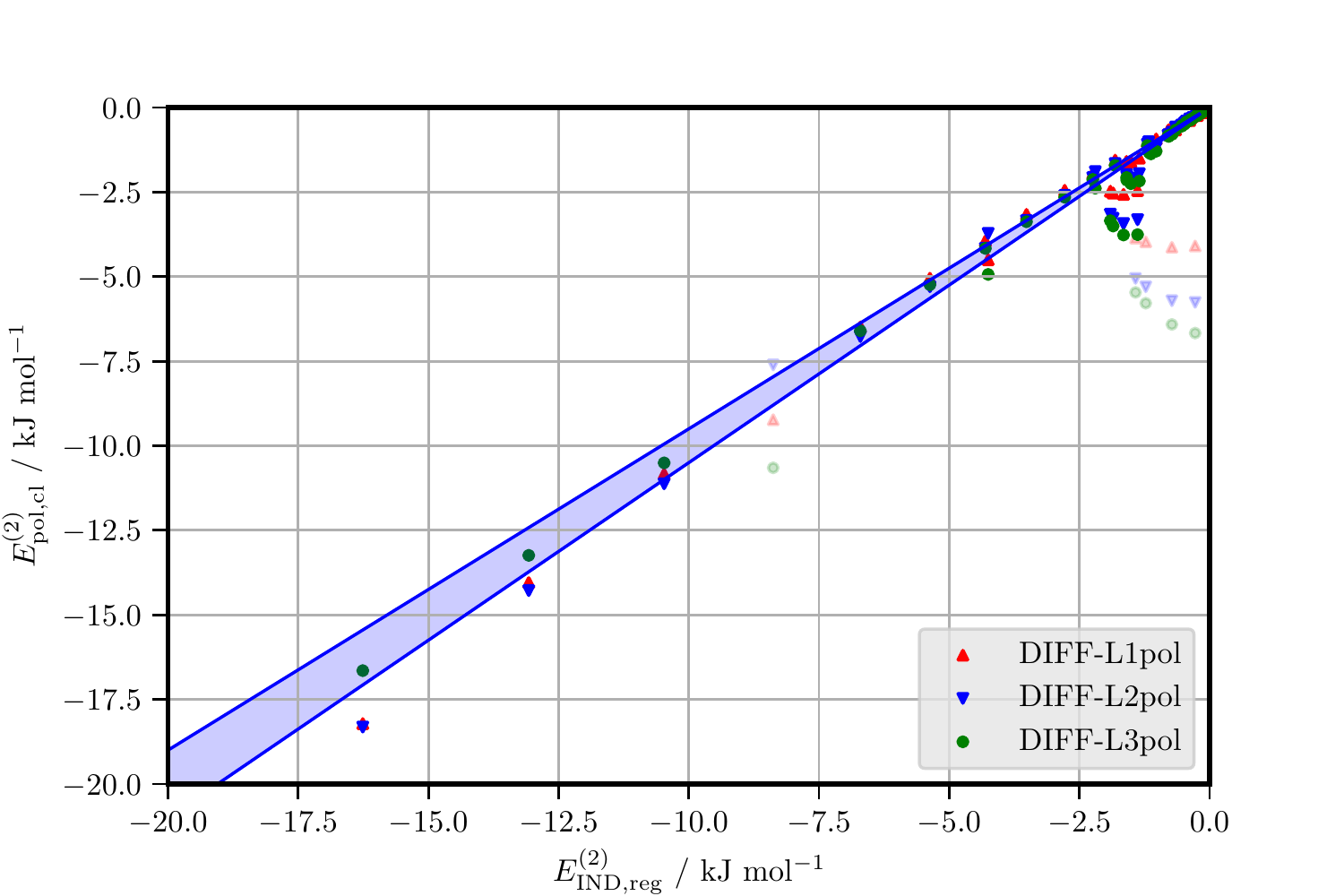}
    \end{center}
    \caption{
       Scatter plot of second-order polarization energies computed using the three DIFF models
       against the reference \EINDreg{2} energies computed for the dimer orientations shown in 
       fig.~\ref{fig:dimers_for_damping}.
       The solid points denote dimers with total energy less than $45.0$ \kJmol 
       and the faded points denote dimers with larger total interaction energies. 
       The latter were not considered when determining the damping parameters for this
       model.
    }
    \label{fig:pol2_scatter}
\end{figure}

There are three unique site-pairs in the water dimer (O\dots O, O\dots H, and H\dots H) and we
associate each of these with a unique damping parameter, consequently we need to select dimers
with suitable close contacts so as to be able to extract the damping information from the \EPOL{2}
energies. 
Following earlier work \cite{Misquitta13a} we use the three dimer orientations shown
in \figrff{fig:dimers_for_damping}, and for each dimer orientation, \EINDreg{2} energies 
are computed at several intermolecular separations. 
The polarization damping parameters, \BETApol{\mathrm{OO}}, \BETApol{\mathrm{OH}}, and \BETApol{\mathrm{HH}} for the three
DIFF models are shown in \tabrff{tab:poldamping}, and in \figrff{fig:pol2_scatter}
we display the scatter plot of \EINDreg{2} energies versus the corresponding energies from the 
three DIFF polarization models. We see that the O\dots O pair needs more damping (a smaller 
damping coefficient) than the H\dots H pair, with the mixed, O\dots H damping coefficient 
intermediate. 
Further, as the maximum rank of the polarizability terms decreases, \BETApol{\mathrm{OH}} increases, 
presumably to make up for the missing effects. 
The polarization energies are most sensitive to the O\dots H cross term and least sensitive to 
the H\dots H term. In fact, with the data available to us, we were unable to precisely determine the 
H\dots H polarization damping parameter other than to state that it is large and close to the 
chosen value of $\BETApol{\mathrm{HH}}=2.0$ a.u.\ 
To determine the damping for the O\dots O site pair we needed to compromise by focusing only on
dimers with interaction energy less than $45.0$ \kJmol, or just over twice the absolute binding
energy of the water dimer. 
It was not possible to find a damping model that worked for the more repulsive configurations,
presumably because the anisotropy of the oxygen atom in water requires an angular dependence
in the \BETApol{\mathrm{OO}} parameter. 

These damping models combined with the polarizabilities and the appropriate two-body models 
result in the DIFF-L$n$pol models, where $n=1,2,3$ indicates the maximum rank of the polarizability
tensors included in the model. 

\begin{table}[ht]
	\begin{tabular}{lll|l|l}
        \toprule
                     & IP    &         & DIFF  &         \\
		             & $x=0$ & $x=0.5$ & $x=1$ & $x=1.5$ \\
        \midrule
		L3:          &       &         &       &         \\
		$\beta_{OO}$ & 1.926 & 1.588   & 1.25  & 0.912   \\
		$\beta_{OH}$ & 1.926 & 1.698   & 1.47  & 1.242   \\
		$\beta_{HH}$ & 1.926 & 1.963   & 2.00  & 2.037   \\ \hline
		L2:          &       &         &       &         \\
		$\beta_{OO}$ & --    & --      & 1.25  & --      \\
		$\beta_{OH}$ & --    & --      & 1.57  & --      \\
		$\beta_{HH}$ & --    & --      & 2.00  & --      \\ \hline
		L1:          &       &         &       &         \\
		$\beta_{OO}$ & 1.926 & 1.588   & 1.25  & 0.912   \\
		$\beta_{OH}$ & 1.926 & 1.803   & 1.68  & 1.557   \\
		$\beta_{HH}$ & 1.926 & 1.963   & 2.00  & 2.037   \\
        \bottomrule
	\end{tabular}
	\caption{
        Polarisation damping parameters used for each model used in this work.
        The column titled ``IP'' indicates the damping based on the ionization 
        potential of water (see text for details), and ``DIFF'' indicates the 
        optimized damping for the DIFF models. 
    }
	\label{tab:poldamping}
\end{table}

\subsubsection{Alternative polarization damping models}
\label{sec:alternative_pol_damping}

In addition to the DIFF-L$n$pol models described above we have explored models with alternative
polarization damping in order to shed light on the importance of the choice of damping
on the many-body energies and energy landscape. 

In an early work on what were then termed ``induction models'', Misquitta, Stone \& Price
recommended that the induction damping be determined from the molecular ionization potentials
according to the formula \cite{MisquittaSP08}:
\begin{eqnarray}
  \betapolIP = \sqrt{2I_A} + \sqrt{2I_B},
    \label{eq:IP-damping}
\end{eqnarray}
where $I_{A/B}$ (in a.u.) are the vertical ionization potentials for monomers A and B. 
Notice that unlike the three-parameter DIFF damping models, this procedure results in a single
parameter for all site pairs: that is, the damping depends on the interacting molecules as a whole.
Despite this simplicity this damping model, which we will call the IP-based model, was shown
to work remarkably well for a set of systems, but with this model the damped polarization energy
was always close to \EIND{2}; that is, the total second-order induction energy which includes
both the polarization and charge-delocalization energies at second-order. 

\begin{figure}
    \begin{center}
        \includegraphics[width=0.48\textwidth]{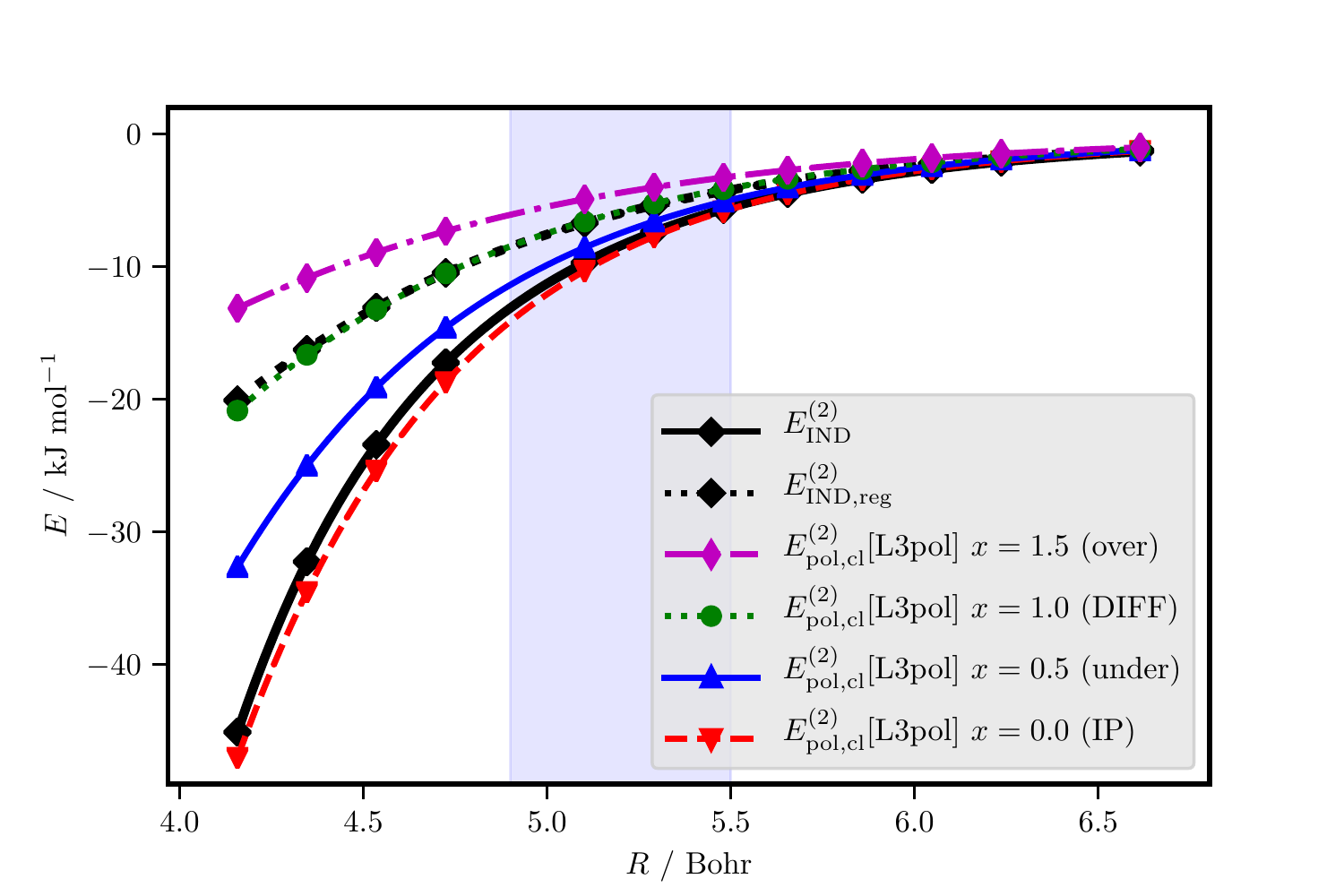}
    \end{center}
    \caption{
       The second-order induction energy, \EIND{2}, regularised induction, \EINDreg{2},
       and second-order polarization energies from the \ISApolLn{3} polarizabilities with various
       damping models, all plotted as a function of the O..O separation in the 
       water dimer in its hydrogen-bonded configuration (dimer (a) in 
       fig.~\ref{fig:dimers_for_damping}).
       As explained in the text, $x=1.5$ is the over-damped model, $x=1.0$ is the DIFF model
       with damping obtained from the regularised induction energies, $x=0.5$ is an underdamped
       model, and $x=0.0$ is an even more underdamped model derived from the ionisation energies
       of the water molecules.
    }
    \label{fig:IND_INDreg_and_models}
\end{figure}

In \figrff{fig:IND_INDreg_and_models} we display \EIND{2} and \EINDreg{2} for the water dimer
in the hydrogen-bonded configuration. Also displayed are polarization models all with 
maximum rank of 3 but with different damping models. 
First of all, the DIFF-L3pol model can be clearly seen to reproduce \EINDreg{2}, but as found by
Misquitta \etal \cite{MisquittaSP08}, the IP-damped model reproduces \EIND{2}.
We also display two additional models, one over-damped and the other under-damped, but intermediate
to the DIFF-L3pol model and the IP-based one. 
The damping coefficients used in these four polarization models are related through linear
interpolation with the parameter $x$ determining the model as follows:
\begin{eqnarray}
   \BETApol{ab}(x) = \betapolIP + x(\BETApol{ab} - \betapolIP),
   \label{eq:pol_damping_interpolation}
\end{eqnarray}
where $\betapolIP=1.926$ a.u.\ is the IP-based damping derived using \eqrff{eq:IP-damping} using 
a vertical ionization of $0.4638$ a.u.\ \cite{Lias00}, and \BETApol{ab} are the optimized
DIFF damping parameters for site-pair $ab$ shown in \tabrff{tab:poldamping}.
Varying $x$ allows us to smoothly interpolate between the significantly underdamped
IP-based damping model ($x=0.0$), through a moderately underdamped $x=0.5$ model, to the 
optimized DIFF model ($x=1.0$), and to an overdamped model with $x=1.5$.
We have created complete interaction models for $x=0.0,0.5,1.0,~\text{and}~1.5$ for 
the polarization models with maximum rank 1 and 3, and can therefore perform energy evaluations
and geometry optimizations with these models.
The second-order polarization energies from the $x=0.5$ and $x=1.5$ models are displayed
in \figrff{fig:IND_INDreg_and_models}.

The range of polarization models available to us (both in rank and in choice of damping model)
will allow us to evaluate the performance of these models in ways not possible before.
In particular, we will be able to assess the predictive power of the DIFF damping models based
on \EINDreg{2} energies as described earlier in this section, and the availability of entire
interaction models with different polarization damping will allow us to assess the importance of the
choice of damping on the structures of water clusters. 

\subsubsection{The infinite-order polarization charge-delocalization}
\label{sec:infinite_order_CD}

Before moving on we note that having obtained the DIFF damping parameters,
we can proceed to define the infinite-order CD and POL energies. 
Following Misquitta \& Stone \cite{MisquittaS16}, this is done by first approximating
the infinite-order induction energy as:
\begin{align}
    \EIND{2-\infty} \approx  \EIND{2} + \deltaHF
      \label{eq:Eind_infinite_order}
\end{align}
and then defining the two-body infinite-order charge-delocalization energy to be
\begin{align}
    \ECD{2-\infty} &= \EIND{2-\infty} - \EPOL{2-\infty} \nonumber \\
                   &\approx \EIND{2} + \deltaHF - \EpolCL{2-\infty},
    \label{eq:CD_infinite_order}
\end{align}
where $\EPOL{2-\infty}$ is the infinite-order polarization energy which is approximated
by $\EpolCL{2-\infty}$ from the classical polarization model iterated to convergence
(see \eqrff{eq:pol_classical} and \eqrff{eq:deltaQ}).
While this expression is readily implemented, it has a drawback in that the
definition depends on the type of polarization model used, but this dependence is relatively 
minor. 

The CD contributions from higher than second order are important and actually dominate at the
H-bonded minimum energy dimer geometry: Here $\ECD{2}=-0.95$ \kJmol, but using $\EpolCL{2-\infty}$
from the DIFF-L3pol polarization model we get, from \eqrff{eq:CD_infinite_order},
$\ECD{2-\infty} = -3.30$ \kJmol, 
which is in good agreement with the ALMO(CCSD) result
of $-3.51$ \kJmol from Azar \& Head-Gordon \cite{azar_energy_2012}.

\subsubsection{Summary of main features of the DIFF polarization models}
\label{sec:pol-summary}

\begin{itemize}
  \item {\em Only 1-body and 2-body information is used.} To construct the polarization models
    we need the molecular multipole moments and the molecular static polarizabilities. Both are
    1-body properties that depend on many-body effects only through the dependence of the 
    molecular conformation on the many-body interactions. 
    The damping is determined using only two-body second-order induction energies calculated
    through \regSAPTDFT. 
  \item {\em \regSAPTDFT allows us to define a definite damping model} whose parameters 
    vary only with the ranks of the multipole and polarizability models. 
  \item {\em Models of arbitrary rank can be created.} The algorithm present here can be used to
    develop models of any rank or complexity. That is, while we will use multipole models with 
    maximum rank 4 and anisotropic, fully coupled
    point-polarizability models with maximum ranks 1, 2 and 3, it should be equally possible
    to use Drude oscillator models with multipoles represented by point charges only.
\end{itemize}

It is perhaps the first of these points which is the most important: In the above procedure,
no many-body information is used. In this respect, the models we present in this paper will differ from 
almost every other previously developed many-body polarization model.

\begin{figure}
	\begin{center}
		\includegraphics[scale=0.20]{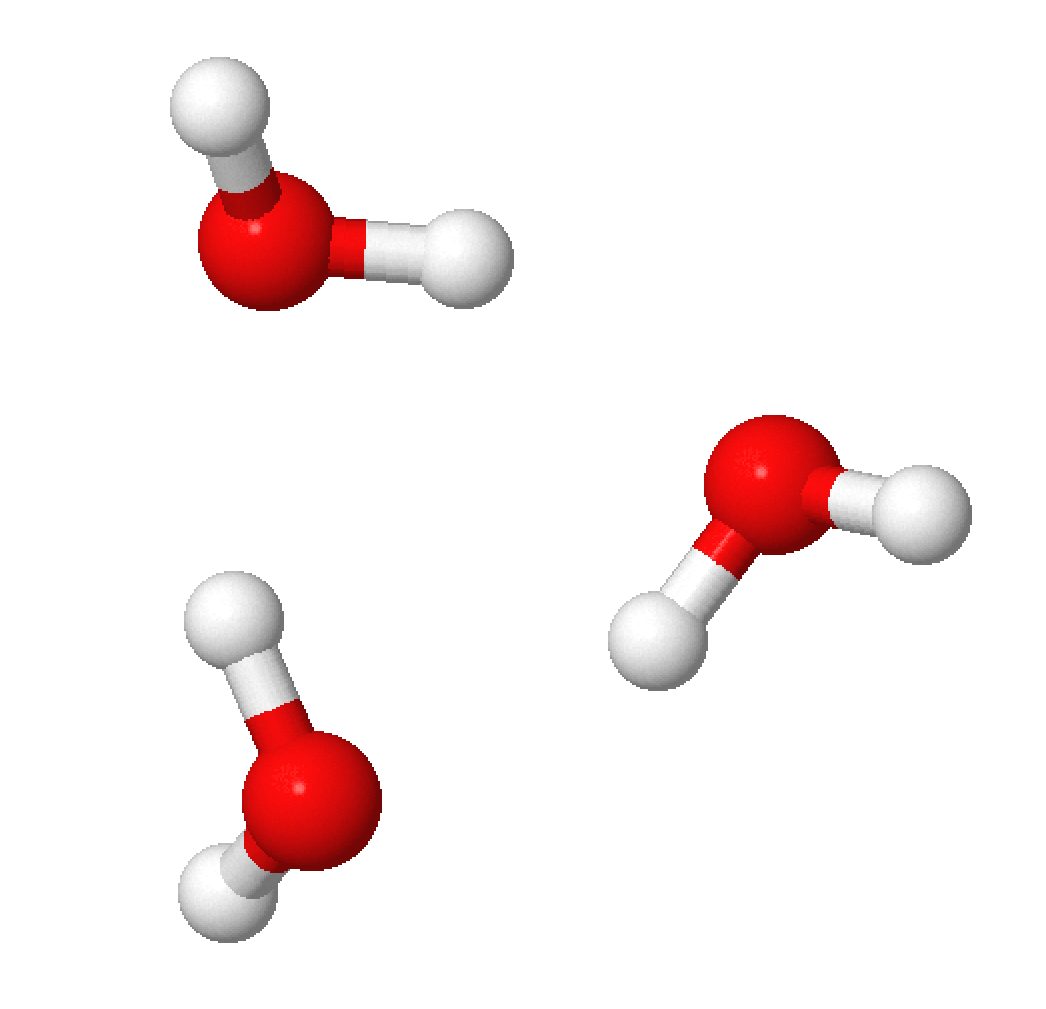}
	\end{center}
	\caption{
        Water trimer structure.
    }
	\label{fig:trimer}
\end{figure}

\section{Assessment of the DIFF models}
\label{sec:assessment}

Ideally we would use the DIFF models to simulate various bulk properties of water, but this
is not yet possible as even the simplest of these models cannot yet be used in mainstream 
simulation programs. While this may be possible soon, at present we are limited to tests on 
water clusters only. 
In some ways, this is advantageous as one of the main aims of this paper is to assess the
predictive power of the DIFF models for the many-body energies, and a detailed analysis of
water clusters of various sizes can allow us to make this assessment in an unambiguous manner
as we have very accurate energies and optimized geometries for a whole range of water clusters
from high-accuracy theory, as well as some information from experiment. 

\subsection{Trimers: Three-body (3B) non-additivity}
\label{sec:assessment-3B}

The water trimers provide us with the first test of the capability of the DIFF models to 
describe the non-additive energies. 
We first use the reference trimer set from Liu \etal \cite{liu2017capturing} which
consists of variations of the water trimer in its minimum energy configuration shown in 
\figrff{fig:trimer} with O\dots O bonds changed in a systematic manner.
The trimer non-additive energies have been computed by Liu \etal using second-order
M{\o}ller--Plesset (MP2) perturbation theory at the complete basis-set (CBS) limit.
Note that the monomer geometries used for these trimers is the same as that used to 
construct the DIFF models and consequently the DIFF multipoles and polarizabilities
are appropriate for these trimers.

\begin{figure}
	\begin{center}
		\includegraphics[width=0.48\textwidth]{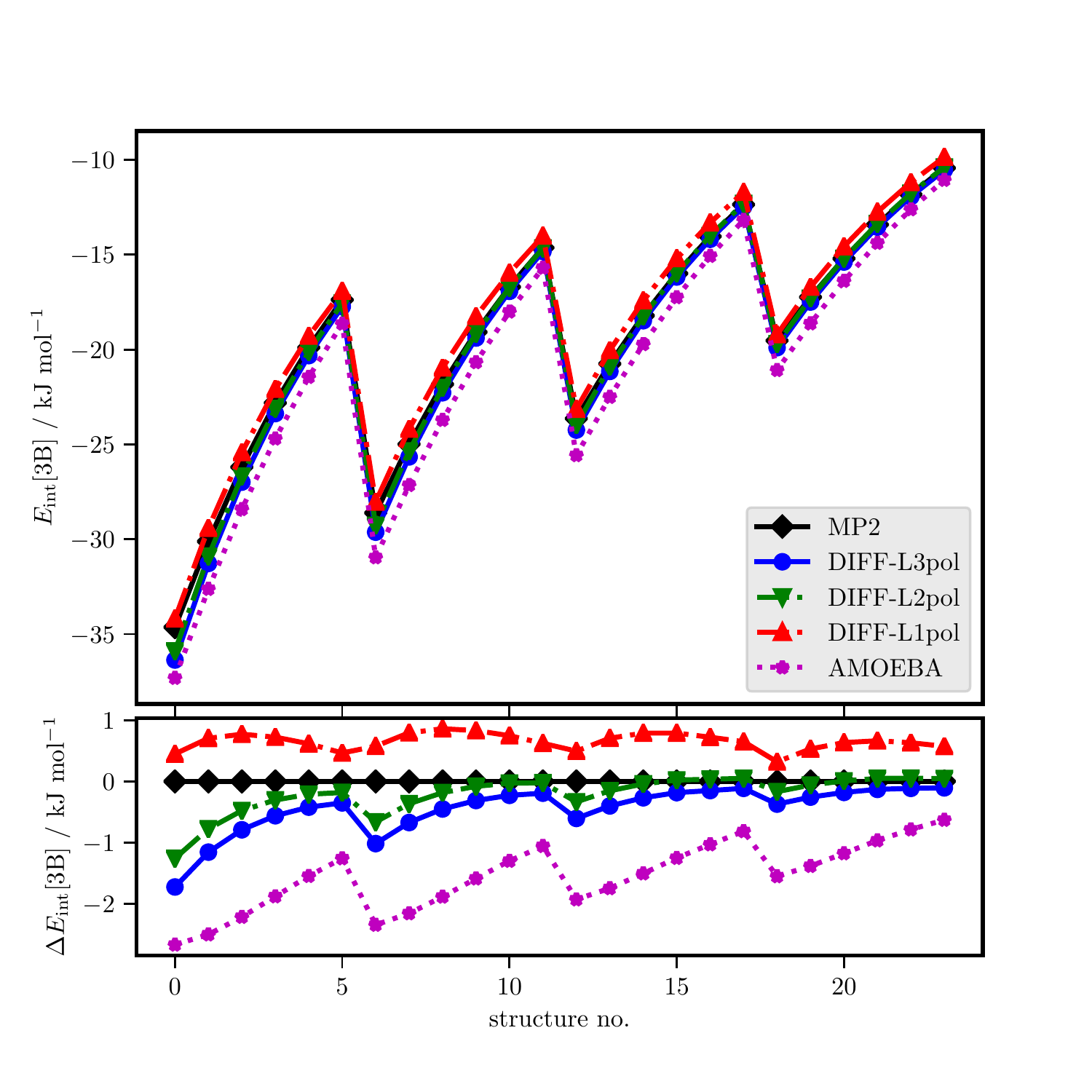}
	\end{center}
	\caption{
        Trimer three-body energies using the three DIFF models as well as
        the AMOEBA model.
	    Trimer geometries, MP2 and AMOEBA energies are from from Liu \etal \cite{liu2017capturing}.
        The lower panel displays three-body non-additive energies relative to those 
        from MP2.
    }
	\label{fig:trimer_3B_energies_all_models}
\end{figure}

In \figrff{fig:trimer_3B_energies_all_models} we display the non-additive 3B energies
for these trimers from the DIFF models as well as from the AMOEBA model studied by
Liu \etal \cite{liu2017capturing} 
The performance of all models is good with energy differences not exceeding 2 \kJmol, but
the DIFF-L2pol and DIFF-L3pol models show the best agreement with the MP2/CBS data with 
most of the differences being substantially less than 0.5 \kJmol. 
The DIFF-L1pol model tends to underestimate the 3B non-additive energies, but usually this 
by less than 1 \kJmol, and for the lowest energy trimers (structures 0 and 1) 
this model comes closer to MP2 reference energies than the others.

\begin{figure}
	\begin{center}
		\includegraphics[width=0.48\textwidth]{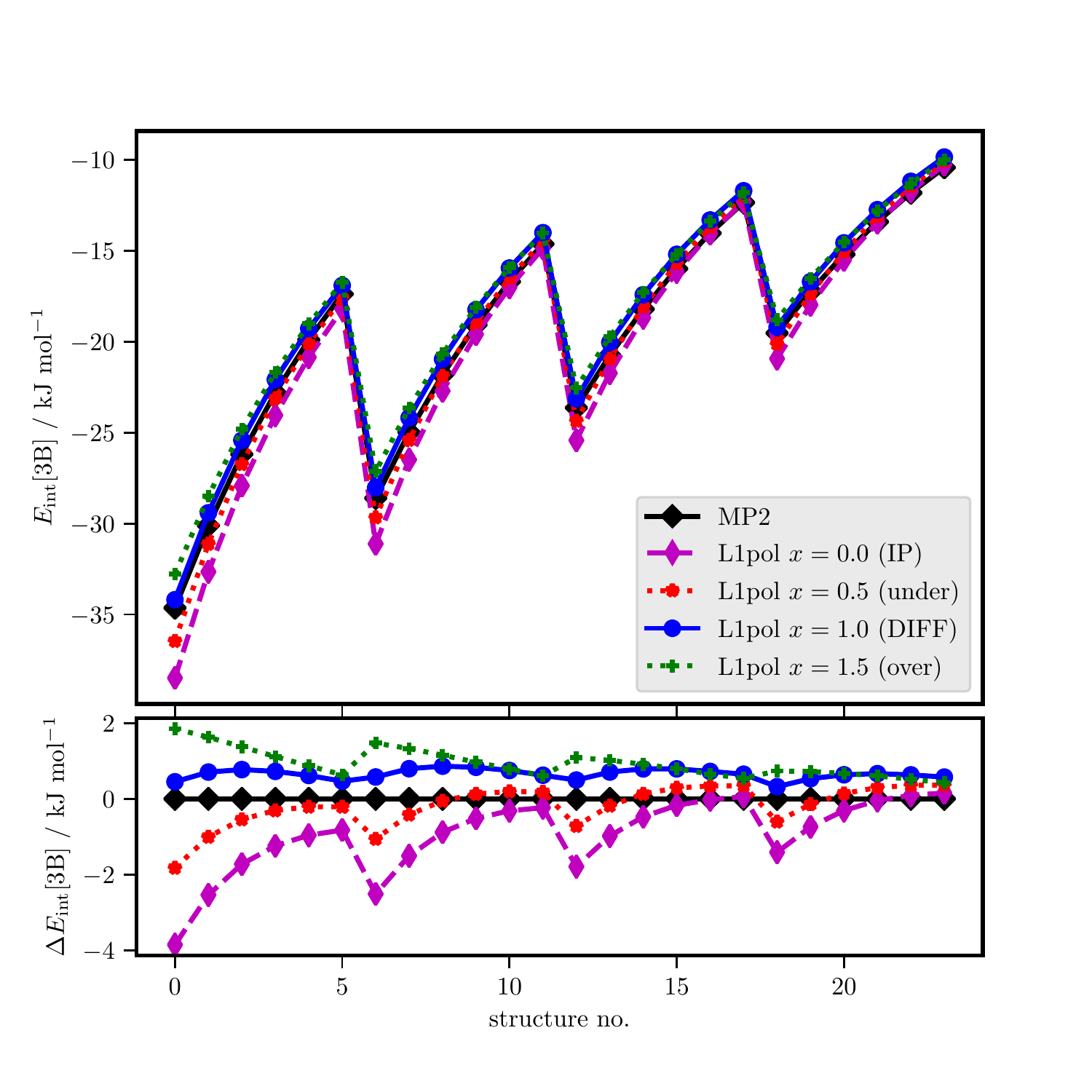}
		\includegraphics[width=0.48\textwidth]{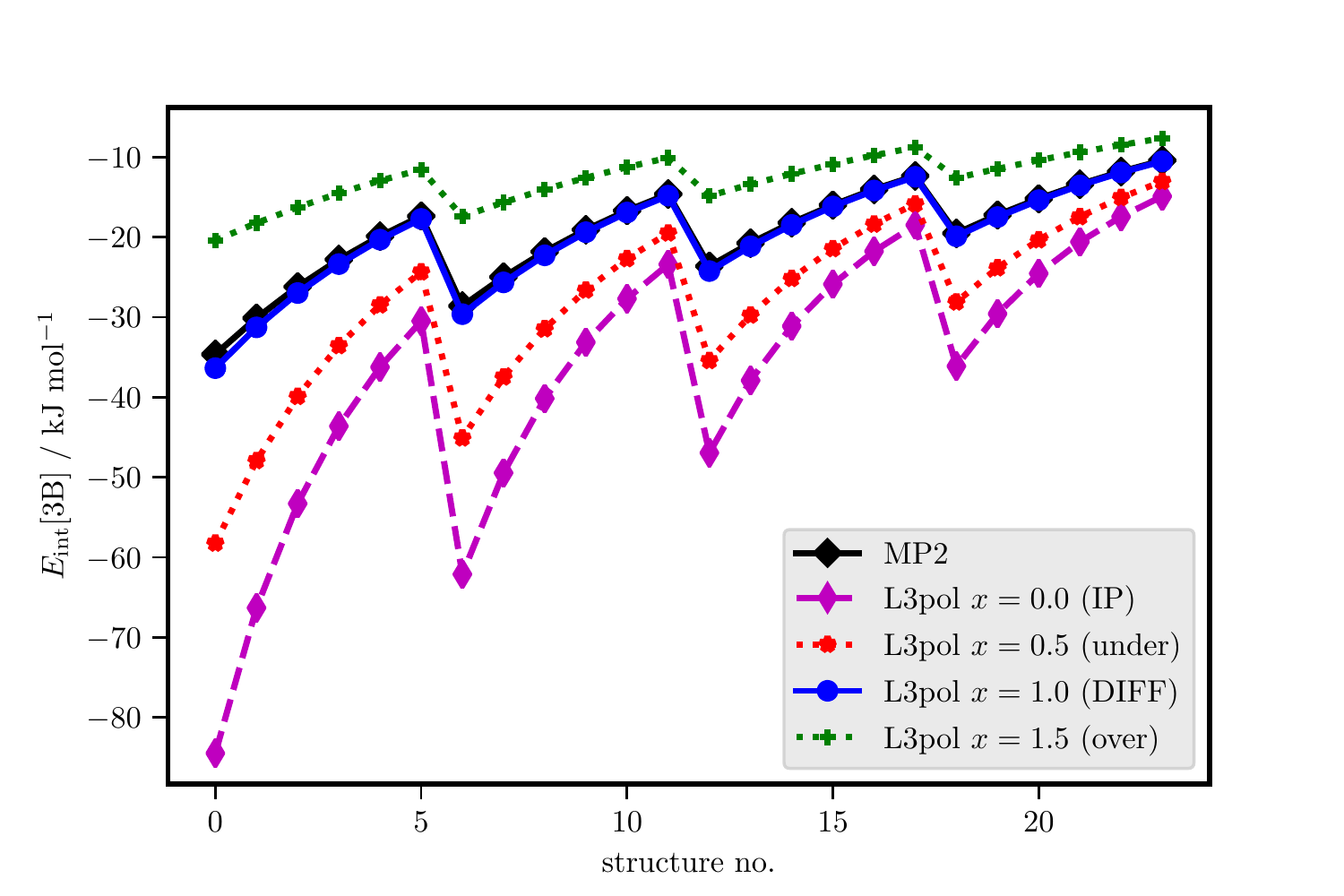}
	\end{center}
	\caption{
        Trimer three-body energies using all
	    seven models. Geometries, MP2 and AMOEBA energies from Liu \etal 
	    \cite{liu2017capturing}.
        For the L1pol models the three-body non-additive energies are
        also displayed relative to those from MP2.
    }
	\label{fig:trimer_3B_energies_var_damping}
\end{figure}

In \figrff{fig:trimer_3B_energies_var_damping} we display the 3B non-additive energies 
from the L1 and L3 polarization models with the alternative damping models
described in \secrff{sec:alternative_pol_damping}. 
The models with maximum rank 1 (L1pol), that is, those that include dipole-dipole polarizabilies
only are not very sensitive to the choice of damping.
The 3B energies for these models vary by just over 5 \kJmol in the worse case, but 
by around 2 \kJmol for the majority of the 24 trimers. 
This shows that, at least for the trimers, the dipole-dipole polarization models are
not too sensitive to the choice of damping model.

On the other hand, also shown in \figrff{fig:trimer_3B_energies_var_damping} is the 
sensitivity of the polarization models with maximum rank 3 (L3pol) to the choice of damping,
and here we see a very different picture. 
The DIFF-L3pol model with its damping determined by fitting to the \EINDreg{2} energies
is the best, with a near perfect agreement with the MP2/CBS reference energies.
All the other damping models result in vastly different 3B non-additive energies.
The IP-based damping model ($x=0.0$), which reproduces \EIND{2} for the dimers
(see \figrff{fig:IND_INDreg_and_models}), results in an overestimation of the 
3B non-additivity by more than a 100\%, and for the $x=0.5$ model the overestimation is
over 30\%. 
At the other end, the over-damped $x=1.5$ model underestimates the non-additivity by
more than 30\%. 

\begin{figure}
	\begin{center}
		\includegraphics[width=0.48\textwidth]{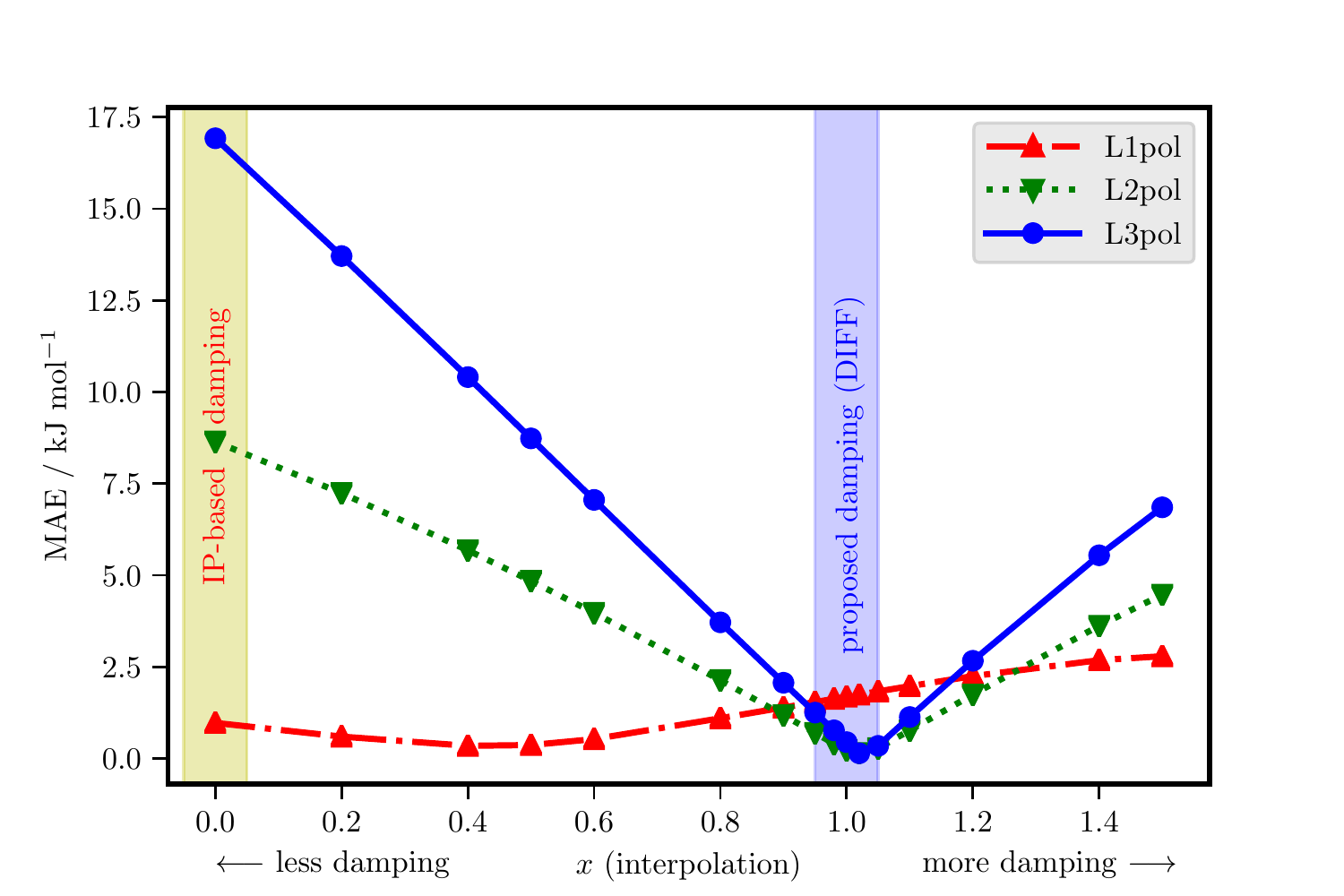}
	\end{center}
	\caption{
        Sensitivity of the 3-body energy of the trimers from the set of Liu \etal
        \cite{liu2017capturing} to changes in the damping model. 
        The proposed (DIFF) damping model ($x=1$) is indicated in blue, and
        the IP-based model ($x=0$) is indicated in yellow.
	}
	\label{fig:trimer_3B_MAEs_var_damping}
\end{figure}

In \figrff{fig:trimer_3B_MAEs_var_damping} we display the mean-absolute errors (MAEs)
of the three polarization models as a function of the interpolation parameter $x$
(see \eqrff{eq:pol_damping_interpolation}). 
We see that for the L1 polarization model the MAE shows a shallow minimum at $x=0.4$, that is,
for this model and for the timers used in this test, the dipole-dipole polarization model 
needs to be under-damped.
However, for the L2 and L3 polarization models there is a very clear and sharp minimum in the
MAE very close to $x=1.0$, which is the (DIFF) damping obtained using the procedure described in 
\secrff{sec:pol_model_theory}.

The water trimers tested thus far are all based on the most stable trimer
configuration and all have negative three-body non-additive energies.
Akin-Ojo and Szalewicz \cite{akin-ojo_how_2013} have provided 
a more extensive set of 600 trimers obtained from snapshots taken from a 
water simulation. These include not only clusters with an attractive 3B contribution,
but also those with a repulsive 3B energy, and consequently present a more challenging
test for the many-body non-additive models.
As with the trimers from Liu \etal (above), the monomer geometries used in this larger
set of timers is the same as that used in the construction of the DIFF models.

\begin{figure}
	\begin{center}
		\includegraphics[width=0.48\textwidth]{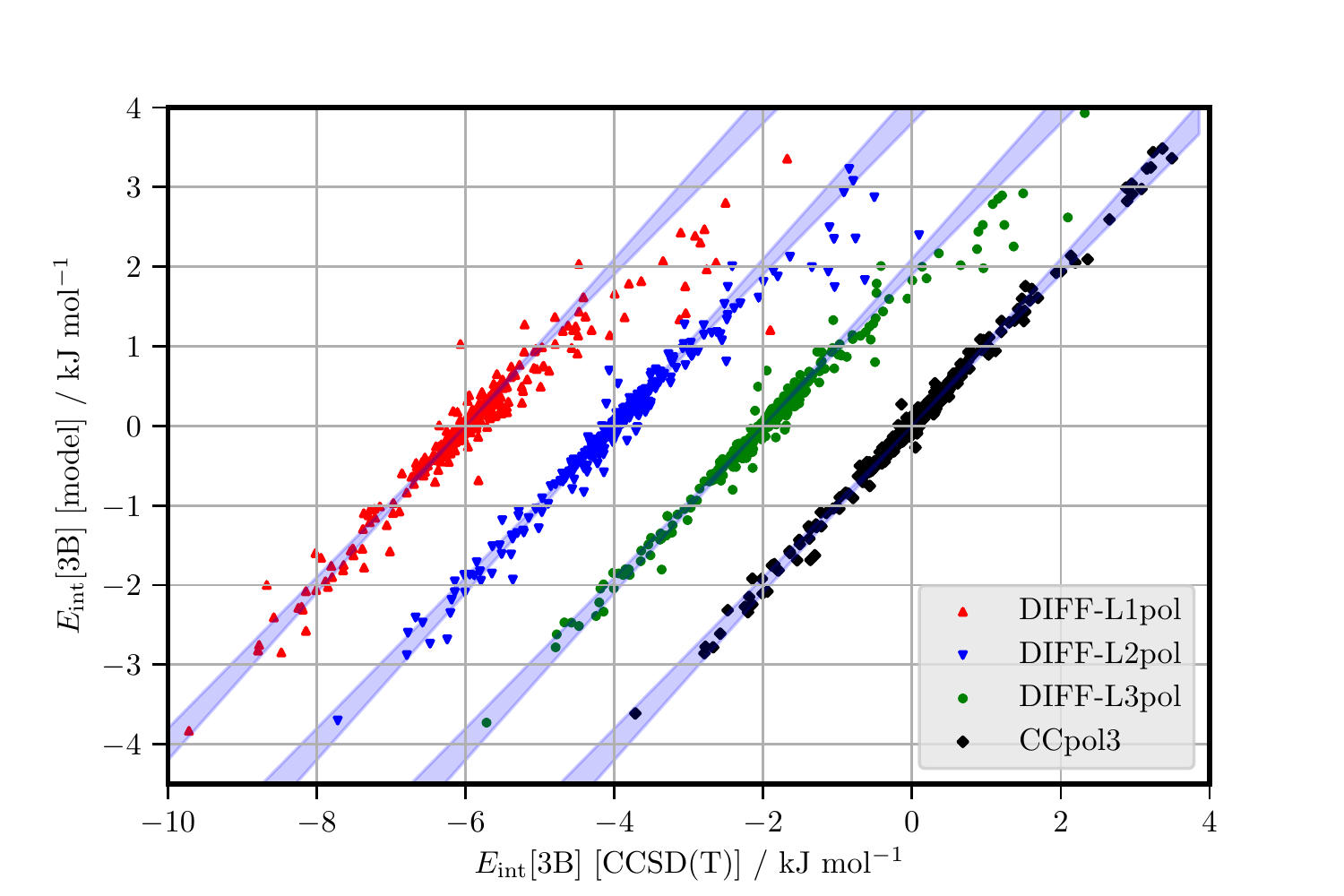}
	\end{center}
	\caption{
        Trimer three-body non-additive energies for the 600 trimer Akin-Ojo \& Szalewicz
        \cite{akin-ojo_how_2013} data set. Comparison is made to the 3B energies from the 
        CCpol23+ model from G{\'o}ra \etal \cite{gora_predictions_2014}
        Starting from the right, the datasets are offset by multiples of $-2$ \kJmol 
        along the $x$-axis.
    }
	\label{fig:trimers_600}
\end{figure}

In \figrff{fig:trimers_600} we display the 3B non-additive energies from the three DIFF
models against the CCSD(T) reference energies from Akin-Ojo and Szalewicz. 
All three DIFF models show a good correlation with the reference energies, with the scatter
progressively decreasing as the maximum rank of the polarizabilities increases.
The DIFF models are able to describe the attractive non-additive energies more accurately,
with errors increasing for the repulsive energies. 
However we clearly see that as the rank of the model increases, so does the ability of the
model to describe the repulsive non-additive energies: DIFF-L1pol shows sizable scatter
and underestimation of the 3B non-additive energy above 1 \kJmol, but the scatter is reduced 
in the L2pol and L3pol models which are in good agreement with the CCSD(T) references even for trimers
with $\EintB{3} \approx 2$ \kJmol.

For comparison we also include the CCpol3 3B model energies in \figrff{fig:trimers_600}.
This model, from G{\'o}ra \etal \cite{gora_predictions_2014} is a 12-dimensional three-body 
non-additive model fitted to 71,000 trimers as described in the Introduction. 
We can see that while CCpol3 is indeed better than all the DIFF models, it is not significantly
better than the DIFF-L3pol model, except for the few trimers with 3B energies larger than 2 \kJmol. 
The mean-absolute and root-mean-square errors (MAE and RMSE) for these models are shown in 
\tabrff{tab:600_trimer_mae_rmse}. 
The differences in MAEs are quite small, particularly between CCpol3 and DIFF-L3pol.
However, the increased overestimation of the 3B energies above 2 \kJmol causes an increase in
the RMSE in DIFF-L3pol compared with CCpol3.
Despite this, it should be evident that if we are concerned primarily with trimers with an
attractive or weakly repulsive three-body non-additivity, then the CCpol3 model is comparable
in accuracy to the DIFF-L3pol and DIFF-L2pol models.
In fact, given that no trimers were used in the determination of any of the DIFF models, their 
predictive power is remarkable.

\begin{figure}
	\begin{center}
		\includegraphics[width=0.48\textwidth]{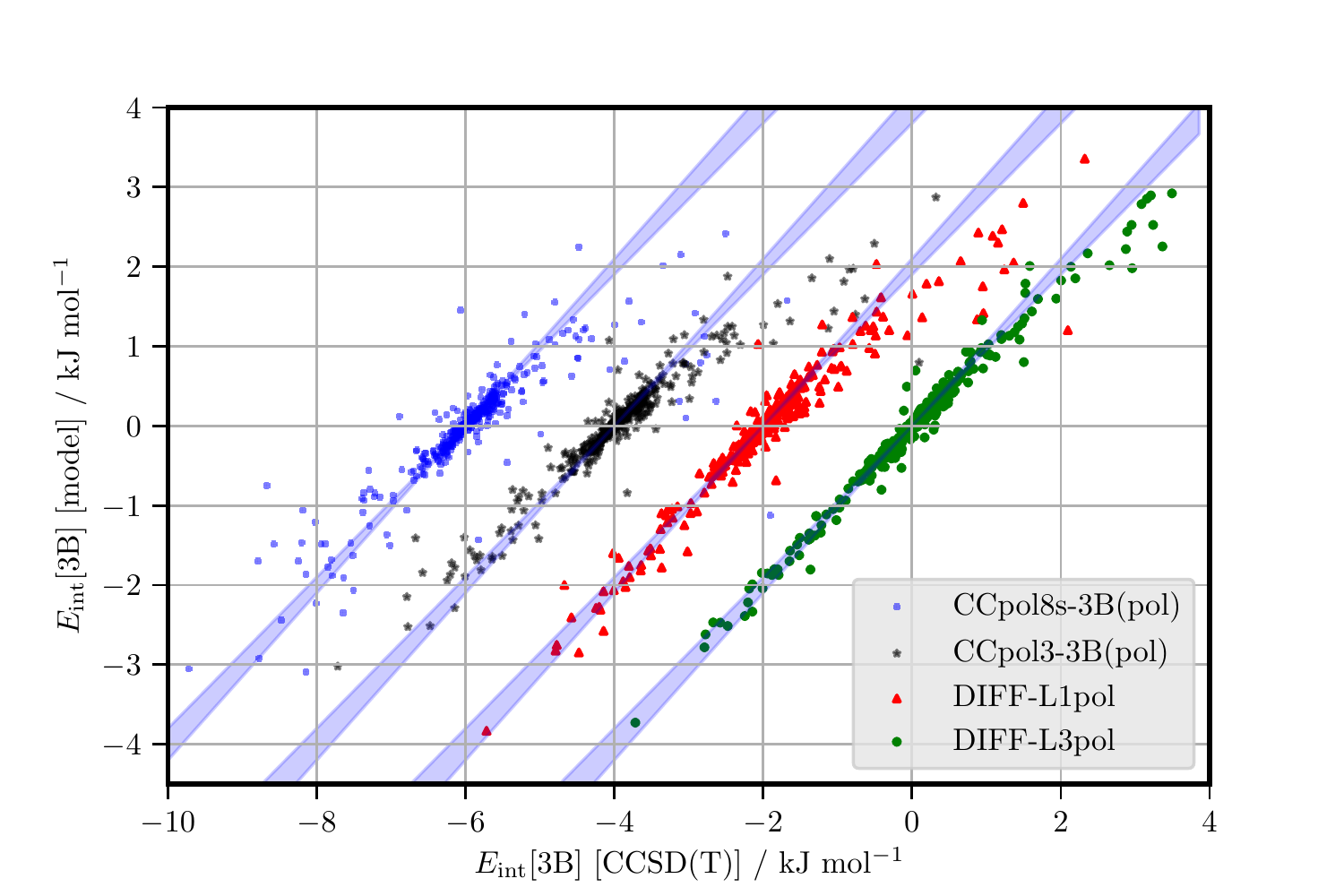}
		\includegraphics[width=0.48\textwidth]{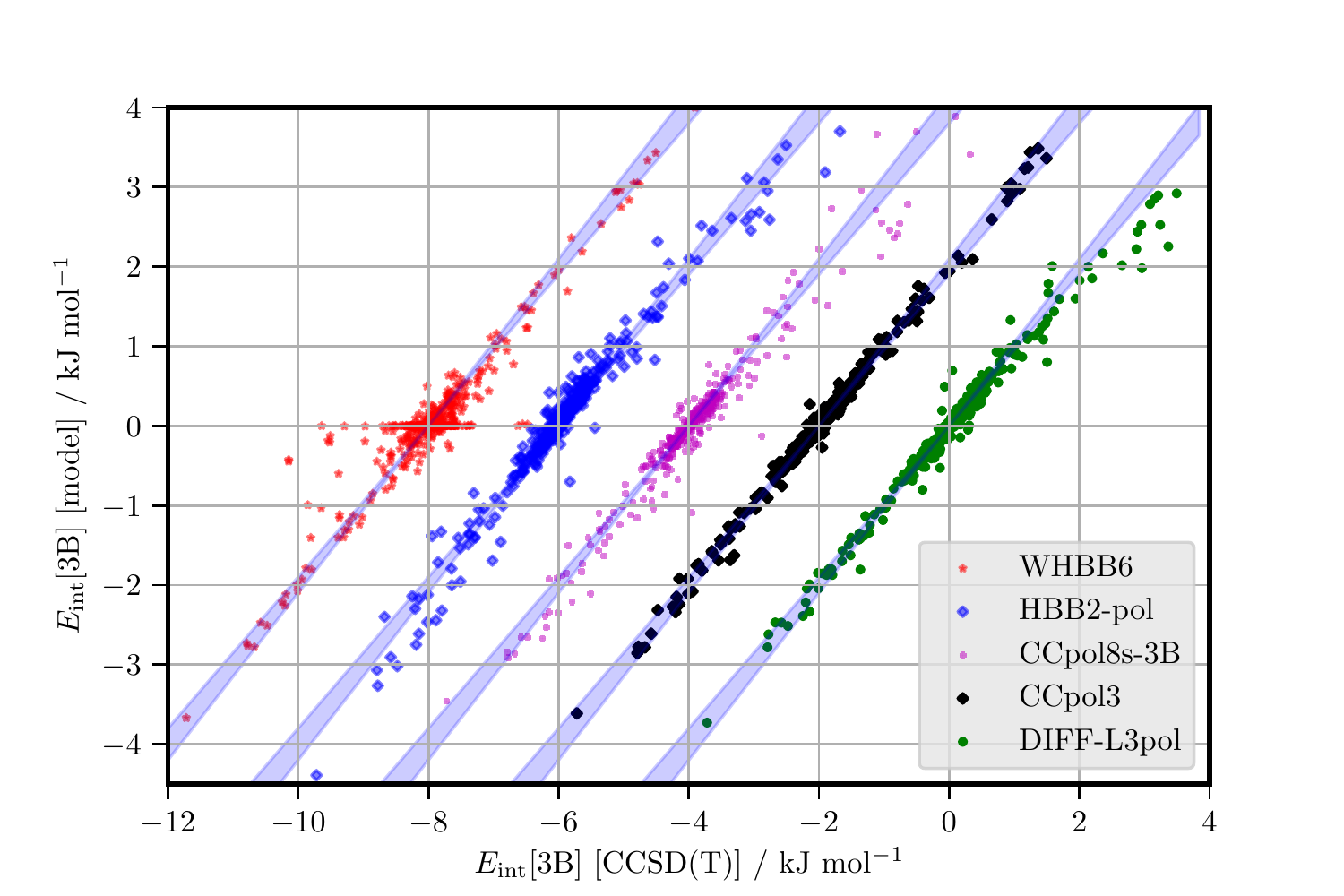}
	\end{center}
	\caption{
        Comparison of the DIFF-L3pol and DIFF-L1pol models with the polarization models from CCpol23+ 
        and CCpol8 (upper panel), and the 3B potentials from HBB2, WHBB6 and CCpol8s
        (lower panel).
        Trimer three-body non-additive energies for the 600 trimer Akin-Ojo \& Szalewicz
        \cite{akin-ojo_how_2013} data set.
        Data for all but the DIFF models was taken from G{\'o}ra \etal \cite{gora_predictions_2014}
        Starting from the right, the datasets are offset by multiples of $-2$ \kJmol 
        along the $x$-axis.
    }
	\label{fig:trimers_600_comparison_CC_HBB2_WHBB6}
\end{figure}

The Akin-Ojo and Szalewicz data set is particularly significant as in a detailed
study of polarization modelling using the damped Drude oscillator approach, they concluded
\cite{akin-ojo_how_2013} that classical polarization models cannot be used to model
the non-additive energies in water, particularly the exchange non-additivity.
Here we have shown that this is not the case: the classical polarization models we have constructed
can all describe the three-body non-additivity, except for the trimers with the most repulsive
\EintB{3} energies. 
Why did Akin-Ojo and Szalewicz come to this perhaps too pessimistic conclusion?
Some insight may be obtained from Fig.~1 from their paper, and also from the polarization 
energy models that are part of CCpol3 and CC-pol-8s. 3B energies from the latter two
are visualised in \figrff{fig:trimers_600_comparison_CC_HBB2_WHBB6}. 
These are both significantly poorer at modelling the reference CCSD(T) 3B energies compared with
DIFF-L3pol or even DIFF-L1pol, which is similar in complexity. 
The MAE/RMSE errors for the CCpol3-3B(pol) and CC-pol-8s-3B(pol) models are shown in 
\tabrff{tab:600_trimer_mae_rmse} and these are much larger than those from the DIFF models, 
in fact, they are substantially larger than those from DIFF-L1pol. 
It is not clear why the polarization models in the CCpol3 and CC-pol-8s models are not
more accurate. It could be the way in which the damping was modelled or the way in which the
molecular polarizabilities and multipoles were chosen. But whatever the cause, the poor 
performance of these models seems to have caused Akin-Ojo and Szalewicz to be over pessimistic
about the accuracies that can the attained by a well constructed polarization model, and 
any of the DIFF models may serve as a counterexample. 

\begin{table}[ht]
	\begin{tabular}{lll}
        \toprule
        Model             & MAE      & RMSE     \\
        \midrule
        DIFF-L1pol        & 0.091    & 0.235    \\
        DIFF-L2pol        & 0.064    & 0.163    \\
        DIFF-L3pol        & 0.058    & 0.141    \\
        \midrule
        \multicolumn{3}{l}{ \small{Polarization part of CCpol models:}} \\
        CCpol3-3B(pol)    & 0.121    & 0.307    \\
        CC-pol-8s-3B(pol) & 0.159    & 0.449    \\
        \midrule
        \multicolumn{3}{l}{ \small{Explicit 3B potentials:}} \\
        CCpol3            & 0.042    & 0.065    \\
        CC-pol-8s         & 0.094    & 0.175    \\
        HBB2-pol          & 0.082    & 0.157    \\
        WHBB6             & 0.148    & 0.269    \\
        \bottomrule
	\end{tabular}
	\caption{
        Mean-absolute errors (MAEs) and root-mean-square errors (RMSEs) in 3-body energies for the water models
        on the 600 trimer data set from Akin-Ojo \& Szalewicz \cite{akin-ojo_how_2013}.
        Errors are computed against CCSD(T) references and are reported in \kJmol. 
        The DIFF models data have been obtained in this work, all other data was obtained from 
        data supplied by G{\'o}ra \etal \cite{gora_predictions_2014}.
    }
	\label{tab:600_trimer_mae_rmse}
\end{table}

CCpol3 is but one of a group of accurate explicit 3-body non-additive models for water; that is,
the 3-body non-additivity is not modelled through a polarization model, but is fit to a functional
form to include terms such as the exchange and dispersion non-additivities that a 
classical polarization model does not include.
The CC-pol-8s \cite{cencek2008accurate}, HBB2-pol \cite{MeddersBP13}, 
and WHBB6 \cite{wang_flexible_2011} models are amongst the others in this class of explicit 
3-body models, and all are fitted to extensive sets of water trimers. 
In \figrff{fig:trimers_600_comparison_CC_HBB2_WHBB6} we compare these models on the 600 water 
trimers, and MAEs and RMSEs are reported in \tabrff{tab:600_trimer_mae_rmse}.
The CCpol3 model is clearly the best able to model the CCSD(T) reference 3B energies, and this
is no doubt testimony to its careful parametrization on the extensive set of trimers. 
With the exception of CCpol3, the DIFF-L3pol and DIFF-L2pol models are better at reproducing
the reference 3B energies than any of CC-pol-8s, HBB2-pol or WHBB6. The latter fares substantially
worse than DIFF-L1pol. 
Note that HBB2-pol has been superseded by the MB-pol model which has been demonstrated to 
significantly outperform HBB2-pol on the water trimers \cite{babin2014development}.

\subsection{Molecular flexibility and the DIFF models}
\label{sec:flexibility_and_DIFF}

The comparisons made with the extensive set of trimers was straightforward because 
both the Liu \etal \cite{liu2017capturing} and the Akin-Ojo \& Szalewicz \cite{akin-ojo_how_2013}
datasets used trimers with a fixed monomer geometry that was the same as that used in the DIFF
models. 
For the larger clusters this is not the case. Now the monomers are allowed to relax and so we
are led to a question: how are the DIFF models to be used on structures with slightly different
molecular conformations?

One solution would be to transform all flexible monomers in a cluster into the rigid monomer
geometry used in developing the models. Another solution would be to adapt the rigid body model
to the new molecular conformation. The latter approach has the merit that the site-site separations
in the cluster are preserved, however, it is not a priori obvious if the interaction model would
be well-behaved if the expansion centres were moved relative to each other.
To know which of these approaches is best we would need to create a model that includes 
intramolecular flexibility, but this is not the goal of this paper. 
Instead we will rely on the fact that the ISA properties --- multipole moments and perhaps 
even the ISA-Pol polarizabilities --- do not alter significantly on modest changes to the
molecular conformation \cite{VerstraelenAvanSW12} and should be transferable onto the 
molecular conformers without alteration. 
Additionally, we also move the short-range parameters from the DIFF models onto the new molecular
site locations. 
All DIFF model parameters can be defined in the local-axis framework of the molecule, and this
includes the anisotropy parameters. Consequently when the molecular conformation alters and the site
locations change, so do the local-axis frames for the three sites in the water molecules. 
The short-range anisotropy in the Born--Mayer term, as well as the site multipoles and
local polarizabilities, all rotate with the local-axis frames. 
Thus there is a well-defined way of transferring the parameters of any DIFF model onto 
molecules with different conformations, and as long as these conformations are not
too far from the one used in the DIFF model parametrizations, we may expect that the resulting
model will result in sensible interaction energies. 

Clearly the above premise needs to be systematically tested, but we will simply adopt it here 
and instead note that we will now expect to see deviations in the cluster energies because of 
this imposed flexibility of the DIFF models. 

One final point: the DIFF models can be used for {\em intermolecular} geometry optimizations
only. Consequently, when we optimize the clusters geometries, we will do so with the molecular
conformations kept fixed. That is, no intramolecular degrees of freedom will be allowed to vary.

\subsection{The water hexamers}
\label{sec:hexamers}

\begin{figure}
	\begin{center}
		\stackunder[-5pt]{\includegraphics[scale=0.20]{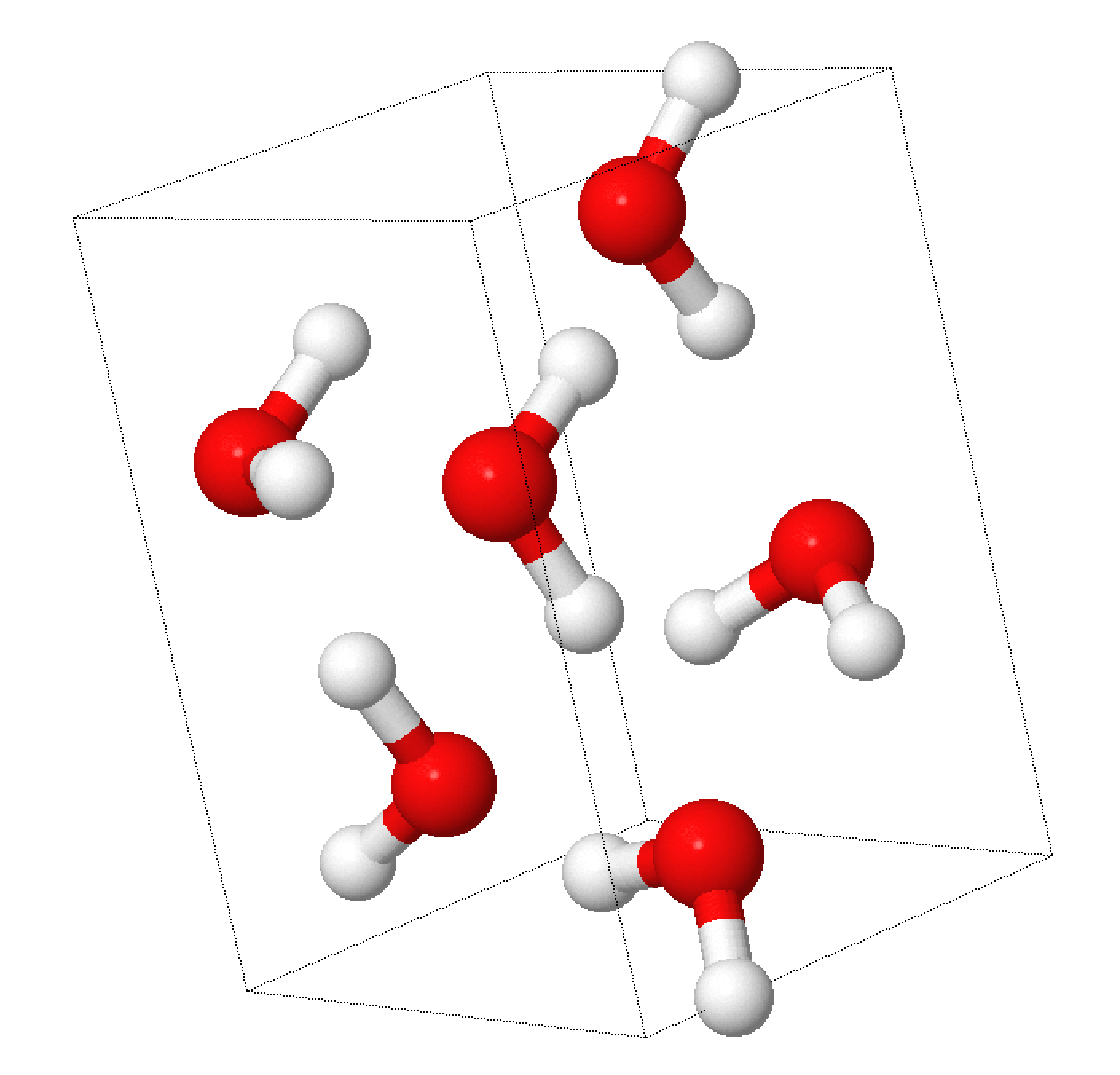}}{(a) prism}
		\stackunder[-5pt]{\includegraphics[scale=0.20]{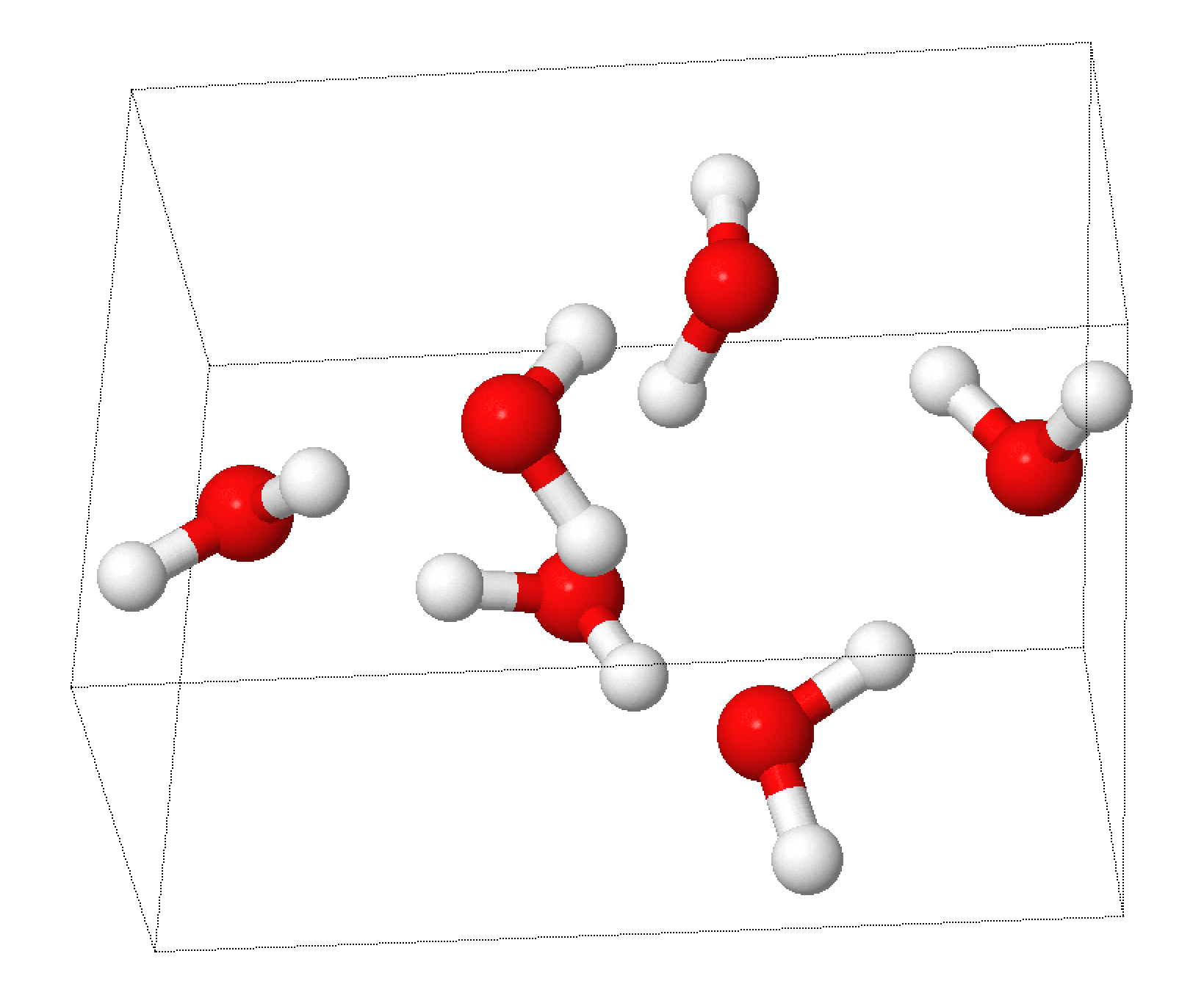}}{(b) cage}
		\stackunder[-5pt]{\includegraphics[scale=0.20]{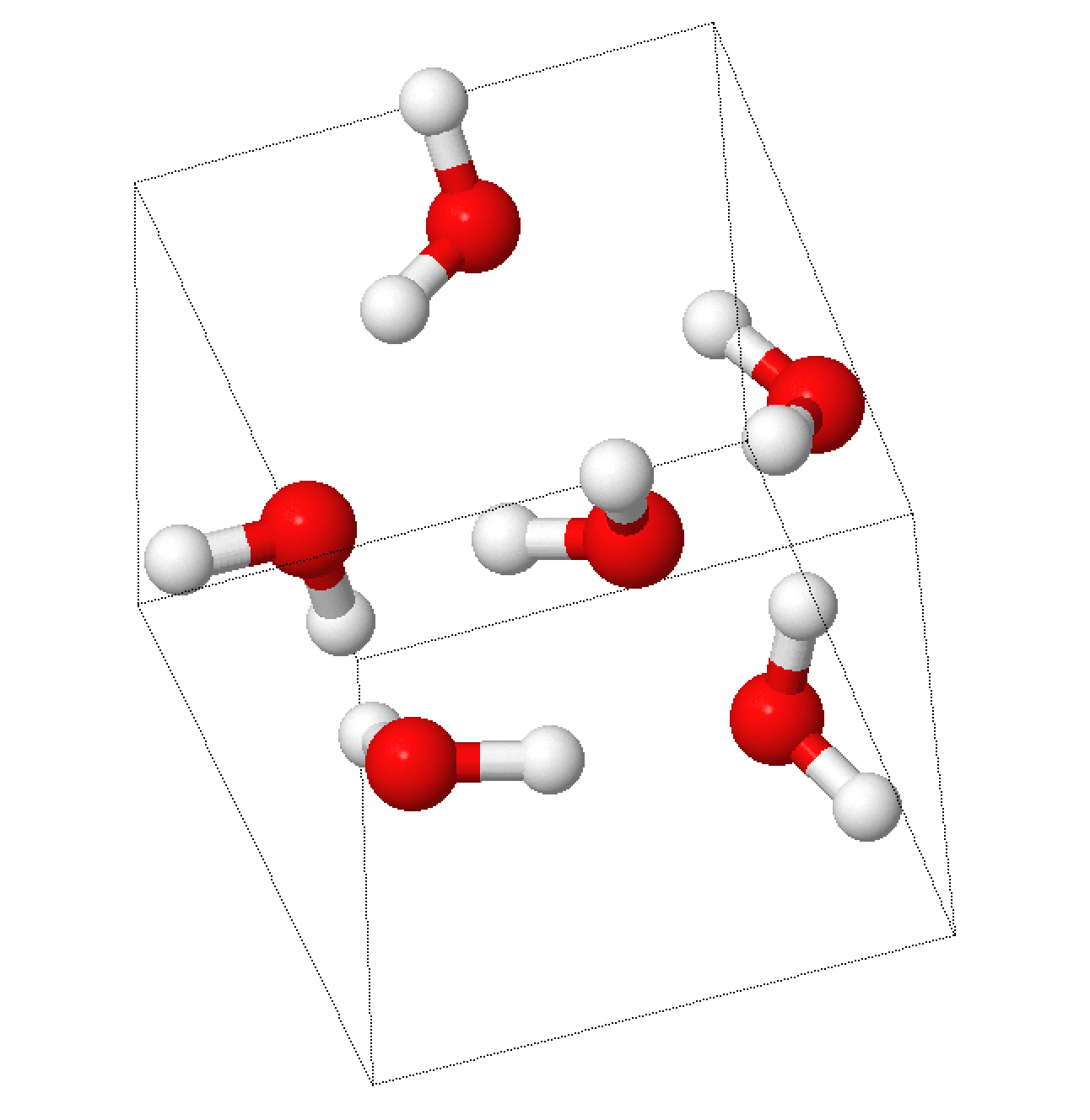}}{(c) bag}
		\stackunder[-5pt]{\includegraphics[scale=0.20]{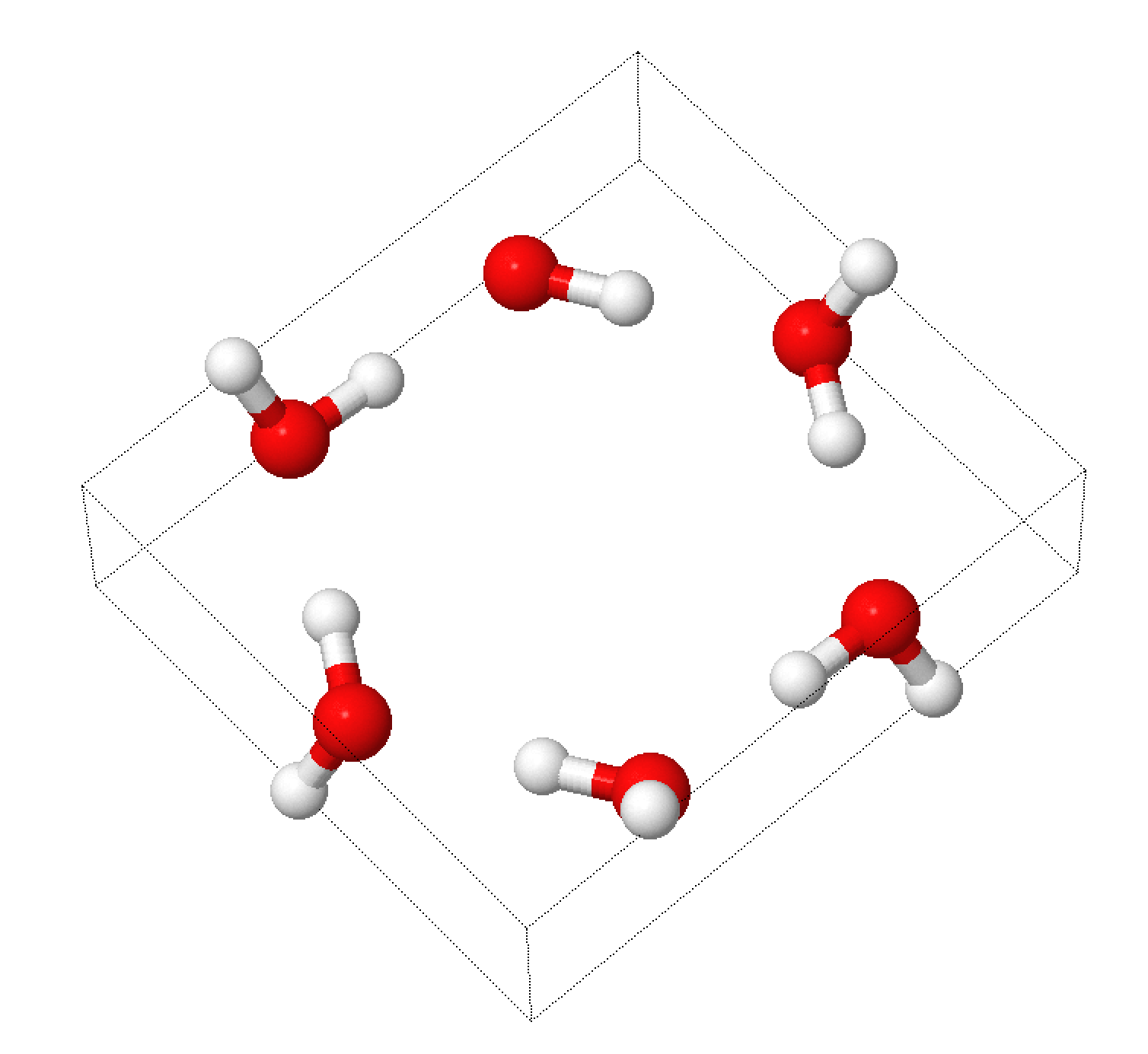}}{(d) cyclic-chair}
		\stackunder[-5pt]{\includegraphics[scale=0.20]{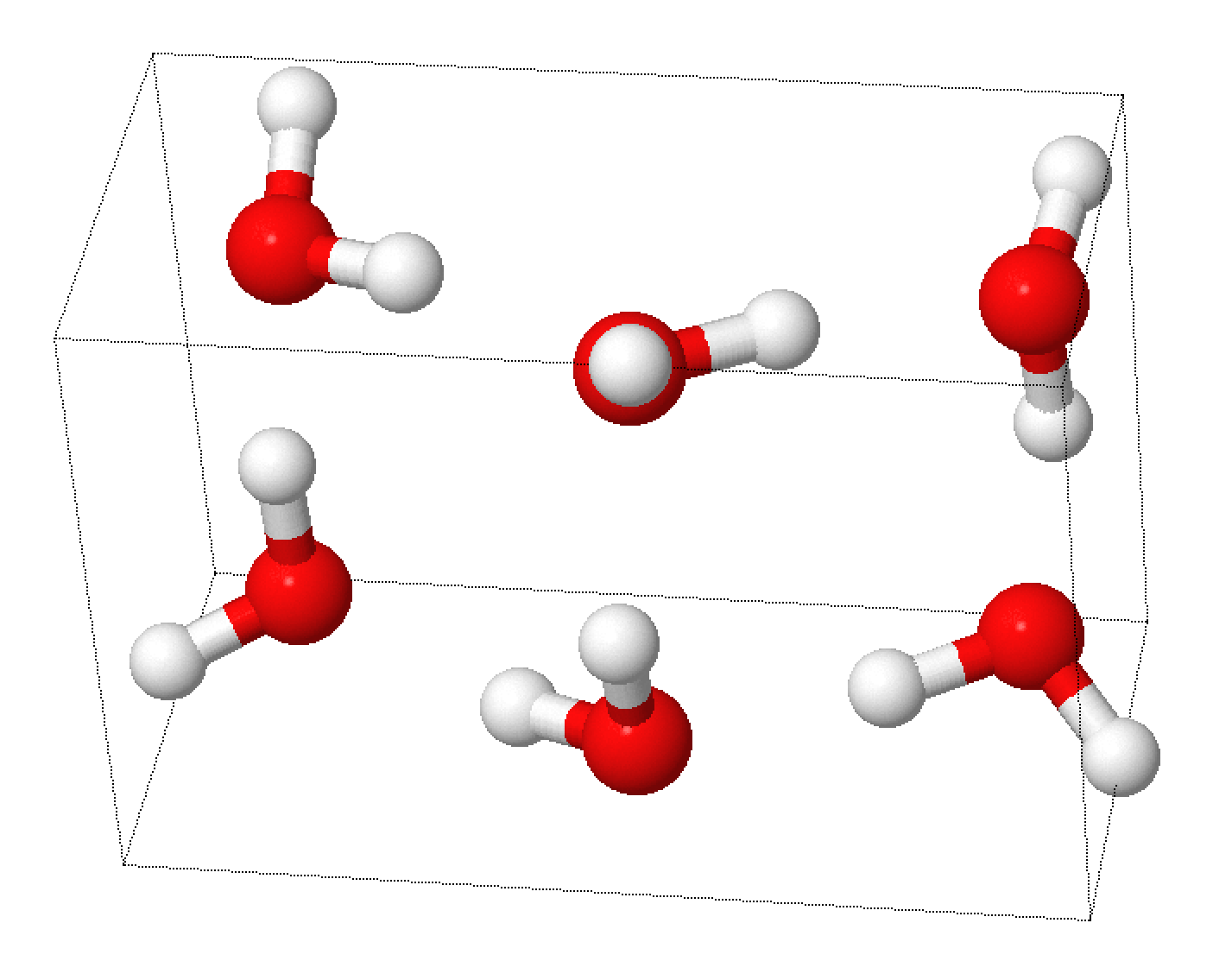}}{(e) book-1}
		\stackunder[-5pt]{\includegraphics[scale=0.20]{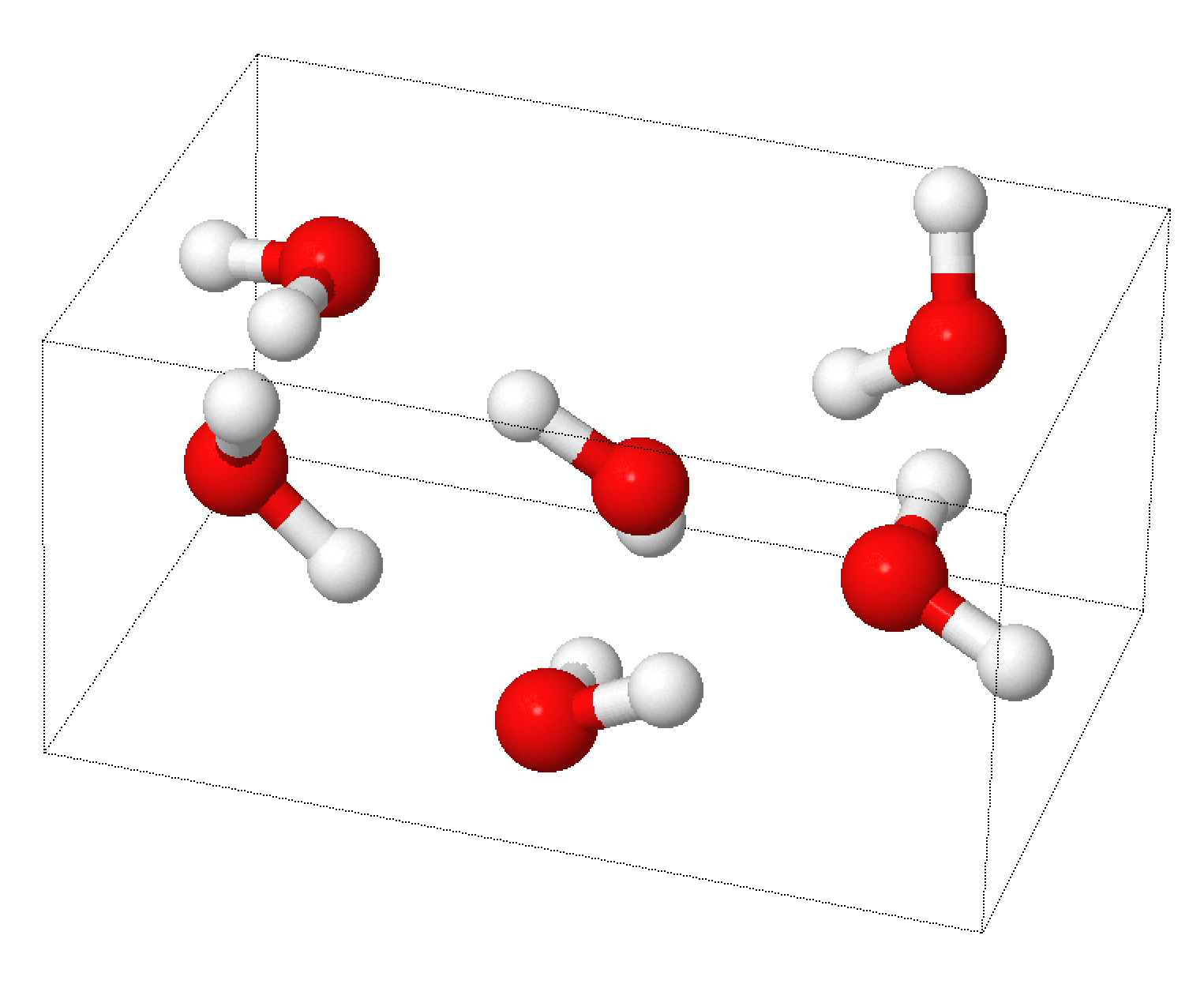}}{(f) book-2}
		\stackunder[-5pt]{\includegraphics[scale=0.20]{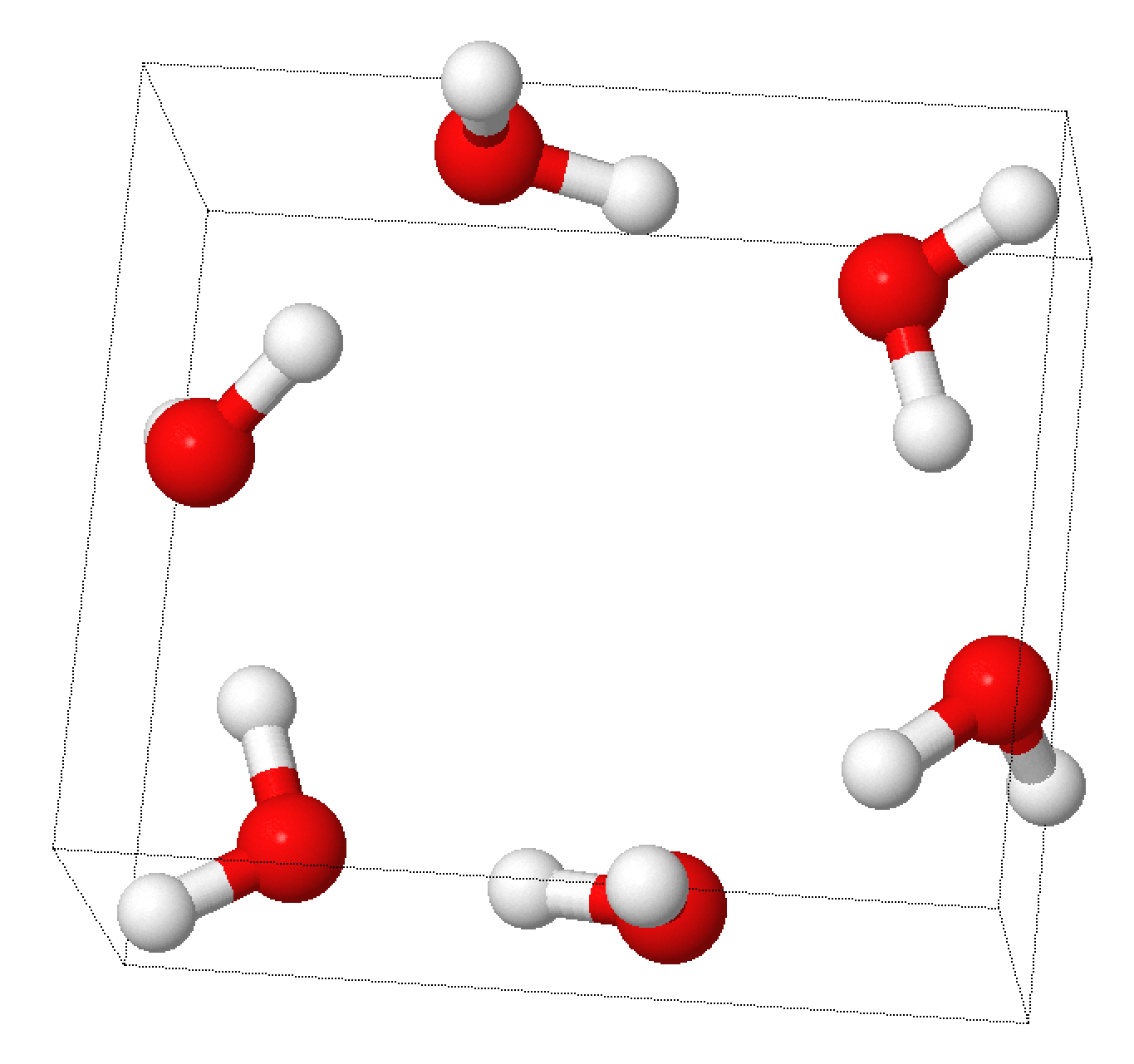}}{(g) cyclic-boat-1}
		\stackunder[-5pt]{\includegraphics[scale=0.20]{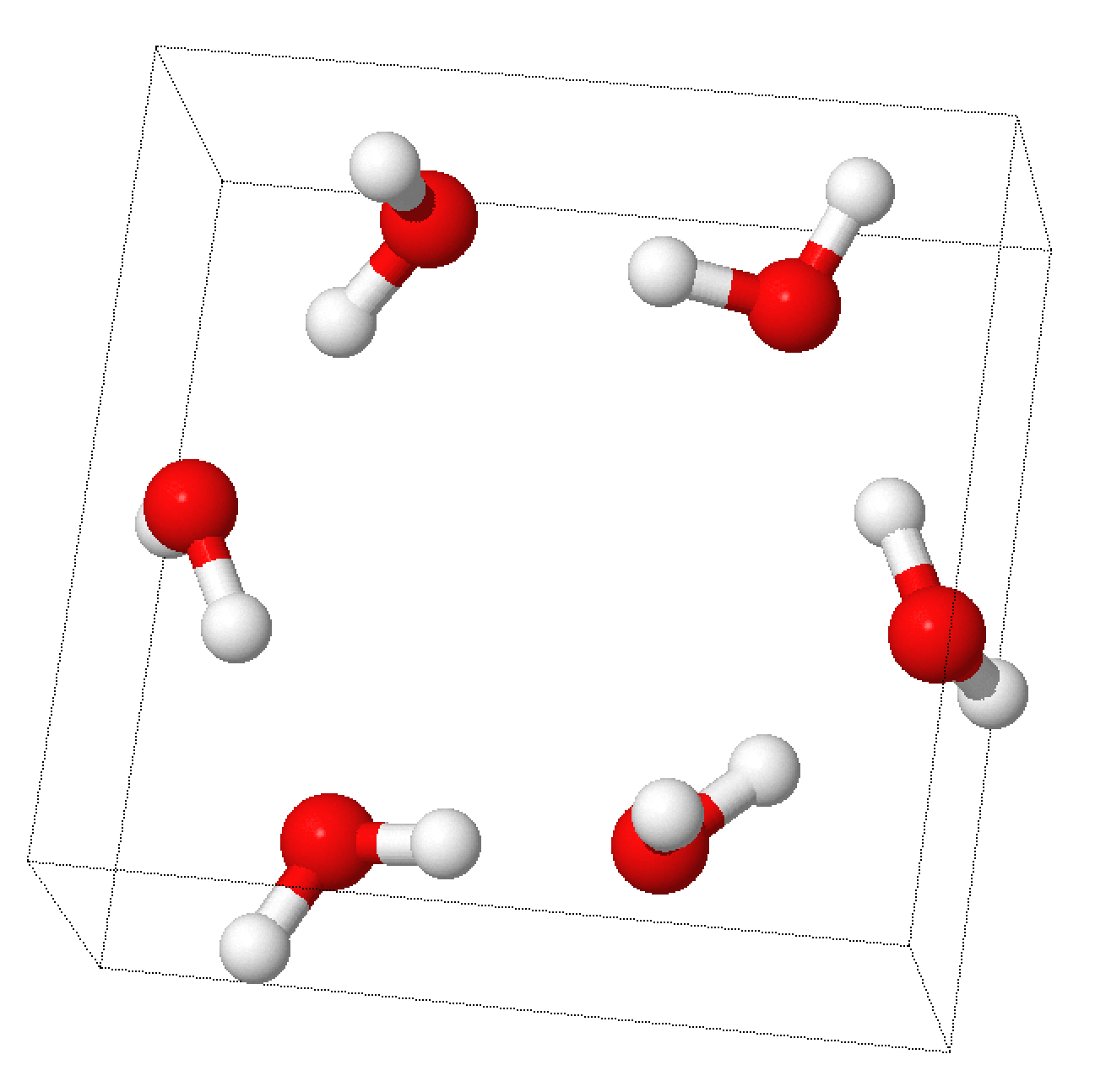}}{(h) cyclic-boat-2}
	\end{center}
	\caption{
        Hexamers used in this study. The structures have been optimized using MP2/CBS by
        Bates \& Tschumper \cite{bates2009ccsd}.
    }
	\label{fig:hexamers_structures}
\end{figure}

The extensive and accurate theoretical data available for the hexamer energies and structures
has meant that every \abinitio model is tested against the hexamers \cite{cisneros_modeling_2016}. 
Additionally, the structural information determined from broadband rotational spectroscopy 
by Perez \etal \cite{perez_structures_2012} means that there is a link to experiment even if
only some structural data (the O\dots O separations) are determined with confidence. 
In this work we use as our references the MP2/CBS optimized hexamer geometries from 
Bates and Tschumper \cite{bates2009ccsd} which are shown in \figrff{fig:hexamers_structures}.
Reference energies are the CCSD(T)-F12 hexamer energies from Medders \etal 
\cite{medders_representation_2015}, who have also computed the many-body decomposition of the total 
intermolecular interaction energies of the hexamer isomers, thus allowing a detailed comparison
of the DIFF models.

\begin{figure}
	\begin{center}
		\includegraphics[width=0.48\textwidth]{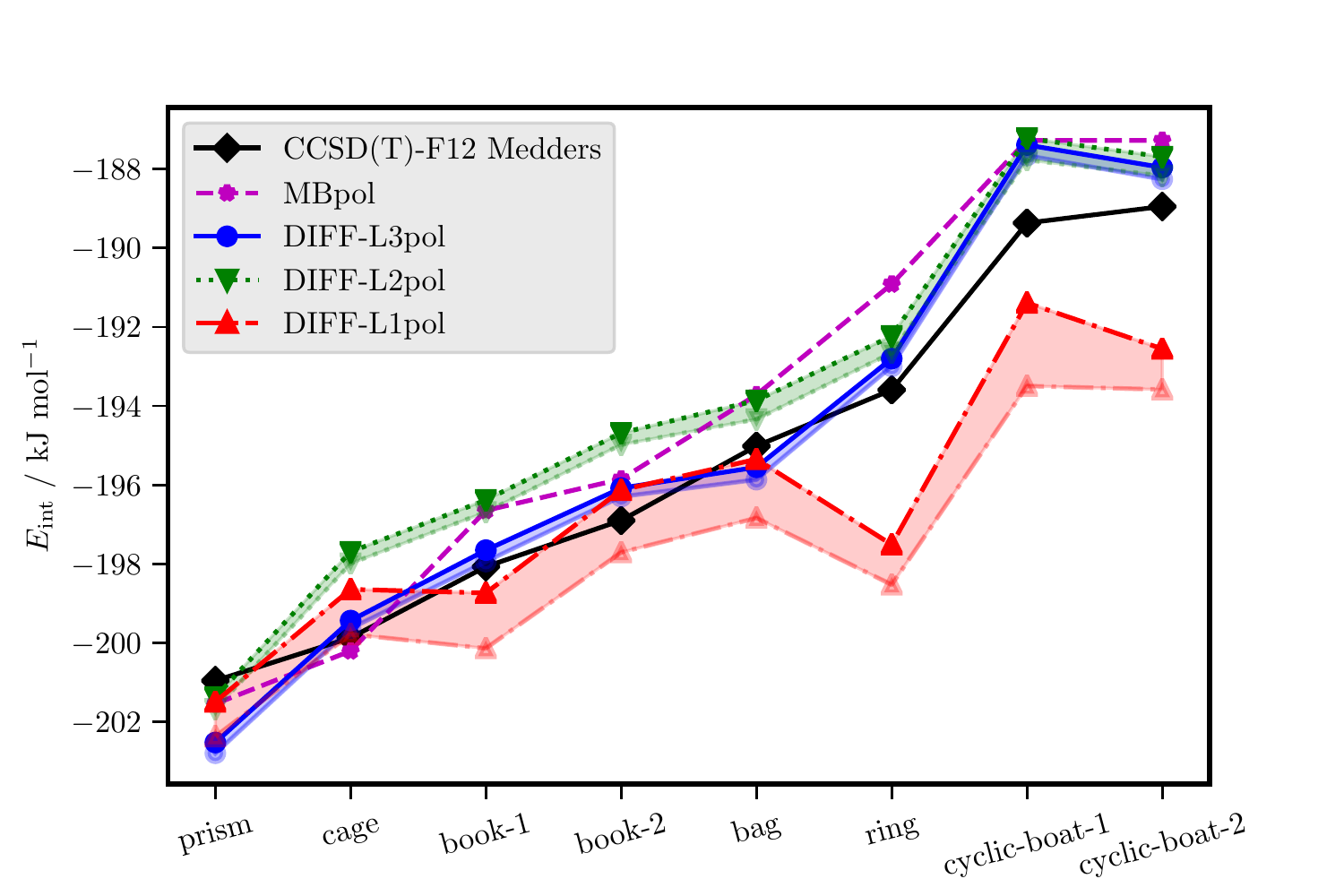}
	\end{center}
	\caption{
        Total interaction energy for hexamers using the DIFF models with
        maximum polarizabilities of maximum rank 1, 2 and 3, all damped with 
        the proposed ($x=1$) DIFF damping models.
        For each of the DIFF models the hexamer geometries have been relaxed
        with the monomers kept rigid. These relaxed interaction energies are indicated
        by the faded points and lines for each of the DIFF models. 
	    Reference interaction energies are the CCSD(T)-F12 energies from Medders \etal
        \cite{medders_representation_2015}.
        Also shown are the MB-pol interaction energies also taken from Medders \etal.
	    The hexamer geometries are from Bates and Tschumper \cite{bates2009ccsd}.
    } 
    \label{fig:hexamers_all_models}
\end{figure}

In \figrff{fig:hexamers_all_models} we display the hexamer intermolecular energies from the 
DIFF models along with the reference CCSD(T)-F12 energies and those from the 
MB-pol model \cite{medders_representation_2015}. 
We do not present the hexamer energies with respect to the prism structure energy as is done
by many authors, but have instead displayed the absolute intermolecular interaction energies.
This is because if the energy of the prism structure happens to be in error, as is the case here,
the relative energies will lead to a falsely pessimistic picture. 
The DIFF-L3pol model gives hexamer energies in good agreement with the CCSD(T)-F12
references with differences in energy of no more than 2 \kJmol at the prism and cyclic-boat-1
structures, and less than 1 \kJmol for the cage to ring structures. 
The DIFF-L2pol model shows somewhat larger errors, but here too the maximum error is only
just over 2 \kJmol. 
On the other hand, while the DIFF-L1pol model shows a remarkable agreement with the reference
energies for the prism to bag structures, this model overestimates the binding energies of
the ring-like structures. 
Both DIFF-L2pol and DIFF-L3pol show the correct energetic ordering of the hexamers up to the
cyclic boat, but this is not the case for DIFF-L1pol.
Notice that both of the higher ranking DIFF models compare favourably to the MB-pol model.

Also shown in \figrff{fig:hexamers_all_models} are the energies of the hexamers with
the intermolecular degrees of freedom relaxed. There are almost no intermolecular geometric changes
on the DIFF-L2pol and DIFF-L3pol surfaces, with O\dots O separations changing no more than $0.03$ \AA.
This is reflected in the small energetic lowering on optimization on these surfaces.
However, for the DIFF-L1pol model the optimized energies can be as much as 2 \kJmol lower, 
and although the O\dots O separations can alter by only $0.04$ \AA\ (similarly to the higher
ranked models), for the more open cyclic-boat structures optimizations using this model
leads to structural torsional changes of around 30\textdegree, that is, the boat structures
invert, while for the DIFF-L2pol and DIFF-L3pol models these changes are only
6\textdegree\ and 3\textdegree. 

\begin{figure}
	\begin{center}
		\includegraphics[width=0.48\textwidth]{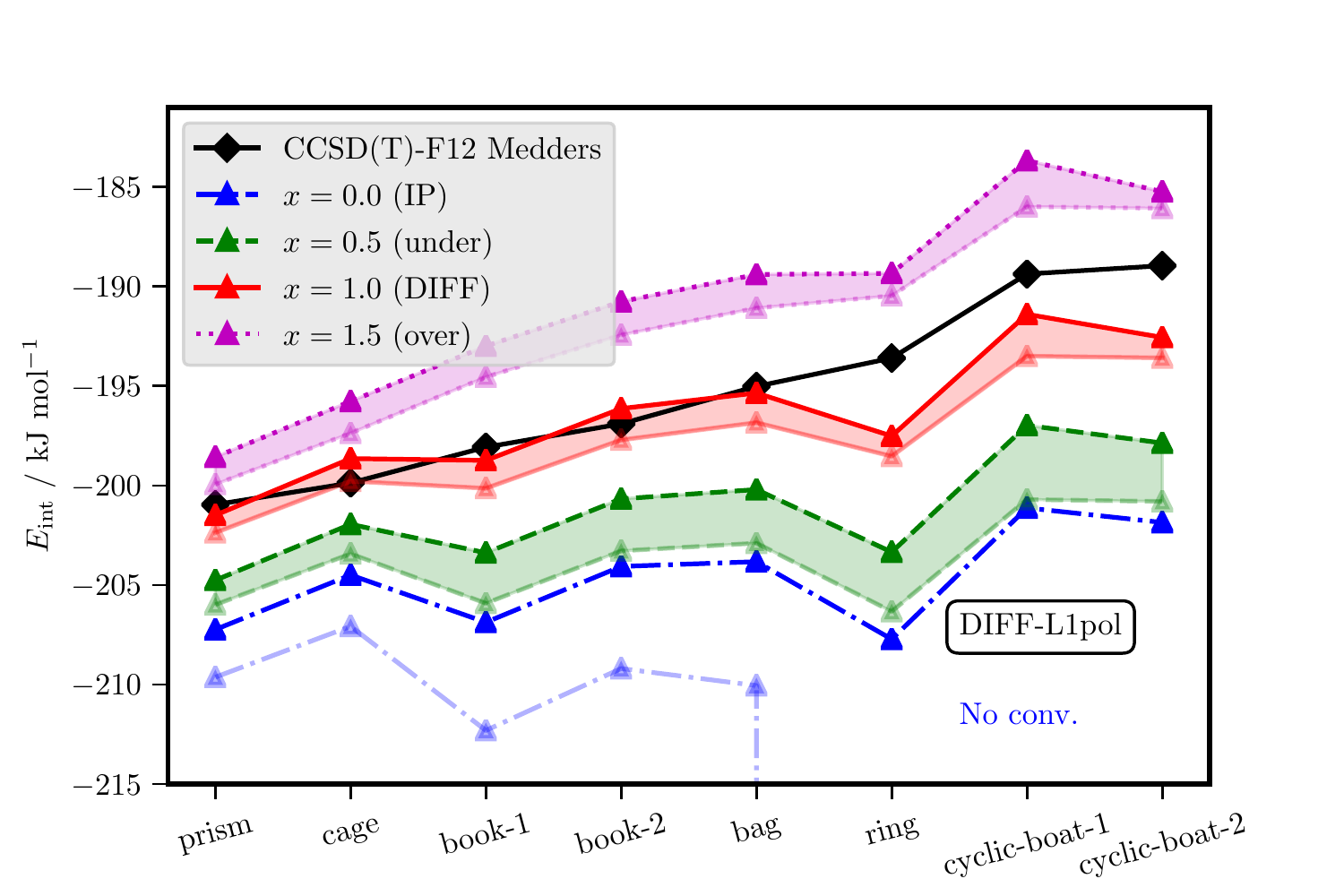}
		\includegraphics[width=0.48\textwidth]{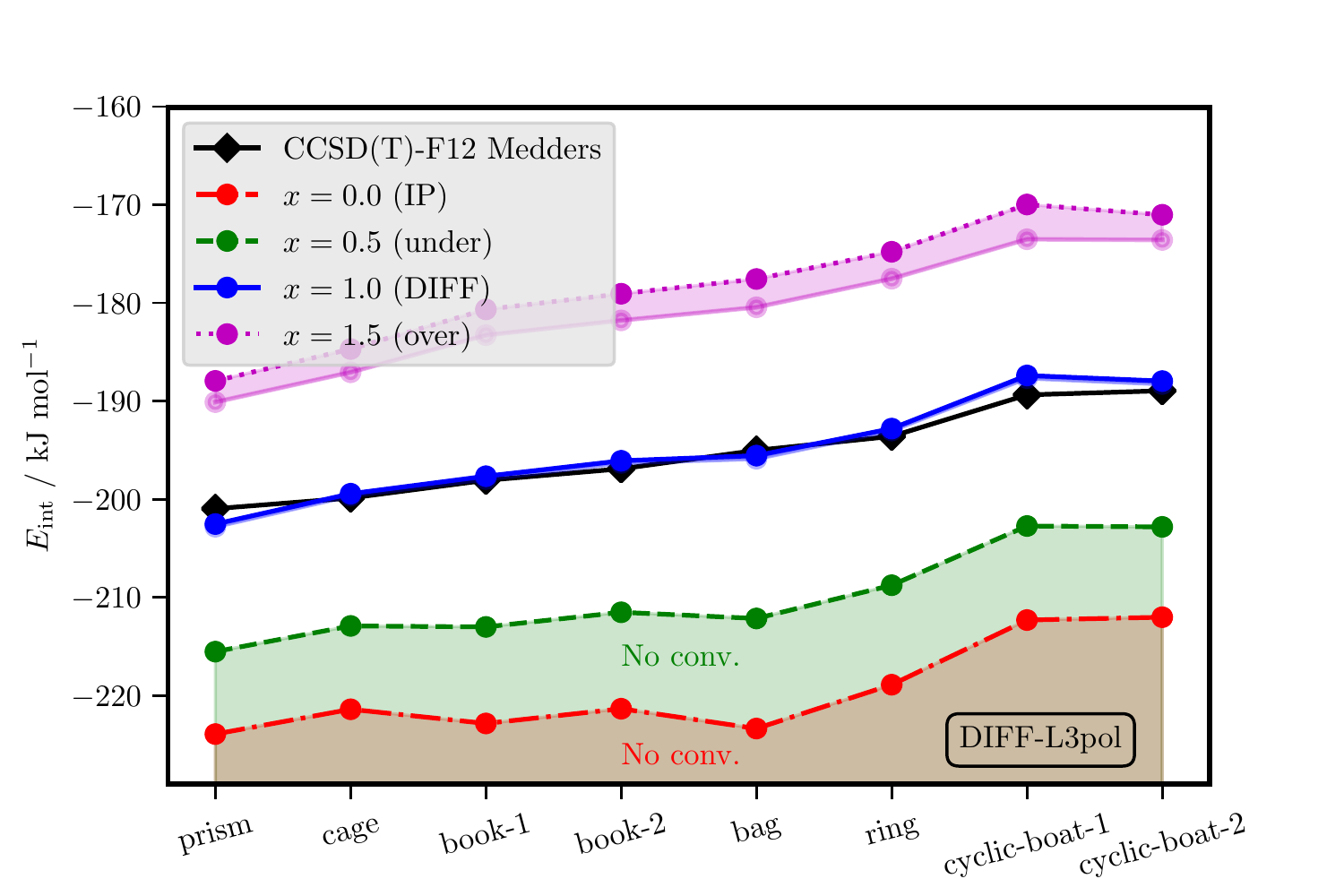}
	\end{center}
	\caption{
        Variations in the hexamer energies as a function of the polarization damping used 
        in the DIFF models.
        The damping models are obtained using the interpolation scheme discussed in the text.
        For all DIFF models the solid points indicate interaction energies computed 
        at the reference geometries from Bates and Tschumper \cite{bates2009ccsd}.
        The faded lines and points indicate the energies of the relaxed structures.
        For some under-damped polarization models the relaxation leads to the
        polarization catastrophe. In these cases geometry convergence is not achieved
        and is indicated with 'No conv.'.
	} 
    \label{fig:hexamers_energies_var_damping}
\end{figure}

\begin{figure}
	\begin{center}
		\includegraphics[width=0.48\textwidth]{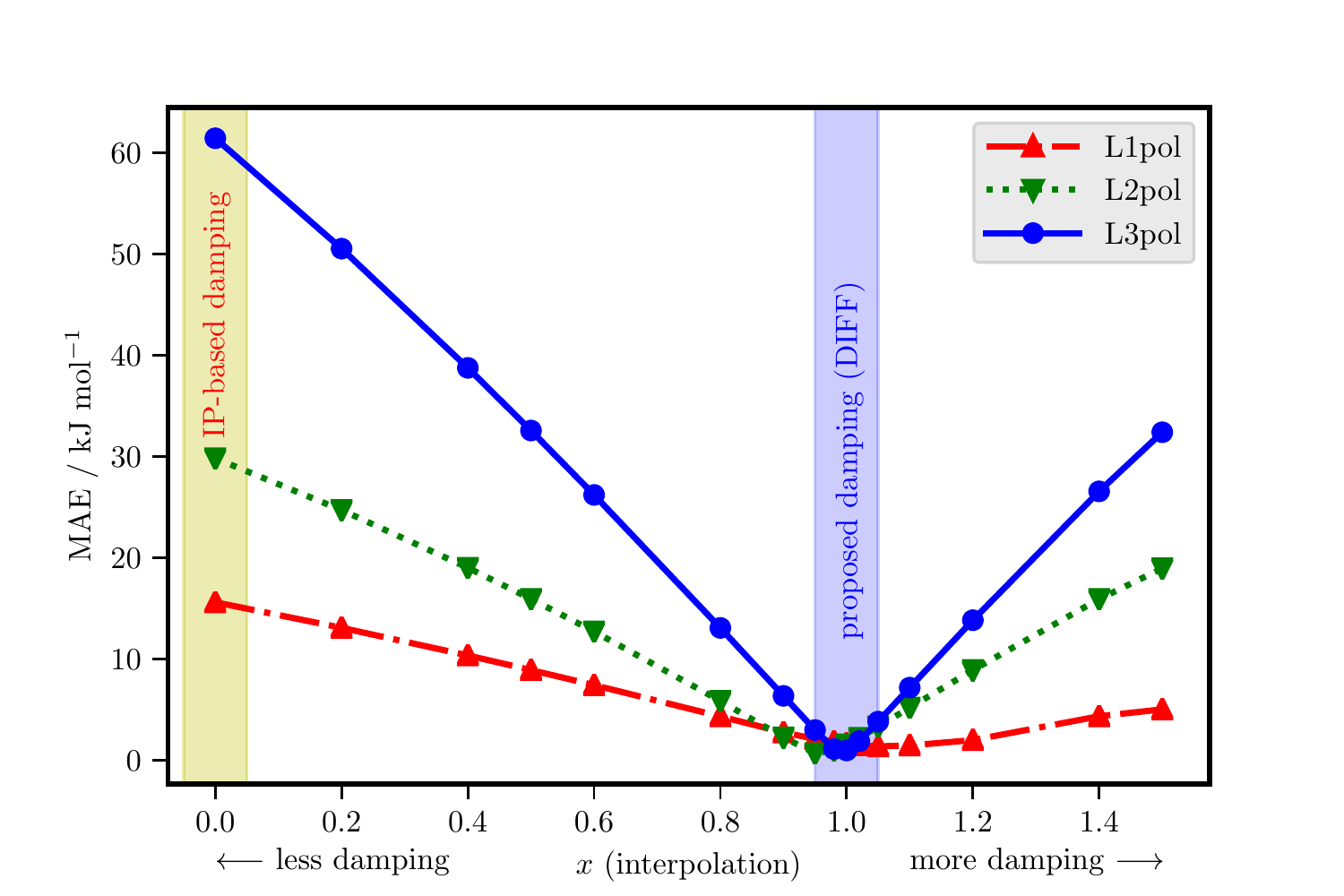}
	\end{center}
	\caption{
        Sensitivity of the many-body polarization energy of the hexamers to changes in the 
        damping model. The proposed damping model ($x=1$) is indicated in blue, and
        the IP-based model ($x=0$) is indicated in yellow.
	}
	\label{fig:hexamers_MAEs_var_damping}
\end{figure}

In \figrff{fig:hexamers_energies_var_damping} we display the hexamer energies for the 
L1pol and L3pol models with the alternative damping models discussed in 
\secrff{sec:alternative_pol_damping}. 
This is done both for the hexamers at the Bates \& Tschumper reference geometries and
for the optimized geometries, where optimizations were possible.
For both the L1 and L3 polarization models, the DIFF damping
models ($x=1.0$) lead to the best agreement with the reference energies.
Unlike the case of the trimers (see \figrff{fig:trimer_3B_energies_var_damping}) where the
changes in damping model made little difference to the energies from the L1 polarization model, 
for the hexamers we see considerably more variations in the energies, with the difference in
energies between structures increasing as the damping model changes.
However, like the case for the trimers, the L3 polarization models show even more sensitivity
to the choice of damping, with the accurate DIFF-L3pol model separated
from the $x=1.5$ and $x=0.5$ models by around 10 \kJmol. 
It is worth recalling that all the models used here were fitted to yield the same two-body
energies: they differ only in how they treat the polarization damping, and this difference 
is magnified substantially in the many-body energies of the clusters.
In \figrff{fig:hexamers_MAEs_var_damping} we display the MAEs for the L1, L2 and L3 polarization
models as a function of interpolation parameter $x$.
The MAE for the L3 model shows a minimum almost exactly at $x=1.0$, while for the L2 model 
the minimum occurs just below 1.0, that is slightly less damping may be needed, and for the
L1 polarization model there is a broad minimum near $x=1.1$, that is more damping is needed.

The alternate damping models not only lead to different cluster energies, but they can also exhibit
very different results for the geometry optimizations of the hexamers.
From \figrff{fig:hexamers_energies_var_damping} we see that the IP-based damping ($x=0.0$)
for both L1 and L3 polarization models leads to the polarization catastrophe:
that is, optimization does not converge and intermolecular bond lengths become arbitrarily small.
The moderately underdamped ($x=0.5$) L3 polarization model also results in 
the polarization catastrophe. 
The DIFF damping model ($x=1.0$) is optimal for the L3 polarization model and close to optimal
for the L1 model. 
While the overdamped ($x=1.5$) L1 model actually shows better hexamer energy ordering and slightly
less structural variations on optimization than the $x=1.0$ DIFF-L1pol model,
for the L3 model the structures tend to expand and show increases in the O\dots O distances
of as much as 0.07 \AA, with large torsional changes of 45\textdegree\ in the cyclic-boat structures. 

\begin{figure}
	\begin{center}
		\includegraphics[width=0.48\textwidth]{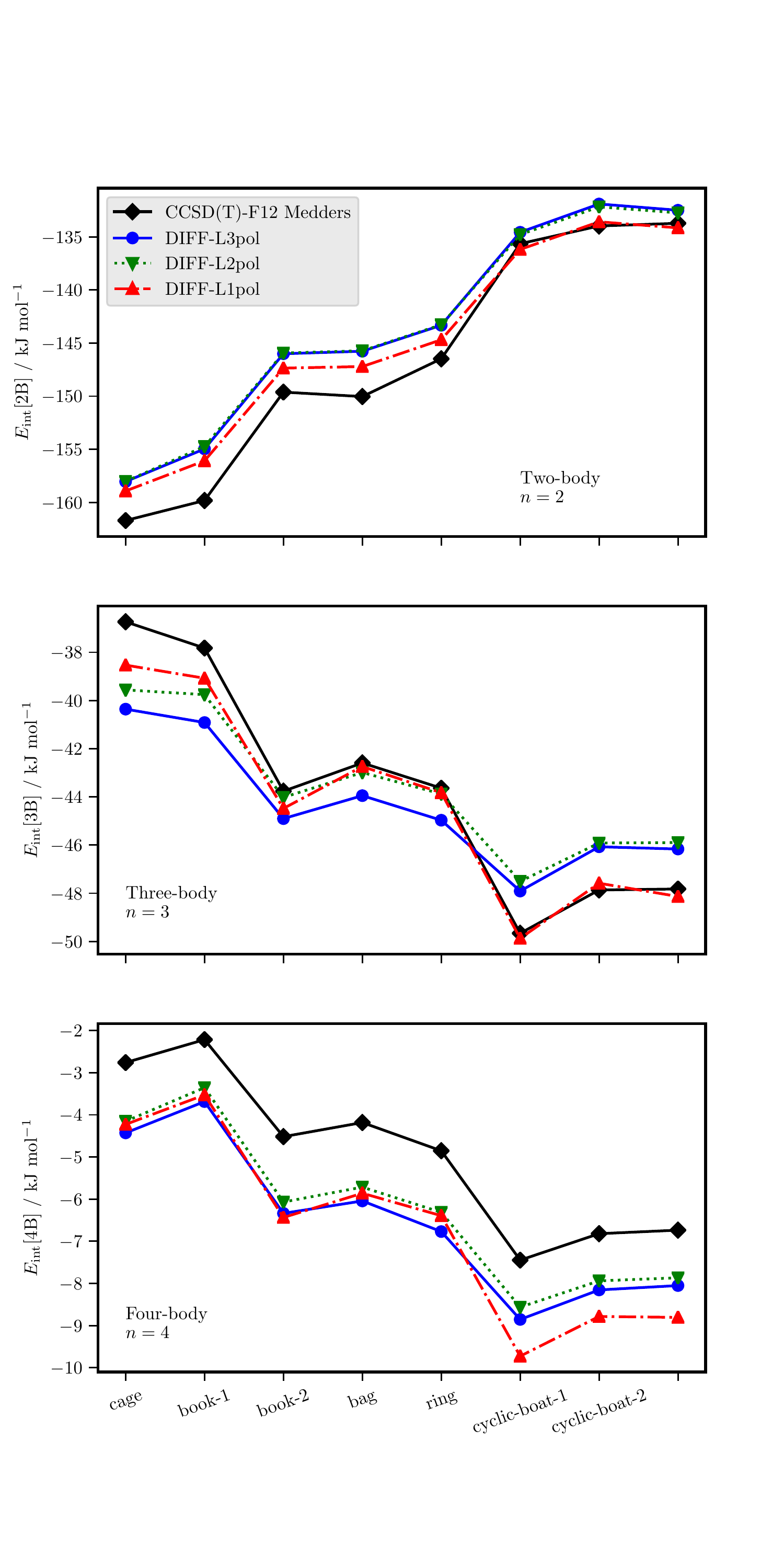}
	\end{center}
	\caption{\small $n$-body decomposition of the hexamer energies using 
		the reference geometries from Bates and Tschumper \cite{bates2009ccsd}.}
	\label{fig:hexamers_many_body_energies}
\end{figure}

Following the discussion in \secrff{sec:flexibility_and_DIFF}, we should expect errors 
in the energies from the DIFF models as the water conformations in the hexamers are not
fixed, but instead vary within and between the hexamer conformers. 
Additionally, no non-additive dispersion is included in the DIFF models, and although these
energies are expected to be small, they are non-negligible \cite{Stone:book:13} for the 
more compact hexamers. 
From \figrff{fig:hexamers_many_body_energies} we see that this good agreement arises from
an error cancellation between the two-body and many-body energies: 
While the DIFF models are able to predict the overall trends of the $n$-body contributions
of the hexamer isomers, they are offset from the reference values. 
We emphasise that this is not necessarily a deficiency of the DIFF models as these were
designed for systems with water molecules held in a fixed geometry. 
What is remarkable is that these deviations in the $n$-body energies cancel to such a large
extent to result in accurate total interaction energies.

\subsection{Larger water clusters: 16-mers and 24-mers}
\label{sec:16_24_mers}

\begin{figure}
	\begin{center}
		\stackunder[-5pt]{\includegraphics[scale=0.20]{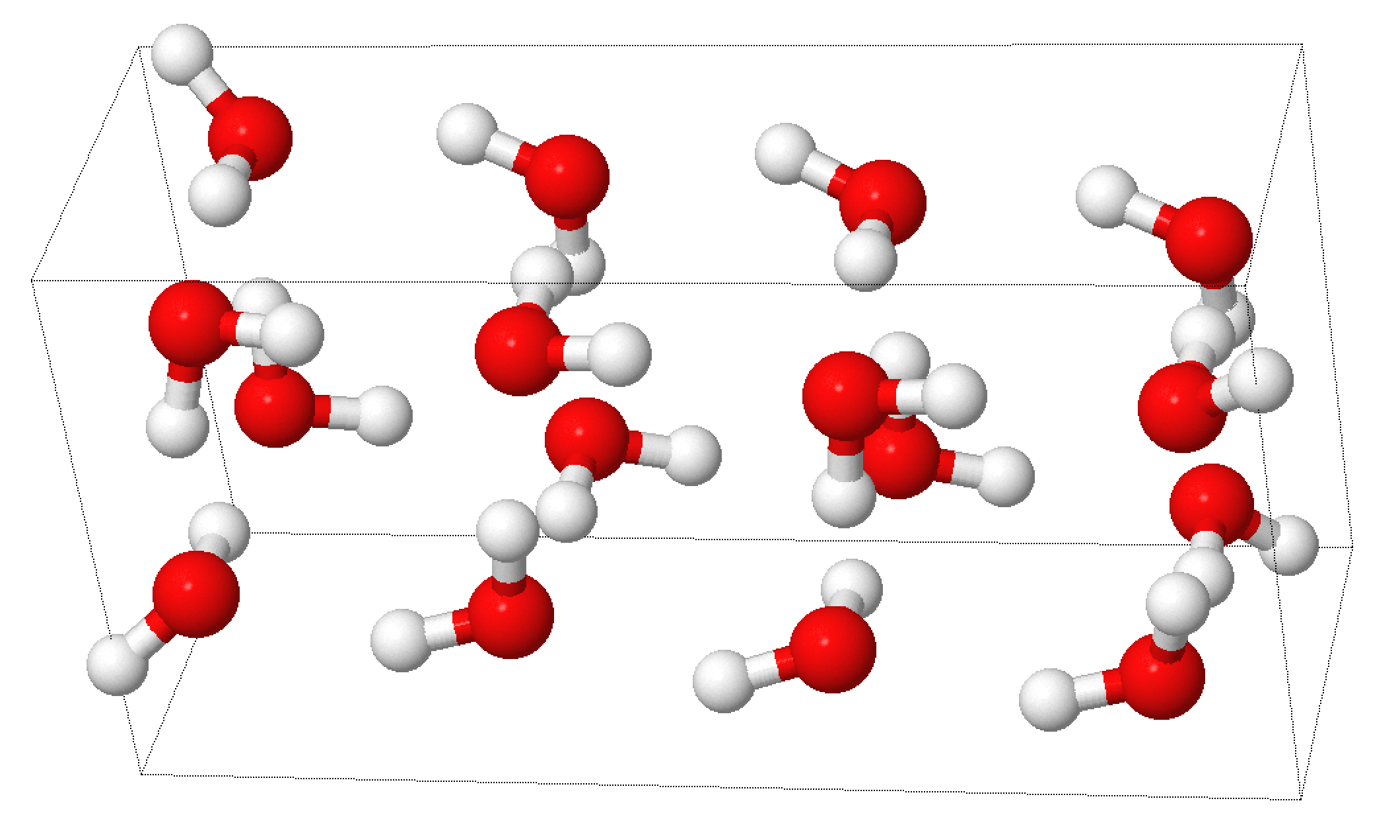}}{(a) 4444-a}
		\stackunder[-5pt]{\includegraphics[scale=0.20]{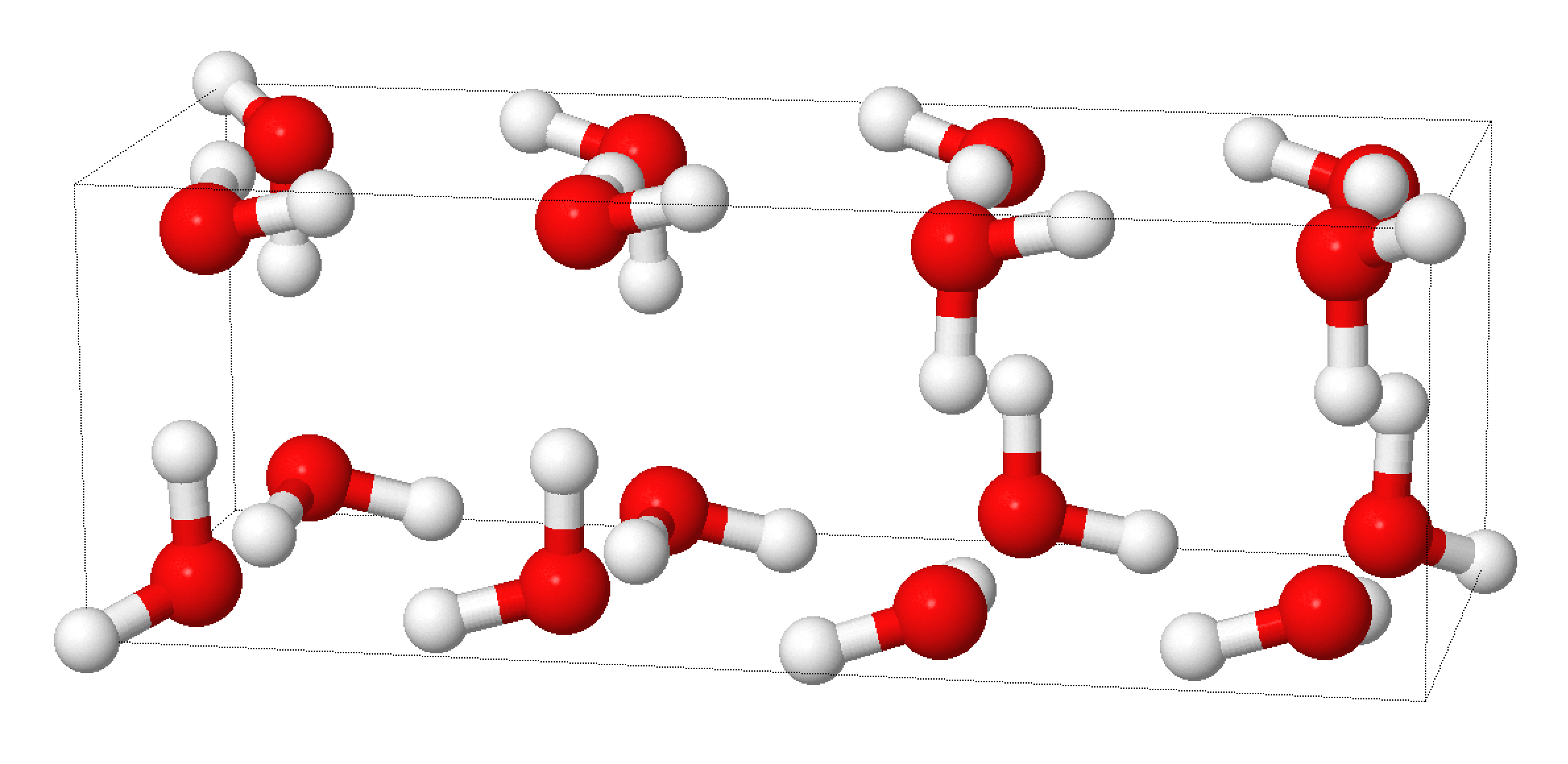}}{(b) 4444-b}
		\stackunder[-5pt]{\includegraphics[scale=0.20]{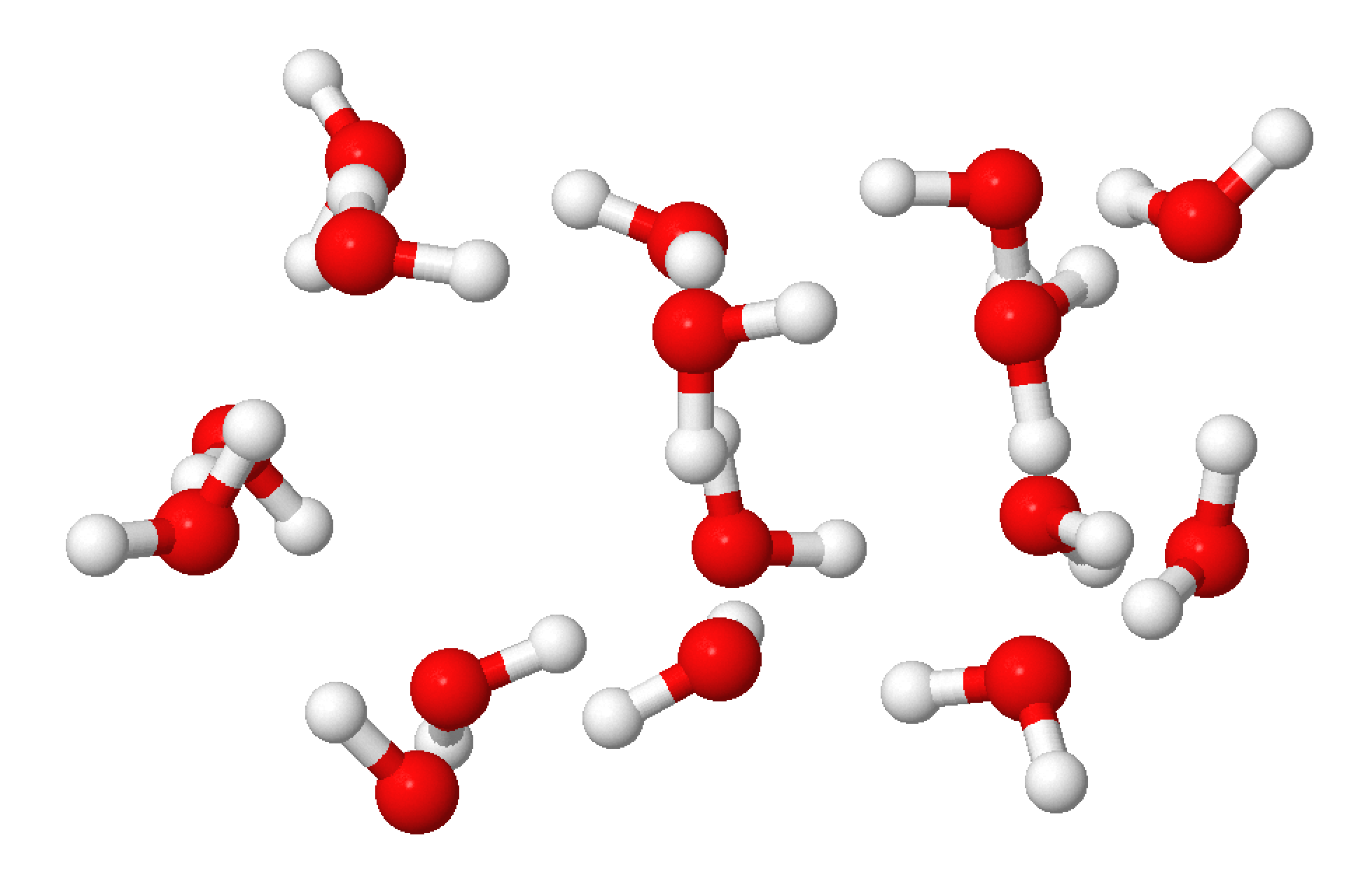}}{(c) anti-boat}
		\stackunder[-5pt]{\includegraphics[scale=0.19]{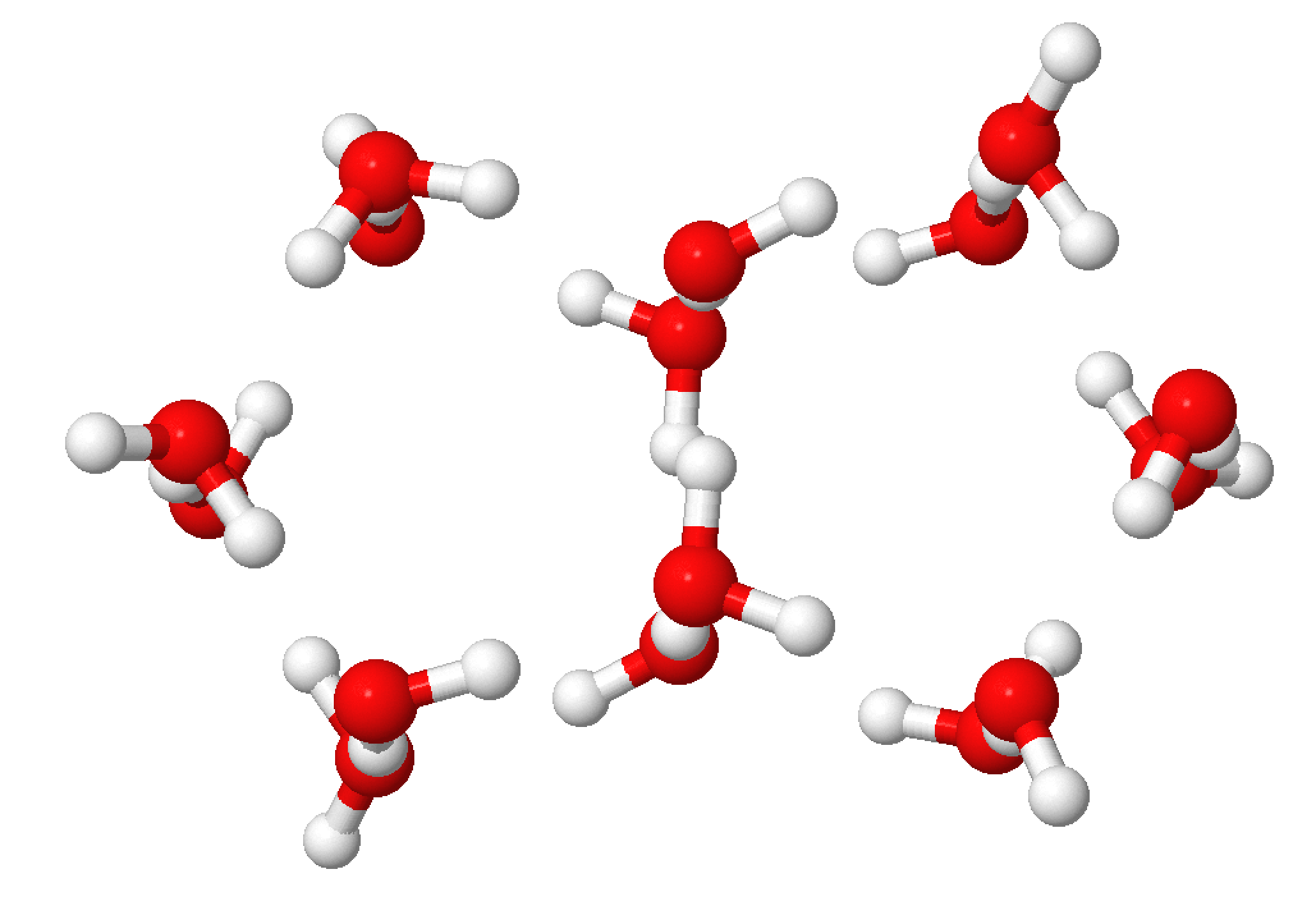}}{(d) boat-a}
		\stackunder[-5pt]{\includegraphics[scale=0.22]{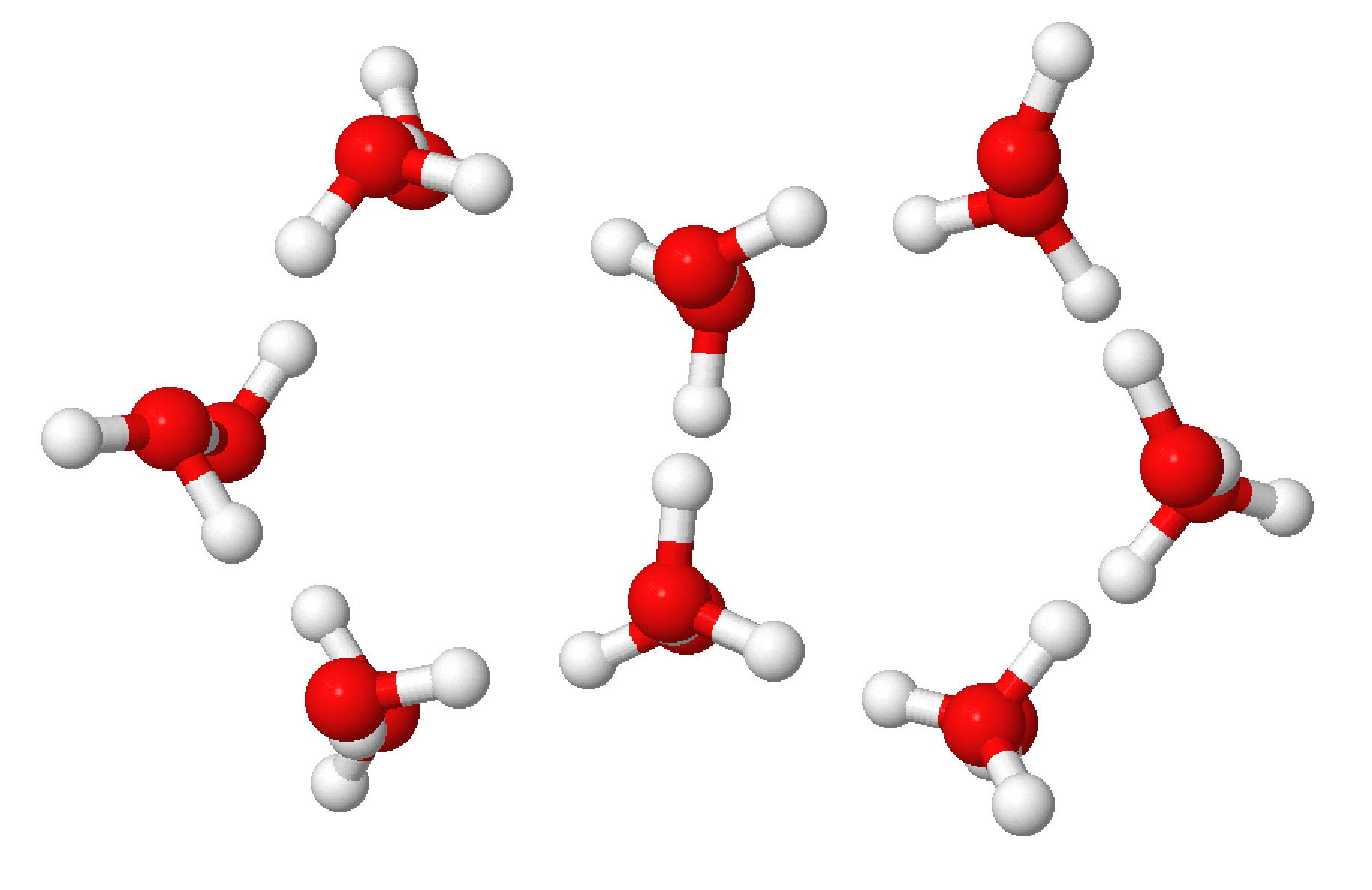}}{(e) boat-b}
	\end{center}
	\caption{
        16mers used in the study. 
    }
	\label{fig:16mers}
\end{figure}

\begin{figure}
	\begin{center}
		\stackunder[-1pt]{\includegraphics[scale=0.20]{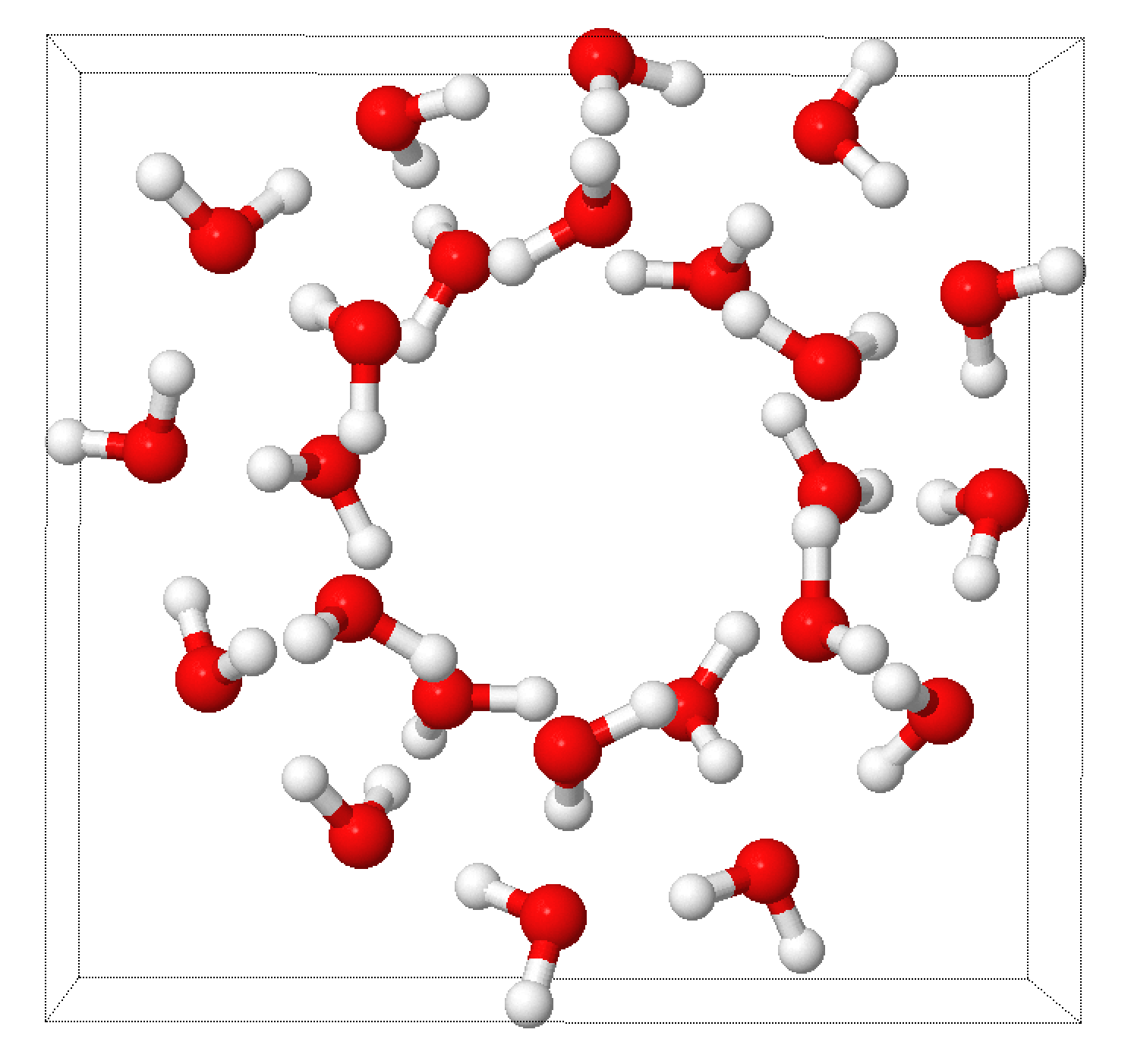}}{(a) 308}
		\stackunder[-1pt]{\includegraphics[scale=0.20]{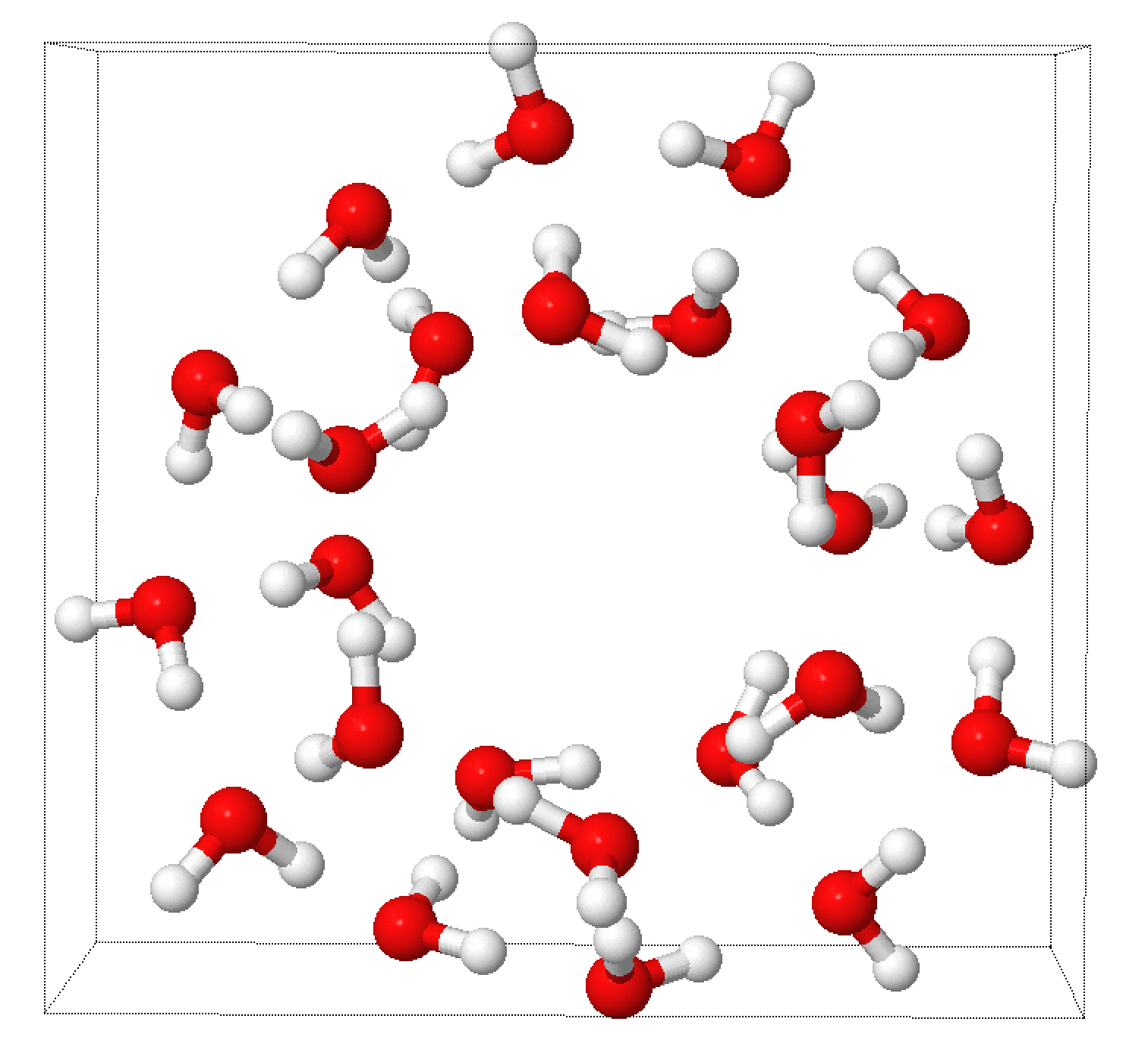}}{(b) 316}
	\end{center}
	\caption{
        24mers used in the study. 
    }
	\label{fig:24mers}
\end{figure}

In this final set of tests, we will use the DIFF models to evaluate the many-body
energies of the larger water clusters shown in \figrff{fig:16mers} and \figrff{fig:24mers}.
In general, for reasons discussed in \secrff{sec:flexibility_and_DIFF} and also 
in \secrff{sec:hexamers} we will focus more on trends than actual interaction energies, 
as these are more likely to be reproduced with models that are not explicitly dependent on the
molecular conformations.

\subsubsection{\waterN{16} isomers}
\label{sec:water16}

\begin{figure}
	\begin{center}
		\includegraphics[width=0.48\textwidth]{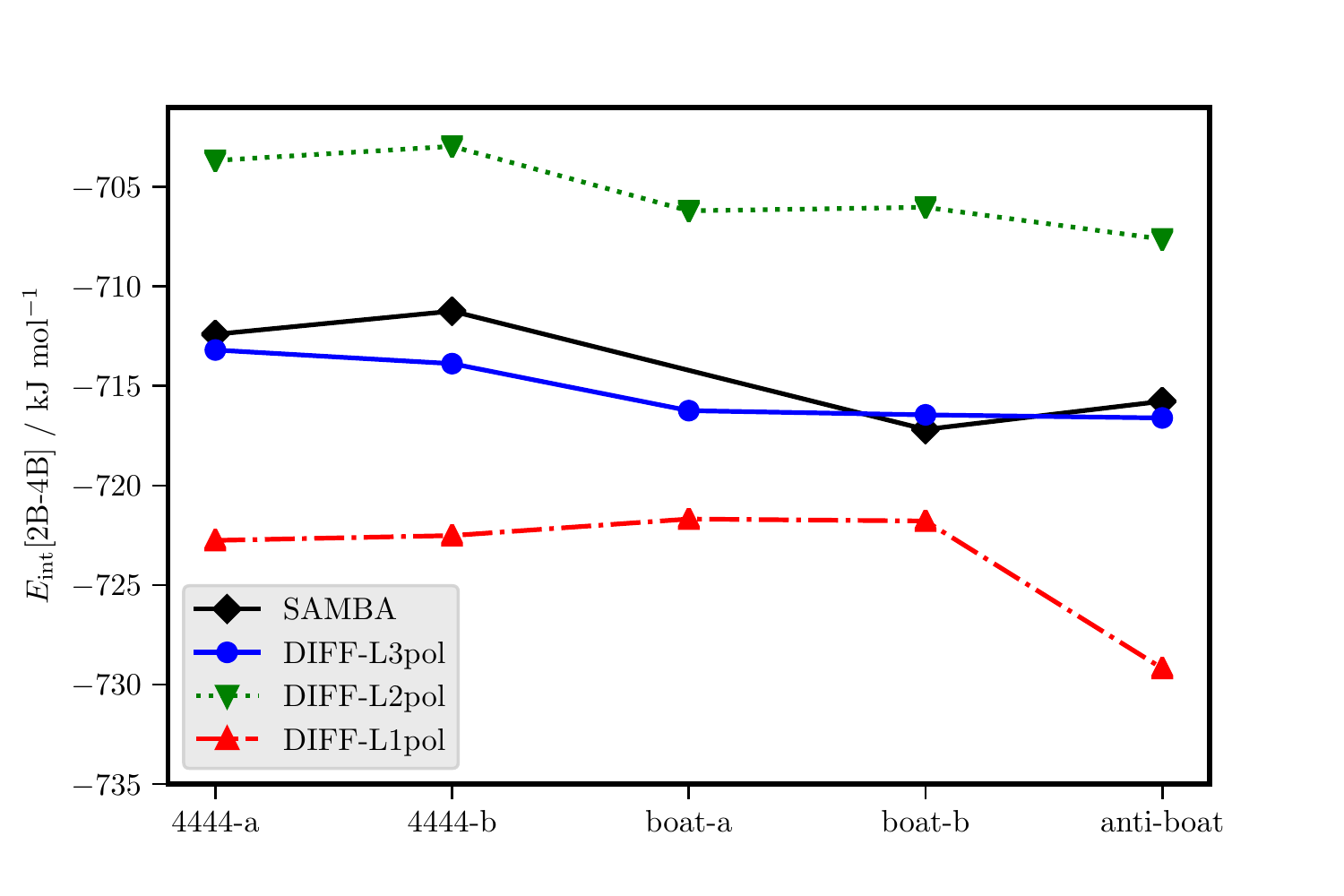}
	\end{center}
	\caption{
        \waterN{16} isomer interaction energies including two-body to four-body contributions,
        \EintS{2}{4},
        computed with the three DIFF models and the reference SAMBA energies from G{\'o}ra \etal
        \cite{gora_predictions_2014}. There are no SAMBA references for the
        boat-a structure. 
        All energies are computed at the reference geometries from 
        Yoo and Xantheas \cite{yoo2017structures}.
    }
	\label{fig:n16_energies}
\end{figure}

The \waterN{16} clusters have been optimized by Yoo and Xantheas \cite{yoo2017structures} and
this set includes two bonding variants of the ``4444'' structure and two of the 
``boat'' structure. 
The fifth structure, the ``anti-boat'', was estimated \cite{yoo2017structures} to have
an energy lying between the two boat structures. 
The best estimates of the energies (total and many-body decomposition) of these clusters
(excluding the boat-a isomer) have been obtained by G{\'o}ra \etal \cite{gora_predictions_2014}
using the SAMBA algorithm.
The SAMBA method avoids energy calculations of the cluster as a whole, but instead makes
use of the many-body expansion to arrive at the total interaction energy from a sum of
dimer, trimer, and higher-body contributions. In this way, appropriate levels of theory
and basis sets can be used for each level, and numerical issues like the basis-set superposition
error (BSSE) can be corrected. 
The intermolecular interaction energies for the \waterN{16} isomers are given in 
\tabrff{tab:n16_energy_optimized} and are terms including 2-body to 4-body contributions,
\EintS{2B-4B}, are displayed in \figrff{fig:n16_energies}. 
Note that SAMBA reference energies are not available for the boat-a isomer.

The \waterN{16} isomers have reference energies within only 6 \kJmol, which is half the 
energy range of the water hexamers. 
The 4444-a and 4444-b isomers differ in energy by only 1.2 \kJmol and the boat-b and anti-boat
isomers differ by 1.4 \kJmol. However there is a wider gap between the 
4444-a/b and boat-b/anti-boat sets which are separated by around 3.4 \kJmol. 
Consequently when looking for trends we focus on this larger energy separation as it is less
likely to be an artefact of the SAMBA reference energies. 
First of all, the DIFF-L3pol energies are within 2 \kJmol of the SAMBA energies, and
the DIFF-L2pol and DIFF-L1pol models are within 10 \kJmol, with the former underestimating
the cluster binding, and the latter leading to an overestimation.
Both higher-ranking DIFF models give an acceptable separation of the 4444-a/b and boat-b/anti-boat
groups of isomers: for DIFF-L3pol and DIFF-L2pol the separation is 2.6 \kJmol and 3.1 \kJmol,
resp., in good agreement with the 3.4 \kJmol SAMBA reference. 
However, the DIFF-L1pol model results in 4444-a/b and boat-b being nearly iso-energetic, and 
the anti-boat structure being more stable than the boat-b isomer by 7.4 \kJmol. 
It is not clear why this is the case, but the problem appears similar to the inability of the 
DIFF-L1pol model to describe the energetic ordering of the ring-like water hexamers.

Also shown in \tabrff{tab:n16_energy_optimized} are the interaction energies of the 
optimized \waterN{16} isomers on the DIFF surfaces. 
All isomers are minima on the DIFF-L3pol and DIFF-L2pol surfaces, with very little energetic
reduction on optimization, and geometric changes in the O\dots O separations of order 0.01  \AA\ only.
However for the DIFF-L1pol model there are larger geometric changes on optimization with 
O\dots O separations changing by as much as 0.04 \AA, and energies lowering by between 4 to 8 \kJmol.
This is again similar to the performance of these models on the water hexamers, and once again
we observe that the higher ranking polarization models seem better able to describe both the
energies and structures of the water clusters.

\begin{figure*}
	\begin{center}
		\includegraphics[width=0.96\textwidth]{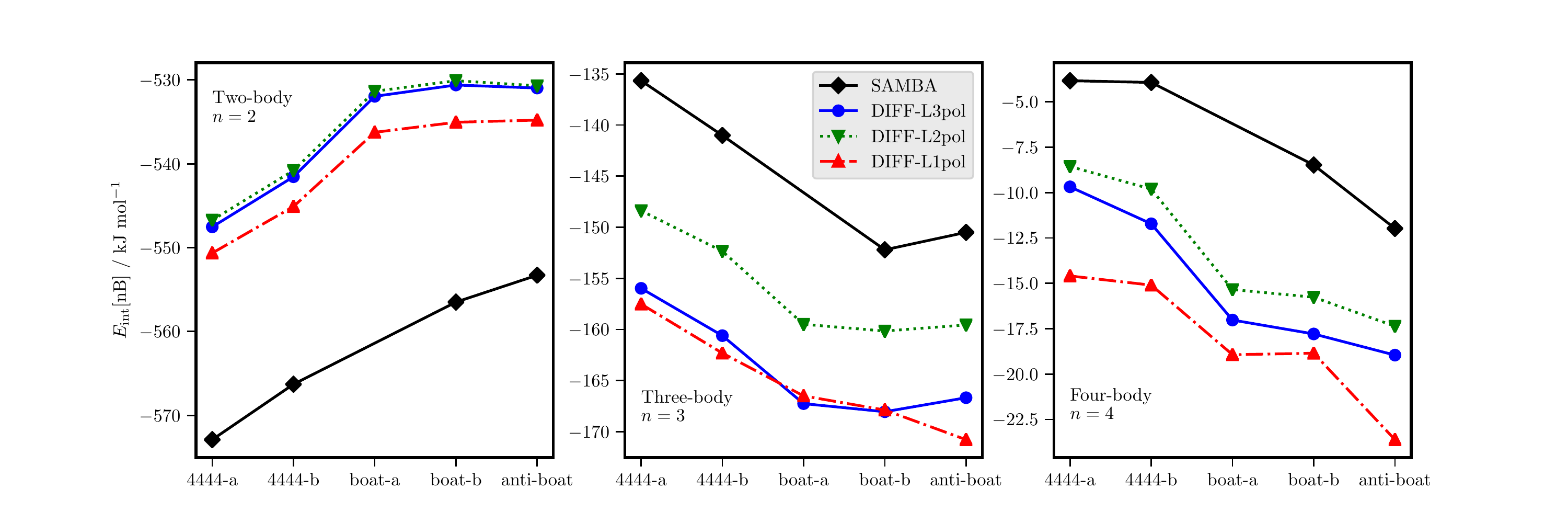}
	\end{center}
	\caption{
        Comparison of the $n$-body energies from the DIFF models with the
        SAMBA references from G{\'o}ra \etal \cite{gora_predictions_2014}. 
        There are no SAMBA references for the boat-a structure.
    }
	\label{fig:n16_energy_decomposition}
\end{figure*}

In \figrff{fig:n16_energy_decomposition} we display the many-body energy decomposition of the
\waterN{16} clusters. As with the hexamers (see \figrff{fig:hexamers_many_body_energies}) we
see that the excellent total interaction energies from the DIFF models results from a 
cancellation of errors made in the 2-body and many-body energies. 
The general trends in the DIFF-L2pol and DIFF-L3pol $n$-body energies are very similar to
those from the SAMBA references, but the DIFF-L1pol model shows a comparative
over-estimation of the anti-boat many-body energies. 
Part of the overestimation of the many-body energies results from the use of molecular 
properties evaluated at the reference monomer geometry and transferred to the flexible
monomers in the clusters. We have made an estimate of the magnitude of this effect
by evaluating the molecular properties at the water equilibrium geometry rather than
at the vibrationally averaged geometry used in the DIFF models. This change results in 3-body
energies that are around 5 \kJmol smaller (in magnitude) than those from the DIFF models,
and brings them closer to the SAMBA references.
The main effect arises from the change in the water multipoles, with the changes in the 
polarizabilities having a smaller impact on the energies.
A more systematic study of this effect is needed and is in progress in our group.

\subsubsection{\waterN{24} isomers}
\label{sec:water24}

The two \waterN{24} isomers shown in \figrff{fig:24mers} are the largest clusters 
considered in this study. 
SAMBA reference energies for these clusters \cite{gora_interaction_2011} show that
the 308 and 316 isomers are nearly isoenergetic, with an \EintS{2}{4} energy difference
of less than 1 \kJmol, and the 308 isomer the more stable.
This small difference arises from a near cancellation of energy differences in the 2B and 
3B energies of the two isomers. 
From Table 20 in the SI we see that at the 2B level, 308 is more 
stable by around 7 \kJmol, but at the 3B and 4B levels it is less stable by around 5 \kJmol
and 1 \kJmol. 
Therefore, these clusters present a good test of the balance between the two-body and 
many-body parts of the interaction models. 

\begin{figure}
	\begin{center}
		\includegraphics[width=0.48\textwidth]{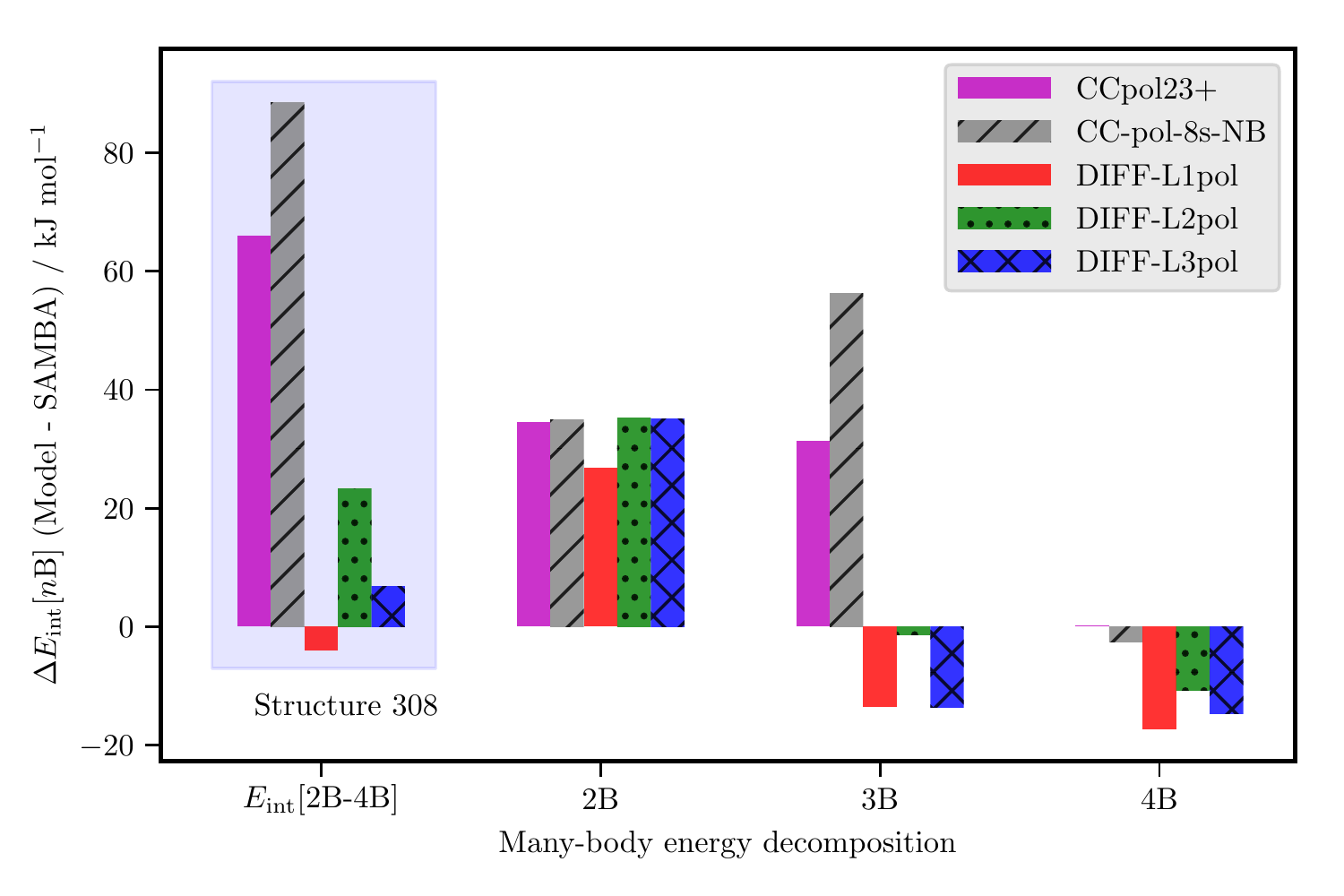}
		\includegraphics[width=0.48\textwidth]{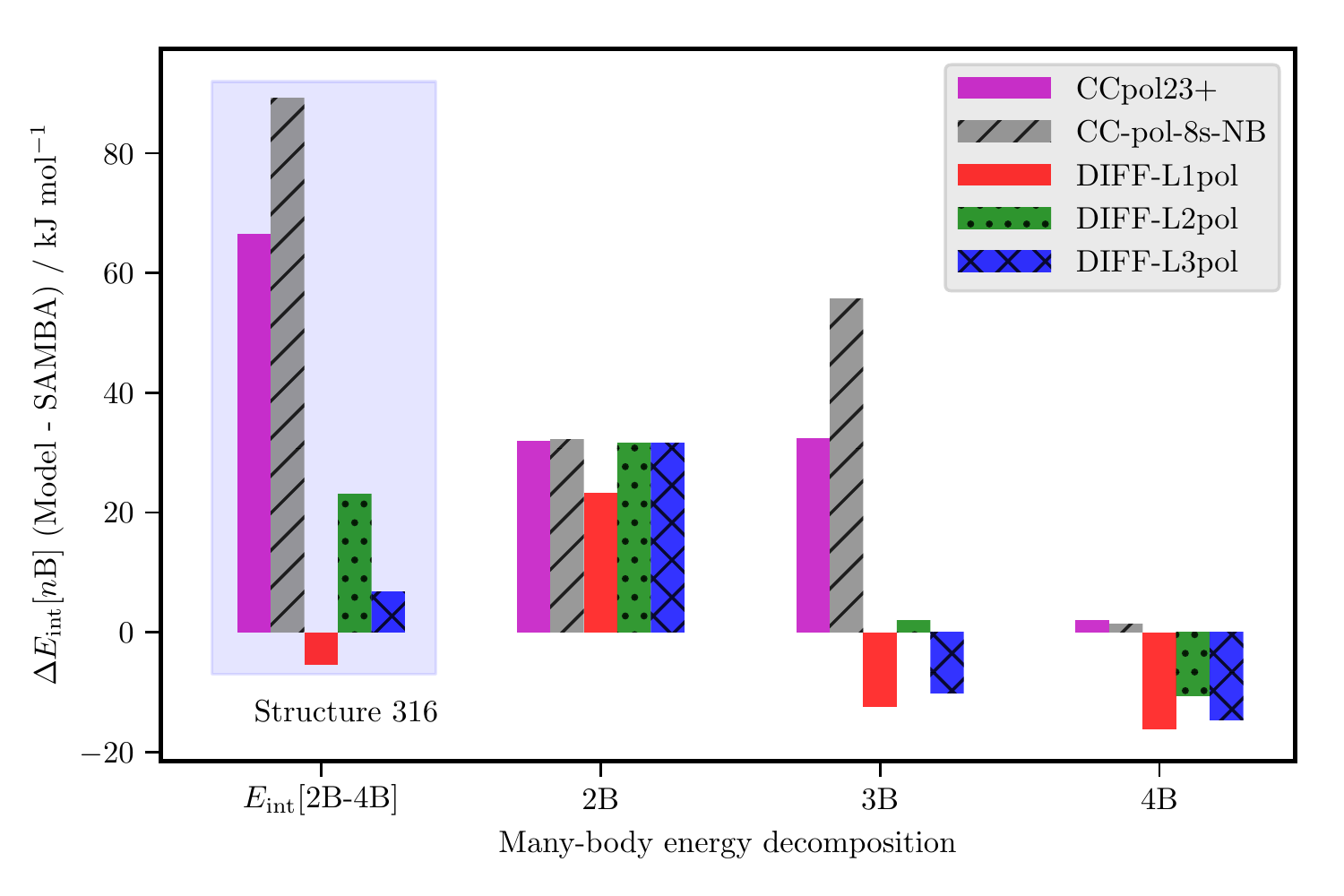}
		\includegraphics[width=0.48\textwidth]{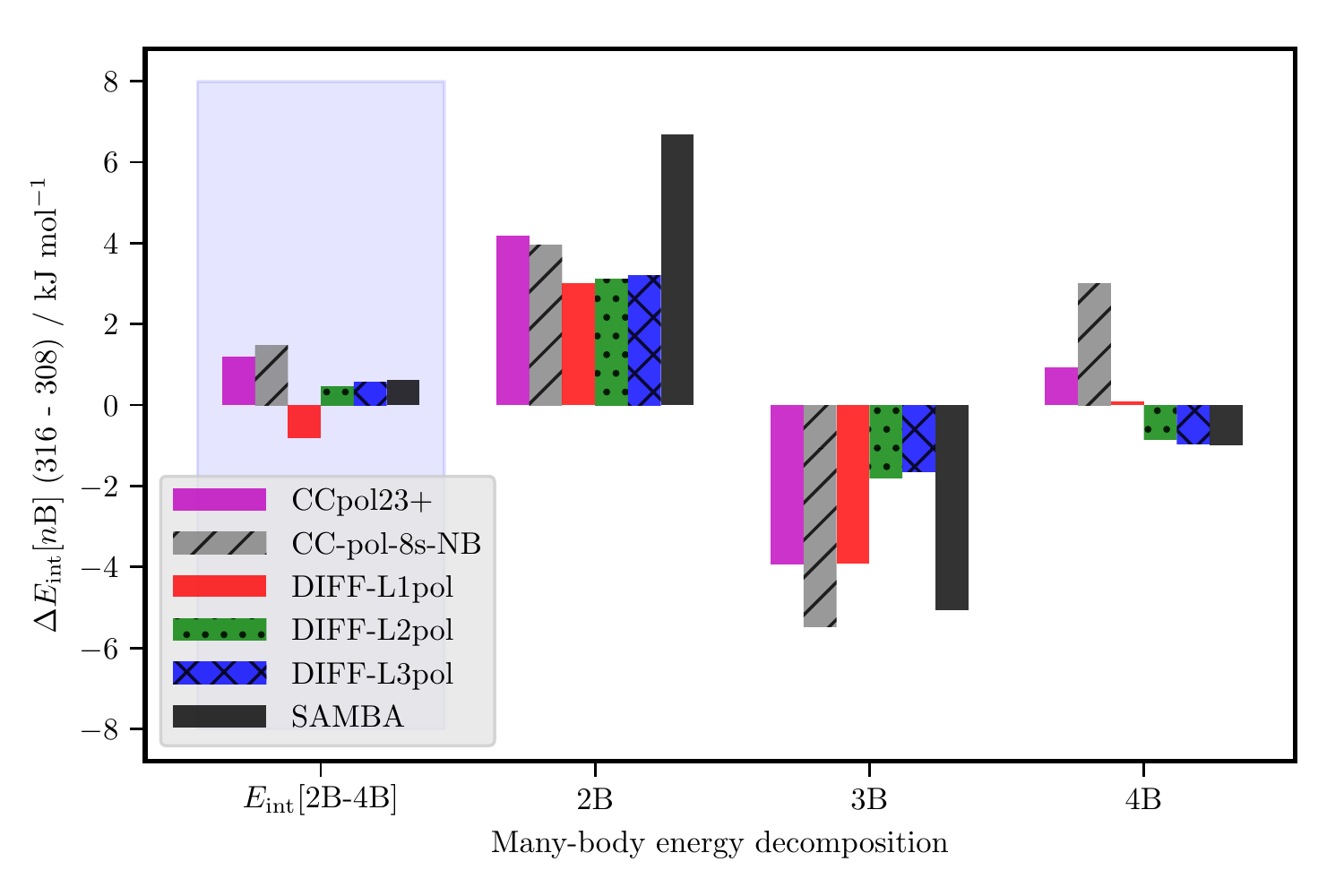}
	\end{center}
	\caption{\small Two-, three- and four-body contributions to the
	\waterN{24} dodecahedra 316 and 308 measured relative to
	SAMBA reference energies from \cite{gora_interaction_2011}, and
	the difference in energy between them.
    Note that there is an error in the SAMBA estimate of $\Delta\EintS{2}{4}$ 
    as reported in Table.~VI from G{\'o}ra \etal \cite{gora_predictions_2014}. 
    This has been corrected in this figure.
    }
	\label{fig:n24_energies}
\end{figure}

In \figrff{fig:n24_energies} we display the differences in the model energies from the SAMBA
references. Here we consider the DIFF as well as CCpol23+ and CC-pol-8s-NB models with data
for the latter two taken from G{\'o}ra \etal \cite{gora_predictions_2014}.
All models show almost exactly the same errors for the 2-body energies, but the DIFF models
show smaller errors for the 3-body non-additivity, and the CC models show smaller errors for
the 4-body non-additivity. 
As the errors made by the DIFF models for the 2-body and many-body energies are of opposite 
signs, these largely cancel out to result in better agreement than the two CC models
with the SAMBA references for the \EintS{2}{4} estimate of the cluster interaction energies. 
This was not expected as CCpol23+ is based on CCSD(T) reference energies as are the
SAMBA references, but the DIFF models are based on SAPT(DFT) references.

Also shown in \figrff{fig:n24_energies} are the relative energy differences, 
$\Delta E_{\mathrm{int}}$, between the 316 and 308 isomers from the SAMBA reference calculations
and the DIFF and CC models. These are displayed both for \EintS{2}{4} as well as for the 
individual contributions from the many-body expansion.
All models underestimate the 2-body contribution to $\Delta E_{\mathrm{int}}$, but the CC models
and the DIFF-L1pol model show better agreement with the 3-body contribution to the energy difference,
while the two higher-ranking DIFF models underestimate this difference.
On the other hand, at the 4-body level the DIFF-L2pol and DIFF-L3pol models show an excellent agreement
with the SAMBA references, while the CC and DIFF-L1pol models predict an energy difference of the
wrong sign.
All models show excellent agreement with the very small total isomer interaction energy difference,
$\Delta\EintS{2}{4}$, with the CCpol23+ and CC-pol-8s-NB models closer than the DIFF-L2pol and 
DIFF-L3pol models.
G{\'o}ra \etal \cite{gora_predictions_2014} have made comparisons of the WHBB and HBB2-pol
models for these isomers and while HBB2-pol gives energies in reasonably good agreement
with SAMBA, the WHBB models (of order 5 and 6) both result in very large errors of around 50 \kJmol
in the 3-body contribution to $\Delta E_{\mathrm{int}}$. This is well in excess of the 
2--3 \kJmol errors made by the CC and DIFF models.

We must not place too much importance to the relatively small differences in the CCpol23+, 
CC-pol-8s-NB and the three DIFF models as all have been {\em transferred} onto the 
flexible water conformers in these \waterN{24} clusters, so some differences must be expected. 
Indeed, a perfect agreement would be surprising. 
Nevertheless, the fact that these differences are so small is remarkable, particularly
for the DIFF models which have bee constructed from 1-body and 2-body data only.

\begin{table}[ht]
	\begin{tabular}{llllll}
		\toprule
		Structure       & \multicolumn{1}{c}{\text{4444-a}} & \multicolumn{1}{c}{\text{4444-b}} & \multicolumn{1}{c}{\text{boat-a}} & \multicolumn{1}{c}{\text{boat-b}} & \multicolumn{1}{c}{\text{anti-boat}} \\ \midrule
		\multicolumn{6}{l}{SAMBA}                                                                                                                                                                              \\
		\EintS{2}{4}    & -712.397                          & -711.234                          & \text{---}                        & -717.179                          & -715.765                             \\ \hline
		\multicolumn{6}{l}{DIFF-L3pol}                                                                                                                                                                         \\
		\EintS{2}{4}    & -713.187                          & -713.873                          & -716.235                          & -716.447                          & -716.603                             \\ \cline{2-6}
		\EINT(Ref geom) & -710.068                          & -711.449                          & -714.616                          & -715.063                          & -715.107                             \\
		\EINT(Opt geom) & -711.239                          & -712.59                           & -715.833                          & -716.342                          & -716.846                             \\
		$\Delta\EINT$   & -1.171                            & -1.141                            & -1.217                            & -1.279                            & -1.739                               \\ \hline
		\multicolumn{6}{l}{DIFF-L2pol}                                                                                                                                                                         \\
		\EintS{2}{4}    & -703.661                          & -702.952                          & -706.192                          & -706.014                          & -707.615                             \\ \cline{2-6}
		\EINT(Ref geom) & -700.786                          & -700.355                          & -704.446                          & -704.014                          & -706.119                             \\
		\EINT(Opt geom) & -701.301                          & -701.364                          & -704.971                          & -704.993                          & -707.125                             \\
		$\Delta\EINT$   & -0.515                            & -1.009                            & -0.525                            & -0.979                            & -1.006                               \\ \hline
		\multicolumn{6}{l}{DIFF-L1pol}                                                                                                                                                                         \\
		\EintS{2}{4}    & -722.757                          & -722.513                          & -721.682                          & -721.782                          & -729.189                             \\ \cline{2-6}
		\EINT(Ref geom) & -721.556                          & -721.494                          & -721.174                          & -721.041                          & -729.827                             \\
		\EINT(Opt geom) & -725.641                          & -725.865                          & -727.392                          & -728.978                          & -737.026                             \\
		$\Delta\EINT$   & -4.085                            & -4.371                            & -6.218                            & -7.937                            & -7.199                               \\ \bottomrule
	\end{tabular}
	\caption{
        Many-body intermolecular interaction energies for \waterN{16} clusters.
        The SAMBA reference energies including contributions from 2-body to 4-body
        interactions, \EintS{2}{4}, are from G{\'o}ra \etal \cite{gora_interaction_2011}.
        For the DIFF models we report both \EintS{2}{4} as well as the total 
        intermolecular interaction energy \EINT at the reference geometries (``Ref geom''),
        as well as \EINT for the optimized cluster geometries with intramolecular conformations
        fixed as described in \secrff{sec:flexibility_and_DIFF} (``Opt geom'').
        $\Delta\EINT$ is the difference in energies of the optimized and reference 
        structures.
        All energies are in \kJmol.
    }
	\label{tab:n16_energy_optimized}
\end{table}	

\section{Analysis}
\label{sec:analysis}

We began this paper with three questions that can now be answered:
\begin{itemize}
\item {\em Q1: Can many-body polarization models be developed from the 
  properties of the monomer and dimer energy calculations only?}\\
  Many-body polarization models can be developed from monomer properties and
  dimer interaction energies alone. The route to this is to use monomer distributed
  multipoles computed using the basis-space implementation of the iterated stockholder atoms 
  (BS-ISA) algorithm \cite{MisquittaSF14}, and the monomer distributed polarizabilities 
  computed using the ISA-Pol algorithm \cite{misquitta_isa-pol_2018}. Additionally and perhaps
  most importantly, the damping needed for the polarization model needs to be determined using 
  the true polarization energies, as defined using \regSAPTDFT \cite{Misquitta13a}, 
  that are free from charge-delocalization contributions. These models have been 
  demonstrated to reproduce the geometries and energies of a large number of water clusters
  to an accuracy that rivals that attained by models fitted to extensive sets of water clusters.
\item {\em Q2: Can we develop a systematic hierarchy of polarizability model of increasing rank,
  and how do the accuracies of these models vary with rank?}\\
  The ISA-Pol polarizabilities can be defined at any rank (currently to a maximum 
  of rank 3) and include anisotropy and full coupling within the molecule (all many-body interactions
  within the molecule are accounted for). We have developed three DIFF models using these
  polarizabilities of maximum ranks 1, 2 and 3, and have demonstrated that all result in 
  accurate predictions for the energetics of the water clusters, with the higher ranking models
  also able to accurately reproduce the geometries of the water clusters.
  Although we have used terms of the same rank on all sites, it is possible to create ISA-Pol
  models of lower ranks on some sites (say the hydrogen ions), and even to make the 
  polarizability tensors isotropic. 
\item {\em Q3: How accurate are the damping models obtained from \regSAPTDFT and 
  how sensitive are the many-body energies and cluster geometries to deviations from them?}\\
  We have shown that the polarization damping models determined using
  \regSAPTDFT are close to optimal in the sense that they lead to 
  three-body non-additive polarization energies which are able to describe the  
  MP2 and CCSD(T) reference energies for more than 600 trimers. Additionally, these models
  show very small errors for the total interaction energies of the water hexamers.
  Deviating from these proposed damping models leads to increases in errors that are most notable
  in the models with rank 2 and 3 polarizabilities. 
\end{itemize}

We have demonstrated that the DIFF models, in particular the DIFF-L2pol and DIFF-L3pol models,
lead to many-body interaction energies that are comparable in accuracy to the best \abinitio
water models currently available. These include the recently developed 
CCpol23+ model \cite{gora_predictions_2014} and HBB2-pol \cite{medders_representation_2015} models,
as well as more established WHBB model \cite{wang_flexible_2011}.
These three have dedicated three-body models, with additional many-body non-additive
effects described through a polarization model. 
The three-body models in these potentials are fitted to sizeable sets of water trimers, and
contain considerable numbers of fitted parameters. 
Nevertheless, using a set of 600 water trimers extracted from simulations of liquid water
\cite{akin-ojo_how_2013} we have demonstrated that the DIFF-L2pol and DIFF-L3pol models are
second in accuracy only to the CCpol23+ model (MB-pol was not considered in this comparison
and we expect it to equal or surpass CCpol23+ in accuracy).
All DIFF models show an increase in errors only for the trimers with the most repulsive
three-body non-additive energies (with $\EintB{3} \ge 2$ \kJmol), for which physical effects not
included in the polarization models are expected to be important.
As this high accuracy is attained with models built largely from properties computed
from the monomers, with only three parameters fitted to a small number of dimer energies,
the DIFF approach may be a better route to the development of high-accuracy many-body 
models in systems with strong many-body polarization.
Additional effects may be absorbed in a {\em correction} model
which could be simpler in structure than the full three-body interaction model. 

We have additionally demonstrated how the DIFF models can maintain a high accuracy even
when {\em transferred} onto systems with flexible water molecules.
This high accuracy in the total interaction energy seems to arise from a systematic
cancellation of errors in the two-body and many-body energies from the DIFF models.
It is not clear why this error cancellation is so systematic, or indeed if it is a feature of
water models only (due to the fairly limited bonding types in the water clusters) or is more general.
It is possible that the high transferability of the ISA-based monomer properties, 
in particular the ISA multipoles, may be the reason for the high degree of transferability
of the DIFF models. 
We have already demonstrated that the ISA multipoles do possess a good degree of transferability
on molecular crystals \cite{aina_charge_2019}, and work is in progress in our group to make
an analysis of the ISA-Pol dispersion and polarizability models. 

There are also questions which have arisen during the course of this investigation that 
require further investigation. Perhaps the most important of these 
\begin{itemize}
\item How should the many-body polarization models of increasing rank be {\em systematically}
  constructed so as to be consistent?
\item What is the minimum rank of the polarizability tensor needed to achieve a high 
  accuracy both in the many-body interaction energies and in the cluster geometries. 
  We have shown that the L1pol model is acceptable for the energetics, but perhaps not for the
  geometries. The L2pol (including quadrupole--quadrupole polarizabilities) were needed 
  for accurate geometries. But is this generally the case, and if so, on which atomic sites
  are these terms needed?
\item The damping models rely on true polarization energies defined using \regSAPTDFT, 
  but this theory relies on a regularization parameter --- which defines a length scale --- 
  to separate the polarization and charge-delocalization components of the (second-order) induction 
  energy. How sensitive are the DIFF models to the choice of this parameter, and how does
  this parameter vary with system and atom type?
\end{itemize}

We have thus far constructed the three
DIFF models of maximum polarizability ranks 1, 2 and 3, independently, with the polarization
damping models determined separately for each.
In this way, the weaker damping for the DIFF-L1pol model makes up for the
missing higher-ranking terms, but it is also possible that the resulting larger induced dipoles
cause the slight over-binding seen in the ring-like hexamers and the 16-mers. 
An alternative would be to determine the polarization damping model for the highest ranking
polarization model, and then to construct the lower ranking models with the damping kept fixed.
This method may have the merit of leading to a more systematic behaviour of the DIFF models
as a function of the rank of the polarizability tensors.
We are currently investigating this possibility.

While the energetics of the models are important, their performance in determining the
structures of the water clusters are equally or perhaps even more important.
Both the DIFF-L2pol and DIFF-L3pol models have been shown to result in cluster geometries
that agree with the theoretical benchmarks to a very high accuracy: both models result in 
optimized (intermolecular coordinates only) clusters with O\dots O separations and 
important structural angles and torsions in very close agreement with the reference geometries.
However the DIFF-L1pol model allows for larger cluster geometry variations when optimizations
are performed: O\dots O separations can change by as much as 0.04 \AA\ and, more importantly,
torsional changes can be as much as 30\textdegree. 
It is possible that the cluster geometries are sensitive to the induced quadrupole that
arise from the rank 2 polarizabilities and the coupling terms between the rank 1 and 2 terms.
If so, this would have important consequences for the development of force-fields.
However it is also possible that this sensitivity has arisen because of the way in which 
the DIFF-L1pol model was constructed, as discussed above. 
An in-depth analysis of the magnitude of the induced dipoles and quadrupoles would be needed
to resolve this puzzle.

\regSAPTDFT, which is crucial in the algorithm used to determine the polarization damping
(see \secrff{sec:pol_model_theory}), involves one free parameter ($\eta$), and while this 
parameter was carefully determined by Misquitta \cite{Misquitta13a},
it is possible that the choice $\eta = 3.0$ a.u.\ is close to, but not optimal. In fact, $\eta$
may vary somewhat with system and even with atom-types. 
Any change in $\eta$ would result in a slightly different partitioning of the \EIND{2} energy
into the second-order polarization and second-order charge-delocalization energies, and
this would alter the polarization model damping, and consequently the many-body 
energies would change.
The already excellent agreement of the DIFF model energies with the reference energies on 
the wide range of water clusters studied here strongly suggests that, at least for water, 
the regularization used is appropriate. Nevertheless, as we seek to reach ever higher 
accuracies, we will also need to determine this parameter with a higher degree of confidence,
and also investigate how it varies with molecules and atomic sites within a molecule.

\section{Conclusions}
\label{sec:conclusions}

We set out to determine whether many-body non-additive polarization effects could be determined
from models constructed using monomer properties and dimer energies only, and have 
demonstrated that this is indeed possible. 
Using ISA-based molecular properties (multipoles and polarizabilities), we have constructed
three DIFF (derived intermolecular force-field) models for water that differ in the maximum
ranks of the polarization tensors and the polarization damping models only, and have demonstrated
that all three models are able to predict the many-body non-additive energies of extensive
sets of water trimers, the water hexamers, 16-mers and 24-mers. Additionally, these models
are able to correctly describe the intermolecular geometries of the water clusters
even when used on clusters with flexible water molecules. 
We have further demonstrated that while the DIFF-L1pol model (with rank 1 polarizabilities)
is relatively less sensitive to the choice in damping model, the two DIFF models with 
higher ranking polazization tensors are very sensitive to the polarization damping.
Further, for all three DIFF models, the optimal polarization damping model was determined
from two-body true polarization energies determined using \regSAPTDFT. 

The three DIFF models presented in this paper were never meant to be reference models for 
water, and indeed we have not yet used any in simulations of the liquid, but we have 
successfully demonstrated that they rival some of the most elaborately parametrized 
water models, including HBB2-pol \cite{MeddersBP13}, WHBB \cite{wang_flexible_2011}, 
MB-pol \cite{babin2013development,babin2014development}, as well as the CC-pol-8s-NB
\cite{cencek2008accurate} and recently developed CCpol23+ \cite{gora_predictions_2014}
models. 
Perhaps more importantly for model building, while these models are all built on extensive
sets of water trimers and even larger clusters, the DIFF models use only 1-body and 2-body 
data, in effect reversing a recent trend to large and computationally demanding data sets
\cite{ouyang_modelling_2015}.
Additionally, the many-body part of the DIFF models for \water contains precisely three independent
fitted parameters that are determined using a small number of dimer calculations using 
\regSAPTDFT \cite{Misquitta13a}. 
The key to the performance of the DIFF models lies in the physical insight used to 
understand the origin of the many-body polarization, and not in simply fitting to 
ever-increasing data sets. 

In demonstrating the above we have used models based entirely on SAPT(DFT) and \regSAPTDFT, 
with molecular properties computed using the BS-ISA and ISA-Pol algorithms.
These methods are computationally efficient and can be applied to systems much larger than 
water, and consequently these many-body polarization models can be constructed for larger system
too. 
Indeed we have already done so for the pyridine complex \cite{MisquittaS16} and have demonstrated
that these models lead to the correct description of the structures of the known polymorphs 
of pyridine, including a new high-pressure phase \cite{aina_dimers_2017}. 
While more systems are needed to be sure of the applicability of the methodology, the 
strong physical underpinning of the methods described in this paper (and in previous papers
from our group) leads us to be optimistic that the approach described here is robust and 
may well be a general route to developing many-body non-additive polarization models that
exceed the accuracy of any proposed so far.
We note here that there is no real difficulty in combining the DIFF methodology with
reference properties and dimer energies from more advanced methods. Indeed the present
DIFF models for water all exhibit a mild underbinding of the dimers due to the use of 
SAPT(DFT) references, and they could be easily improved using CCSD(T) references. 

The price to pay for the methodology used in the DIFF models is that the polarization models
are more complex and necessitate the development of simulation tools that can handle the
many-body polarization in the condensed phase in a computationally efficient manner. 
The theoretical framework needed for the DIFF models is well known \cite{Stone:book:13} and has 
been implemented in the Orient program \cite{Orient}, but the na{\"i}ve implementation of the
classical polarization model is computationally inefficient. 
Considerable progress has been made in improving the efficiency of the polarization models, 
most notably by Lagardere \etal \cite{lagardere_tinker-hp_2017} and by Albaugh \& Head-Gordon
\cite{albaugh_new_2017}. However neither of these approaches is directly suitable for even 
the simplest of the DIFF models. 
Nevertheless there is nothing fundamental that stops us from improving the flexibility and the 
computational efficiency of the classical polarization models in simulation codes. 
And now that we can develop accurate polarization models more readily, it is time we 
addressed this deficiency in our computational tools and made the necessary effort to allow
simulations with theoretically motivated force-fields rather than force theorists to simplify
the models and thereby lose in accuracy.
That said, it may still be possible to use force-matching methods 
\cite{salanne_including_2012,iuchi_are_2007} to construct simpler force-fields that are
rigorously linked to the underlying theoretical models (such as DIFF). 
Clearly there are possibilities for moving the DIFF models to mainstream simulations.

\section{Acknowledgements} 
RAJG thanks Queen Mary University of London and the EPSRC for a studentship that has
funded part of this work. AJM additionally thanks the Thomas Young Centre for Theory and
Simulation of Materials for a stimulating environment.
We thank Krzysztof Szalewicz for helpful comments on parts of this work.
We thank Nohad Gresh and Jean-Philip Piquemal for many useful discussions on polarization,
charge-delocalization and water models.
Finally we acknowledge a considerable gratitude to Anthony J. Stone for a long collaboration
which has led to the models discussed in this paper. 
We thanks the Royal Society for an International Exchange grant \url{IEC\R2\181027}
that was used to fund some of this work.

\section{Additional information}
All developments have been implemented in a developer's version of the
\CamCASP{6.0} \cite{CamCASP} program which may be obtained from the
authors on request. \CamCASP{} has been interfaced to the \Dalton (2006 through to 2015), 
\NWChem, \Gamess, and \PsiF programs.

The specifications of the DIFF models are presented in the Supplementary Information.
Additionally we supply input files suitable for use with the Orient 5.0 \cite{Orient}
program which can be freely obtained. 
We also supply the structures of all water clusters used in this paper so that comparisons
can be made easily should the need arise.

%

\end{document}


\title{Supplementary information for: 
``First-principles many-body non-additive polarization energies from monomer and 
  dimer calculations only : A case study on water''
	  }

\author{Rory A. J. Gilmore}
\affiliation{School of Physics and Astronomy 
	and the Thomas Young Centre for Theory and Simulation of Materials
	at Queen Mary University of London,
	London E1 4NS, U.K.}
\author{Martin T. Dove}
\affiliation{School of Physics and Astronomy 
	and the Thomas Young Centre for Theory and Simulation of Materials
	at Queen Mary University of London,
	London E1 4NS, U.K.}
\author{Alston J. Misquitta}
\affiliation{School of Physics and Astronomy 
	and the Thomas Young Centre for Theory and Simulation of Materials
	at Queen Mary University of London,
	London E1 4NS, U.K.}

\maketitle

\section{Model specifications}

\subsection{DIFF functional form}

The DIFF functional form is not fixed, but instead is determined by the best theoretical
understanding available. At present we utilize an anisotropic Born--Mayer functional 
form \cite{MisquittaS16} but we have also used alternative forms 
\cite{van_vleet_beyond_2016,van_vleet_new_2018} sometimes with better results. 

Following Misquitta \& Stone \cite{MisquittaS16} we represent the potential \VINT as 
\begin{align}
    \VINT &= \sum_{a \in A} \sum_{b \in B} \Vint{ab}(r_{ab},\Omega_{ab}),
\end{align}
where, $a$ and $b$ label sites in the interacting
molecules $A$ and $B$, $r_{ab}$ is the inter-site separation, 
$\Omega_{ab}$ is a suitable set of angular coordinates that describes the relative 
orientation of the local axis systems on these sites 
(see ch.\ 12 in ref.~\citenum{Stone:book:13}), and \Vint{ab} is the site--site
potential defined as 
\begin{equation}
    \Vint{ab} = \Vsr{}{ab} + \Ves{ab} + \Vdisp{ab} + \Vpol{ab}.
    \label{eq:Vtot}
\end{equation}
 
The short-range term \Vsr{}{ab} describes the exchange--repulsion energy, the electrostatic
penetration energy, and all other short-range terms, including the charge-delocalization energy:
\begin{equation}
    \Vsr{}{ab} = G \exp{[-\alpha_{ab}(\Omega_{ab})(r_{ab} - \rho_{ab}(\Omega_{ab}))]},
    \label{eq:Vtot_Vsr}
\end{equation}
where $\rho_{ab}(\Omega_{ab})$ is the shape function for this pair of
sites, which depends on their relative orientation described by $\Omega_{ab}$, and
$\alpha_{ab}$ is the 
hardness parameter which will be taken to be independent of orientation. $G= 10^{-3}$ hartree is
a constant energy which determines the units of \Vsr{}{ab}.
The shape-function $\rho_{ab}(\Omega_{ab})$ for site pair $ab$ is dependent on the relative 
orientation of these sites $\Omega_{ab}$ and is given by 
\begin{equation}
\rho_{ab}(\Omega_{ab}) = \rho_{ab}^a(\Omega_{ab}) + \rho_{ab}^b(\Omega_{ab})
\end{equation}
where 
$\rho_{ab}^a(\Omega_{ab}) = \sum_{l,k} \rho_{lk}^a C_{lk}(\theta_a,\phi_a)$ is the shape function for atom $a$ and
$C_{lk}(\theta,\phi) = \frac{4 \pi}{2l + 1} Y_{l,m}(\theta,\phi)$ is a renormalised spherical harmonic term. 

The shape function $\rho_{ab}(\Omega_{ab})$ is best described in local axis
systems that reflect the local symmetries of the sites $a$ and $b$. These
symmetries could be approximate. For example, a convenient choice for the
local $z$-axis at a carbon atom in a benzene molecule might be to have it
point from the carbon to the bonded hydrogen atom. With this choice of
$z$-axis, an approximate cylindrical symmetry may be imposed. In which case,
the potential parameters would be quite simple. But we now need to transform
from these local axis systems to the global axis as the molecular
configurations are defined in the global, or laboratory frame. This
transformation is done using the $S$-functions defined by eqs.~3.3.7
in ref.~\cite{Stone:book:13}  and is given by (eq.~12.2.6 in
ref.~\cite{Stone:book:13})
\begin{equation}
  \rho_{ab}(\Omega_{ab}) = \sum_{l_a l_b j k_a k_b} 
                           \rho^{k_a k_b}_{l_a l_b j}
                           \bar{S}^{k_a k_b}_{l_a l_b j}.
\end{equation}

We do not use the most general $S$-function in our potentials, but only the
special cases: $\bar{S}^{k 0}_{l 0 l}$ and $\bar{S}^{0 k}_{0 l l}$.
Since we do not use mixed terms in the sum, this leads to a very intuitive
result that the shape function of a pair of sites
is the sum of the shape functions of the individual sites. This is so because
these special $S$-functions can be written quite simply as
\begin{equation}
  \bar{S}^{k 0}_{l 0 l} = C_{l,k}(\theta,\phi)^*,
\end{equation}
where the renormalized spherical harmonics (in the Racah definition) are
defined as 
\begin{equation}
  C_{l,k}(\theta,\phi) = \sqrt{\frac{4\pi}{2l+1}} Y_{lm}(\theta,\phi).
\end{equation}
We can use the real components of the renormalized spherical harmonics
(defined below) to get
\begin{equation}
  \bar{S}^{\kappa 0}_{l 0 l} = C_{l,\kappa}(\theta_a,\phi_a),
\end{equation}
where the Greek letter $\kappa$ has been used in place of $k$ to indicate this
is the real component and the angles now have subscripts $a$ to indicate they
are the polar coordinates describing the site--site vector from $a$ to $b$ 
{\em in the local axis system of site $a$}.
Likewise, we define
\begin{equation}
  \bar{S}^{0 \kappa}_{0 l l} = C_{l,\kappa}(\theta_b,\phi_b).
\end{equation}
Now we can write the (approximate) shape function as
\begin{equation}
 \rho_{ab}(\Omega_{ab}) = \rho^{a}(\theta_a,\phi_a) +
                               \rho^{b}(\theta_b,\phi_b),
\end{equation}
where
\begin{equation}
  \rho^{a}(\theta_a,\phi_a) = \sum_{l \kappa} \rho^a_{l \kappa}
                              C_{l,\kappa}(\theta_a,\phi_a),
\end{equation}
with a similar expression for $\rho^{b}(\theta_b,\phi_b)$.

We can interpret $\rho^{a}$ as the shape function of site $a$. This is a very
useful concept when developing atom--atom potentials with the aim of {\em
transferability}, where it is important to define the parameters in the
potential in terms of the properties of the atomic sites.
However Misquitta \& Stone \cite{MisquittaS16} have argued that this interpretation
is only valid at first order. When second-order terms are included then there
is a coupling between the parameters from sites $a$ and $b$ as would happen, for example, 
if there was a strong charge-delocalization between the sites. 
This happens for the O..H interaction in water. So in the DIFF models we have specific parameters
sets for the O..O, H..H, and O..H interactions; i.e., transferability is not imposed. Indeed,
it cannot be imposed without compromising the accuracy of the models.

$\Ves{ab}$ is the expanded electrostatic energy:
\begin{equation}
    \Ves{ab} = \Ves{ab}\bigl( r_{ab},\Omega_{ab}, Q^{a}_{t}, Q^{b}_{u},
        \BETAelst{ab} \bigr);
    \label{eq:Vtot_Ves}
\end{equation}
$Q^{a}_{t}$ is the multipole moment of rank
$t$ for site $a$, where, using the compact notation of ref.~\citenum{Stone:book:13}, 
$t=00,10,11c,11s,\cdots$, and $\BETAelst{ab}$ is
a damping parameter. 
The dispersion energy $\Vdisp{ab}$ depends on the anisotropic
dispersion coefficients  $C_n^{ab}(\Omega_{ab})$ for the pair of
sites, and on a damping function $f_n$ that we will take to be
the Tang--Toennies \cite{TangT92} incomplete gamma functions of order $n+1$: 
\begin{equation}
    \Vdisp{ab} =  - \sum_{n = 6}^{12} f_n\Bigl(\BETAdisp{ab} r_{ab}\Bigr) 
        C_n^{ab}(\Omega_{ab})r_{ab}^{-n}
    \label{eq:Vtot_Vdisp}
\end{equation}
The final term $\Vpol{ab}$ is the polarization energy, which is the long-range
part of the induction energy \cite{Misquitta13a}. $\Vpol{ab}$  depends on
the multipole moments and the polarizabilities $\alpha^{a}_{tu}$,
which are indexed by pairs of multipole components $tu$ 
(for details see refs.\citenum{MisquittaS08a,Stone:book:13}):
\begin{equation}
    \Vpol{ab} = \Vpol{ab}\bigl(Q^{a}_{t}, Q^{b}_{u}, \alpha^{a}_{tu},
                \alpha^{b}_{tu},\BETApol{ab} \bigr).
    \label{eq:Vtot_Vpol}
\end{equation}
 
There are a few points to note about the particular form of the potential \Vpol{ab}.
Although formally written in the form of a two-body potential, many-body polarization
effects are included through the classical polarization expansion \cite{Stone:book:13}.
Also, we will normally define the multipole moments and polarizabilities to
include \emph{intra}molecular many-body effects implicitly, that is, we use the multipoles and 
polarizabilities of atoms-in-a-molecule, localized appropriately.
To this form of the potential we could add a three-body dispersion model, but this is 
not addressed in this paper.

\subsection{Model parameters}

The DIFF model parameters presented here are defined in local axis frame for each atom in the 
water molecule. In the notation used in the Orient program \cite{Orient} the local axes are 
defined to be as follows:
\begin{verbatim}
Axes
    O  z between H1 and H2  x from H1 to H2
    H1 z from  O to H1      x from H1 to H2
    H2 z from  O to H2      x from H2 to H1
End
\end{verbatim}
This choice places the water molecule in the xz-plane with the $z$-axes on each H-atom pointing
outwards, along the O-H bond, and that for the O-atom bisecting the H-O-H angle and pointing 
from the O towards the H-atoms.

The DIFF models are all created for the water molecule in a fixed geometry given as
(in atomic units):
\begin{verbatim}
    O      0.00000000     0.00000000     0.00000000
    H1    -1.45365196     0.00000000    -1.12168732
    H2     1.45365196     0.00000000    -1.12168732
\end{verbatim}

In some of the water clusters the monomer geometries differ from the one above. In this case
the DIFF model parameters presented below were moved onto the sites and the local axis system
was kept the same. That is, the DIFF model parameters were transferred, without change, to the 
new monomer geometries, with the local axis system and parameters kept the same. 

The polarization model parameters are given in the main paper, but are reproduced below for 
convenience.

\begin{table}[ht]
    \newcolumntype{d}{D{.}{.}{9}}
	\begin{tabular}{llldd}
        \toprule
		Pair & $l_a$ & $l_b$ &  \rho         &  \alpha      \\
        \midrule
		O O  & 00    & 00    & 0.575293E+01  & 0.189637E+01 \\
             & 00    & 10    & -0.405000E-02 &              \\
             & 10    & 00    & -0.405000E-02 &              \\
             & 00    & 20    &  0.166090E-01 &              \\
             & 00    & 22c   & -0.117827E+00 &              \\
             & 20    & 00    &  0.166090E-01 &              \\
             & 22c   & 00    & -0.117827E+00 &              \\
		O H  & 00    & 00    &  0.470612E+01 & 0.193432E+01 \\
             & 00    & 10    & -0.265887E+00 &              \\
             & 00    & 11c   &  0.179670E-01 &              \\
             & 10    & 00    & -0.199970E-01 &              \\
             & 20    & 00    &  0.724500E-02 &              \\
             & 22c   & 00    & -0.169075E+00 &              \\
		H H  & 00    & 00    &  0.376139E+01 & 0.199594E+01 \\
             & 00    & 10    & -0.215258E+00 &              \\
             & 00    & 11c   &  0.483480E-01 &              \\
             & 10    & 00    & -0.215258E+00 &              \\
             & 11c   & 00    &  0.483480E-01 &              \\
        \bottomrule
	\end{tabular}
	\caption{
       L1pol $x=0.0$ (IP)}
	\label{}
\end{table}

\begin{table}[ht]
    \newcolumntype{d}{D{.}{.}{9}}
	\begin{tabular}{llldd}
        \toprule
		Pair & $l_a$ & $l_b$ &  \rho         &  \alpha      \\
        \midrule
		O O  & 00    & 00    &  0.574133E+01 & 0.189882E+01 \\
             & 00    & 10    & -0.651900E-02 &              \\
             & 10    & 00    & -0.651900E-02 &              \\
             & 00    & 20    &  0.164840E-01 &              \\
             & 00    & 22c   & -0.118650E+00 &              \\
             & 20    & 00    &  0.164840E-01 &              \\
             & 22c   & 00    & -0.118650E+00 &              \\
		O H  & 00    & 00    &  0.471851E+01 & 0.192714E+01 \\
             & 00    & 10    & -0.279558E+00 &              \\
             & 00    & 11c   &  0.190500E-01 &              \\
             & 10    & 00    & -0.184210E-01 &              \\
             & 20    & 00    &  0.794900E-02 &              \\
             & 22c   & 00    & -0.167358E+00 &              \\
		H H  & 00    & 00    &  0.375748E+01 & 0.200219E+01 \\
             & 00    & 10    & -0.211388E+00 &              \\
             & 00    & 11c   &  0.468130E-01 &              \\
             & 10    & 00    & -0.211388E+00 &              \\
             & 11c   & 00    &  0.468130E-01 &              \\
        \bottomrule
    \end{tabular}
	\caption{L1pol $x=0.5$}
	\label{}
\end{table}

\begin{table}[ht]
    \newcolumntype{d}{D{.}{.}{9}}
	\begin{tabular}{llldd}
        \toprule
		Pair & $l_a$ & $l_b$ &  \rho         &  \alpha      \\
        \midrule
		O O  & 00    & 00    &  0.572103E+01 & 0.190426E+01 \\
             & 00    & 10    & -0.787800E-02 &              \\
             & 10    & 00    & -0.787800E-02 &              \\
             & 00    & 20    &  0.166740E-01 &              \\
             & 00    & 22c   & -0.118533E+00 &              \\
             & 20    & 00    &  0.166740E-01 &              \\
             & 22c   & 00    & -0.118533E+00 &              \\
		O H  & 00    & 00    &  0.473665E+01 & 0.191820E+01 \\
             & 00    & 10    & -0.296016E+00 &              \\
             & 00    & 11c   &  0.207420E-01 &              \\
             & 10    & 00    & -0.171990E-01 &              \\
             & 20    & 00    &  0.847100E-02 &              \\
             & 22c   & 00    & -0.164328E+00 &              \\
		H H  & 00    & 00    &  0.375798E+01 & 0.199913E+01 \\
             & 00    & 10    & -0.207660E+00 &              \\
             & 00    & 11c   &  0.449220E-01 &              \\
             & 10    & 00    & -0.207660E+00 &              \\
             & 11c   & 00    &  0.449220E-01 &              \\
        \bottomrule
	\end{tabular}
	\caption{\small DIFF-L1pol $x=1.0$}
	\label{}
\end{table}

\begin{table}[ht]
    \newcolumntype{d}{D{.}{.}{9}}
	\begin{tabular}{llldd}
        \toprule
		Pair & $l_a$ & $l_b$ &  \rho         &  \alpha      \\
        \midrule
		O O  & 00    & 00    &  0.566957E+01 & 0.194093E+01 \\
             & 00    & 10    & -0.693800E-02 &              \\
             & 10    & 00    & -0.693800E-02 &              \\
             & 00    & 20    &  0.162530E-01 &              \\
             & 00    & 22c   & -0.116227E+00 &              \\
             & 20    & 00    &  0.162530E-01 &              \\
             & 22c   & 00    & -0.116227E+00 &              \\
		O H  & 00    & 00    &  0.477758E+01 & 0.189670E+01 \\
             & 00    & 10    & -0.315240E+00 &              \\
             & 00    & 11c   &  0.248030E-01 &              \\
             & 10    & 00    & -0.164550E-01 &              \\
             & 20    & 00    &  0.953500E-02 &              \\
             & 22c   & 00    & -0.158684E+00 &              \\
		H H  & 00    & 00    &  0.377554E+01 & 0.196057E+01 \\
             & 00    & 10    & -0.207309E+00 &              \\
             & 00    & 11c   &  0.406770E-01 &              \\
             & 10    & 00    & -0.207309E+00 &              \\
             & 11c   & 00    &  0.406770E-01 &              \\
        \bottomrule
	\end{tabular}
	\caption{L1pol $x=1.5$}
	\label{}
\end{table}

\begin{table}[ht]
    \newcolumntype{d}{D{.}{.}{9}}
	\begin{tabular}{llldd}
        \toprule
		Pair & $l_a$ & $l_b$ &  \rho         &  \alpha      \\
        \midrule
		O O  & 00    & 00    &  0.570240E+01 & 0.188230E+01 \\
             & 00    & 10    & -0.288290E-01 &              \\
             & 10    & 00    & -0.288290E-01 &              \\
             & 00    & 20    &  0.829100E-02 &              \\
             & 00    & 22c   & -0.122959E+00 &              \\
             & 20    & 00    &  0.829100E-02 &              \\
             & 22c   & 00    & -0.122959E+00 &              \\
		O H  & 00    & 00    &  0.481023E+01 & 0.191122E+01 \\
             & 00    & 10    & -0.338792E+00 &              \\
             & 00    & 11c   &  0.258230E-01 &              \\
             & 10    & 00    & -0.721500E-02 &              \\
             & 20    & 00    &  0.115020E-01 &              \\
             & 22c   & 00    & -0.165928E+00 &              \\
		H H  & 00    & 00    &  0.367186E+01 & 0.201785E+01 \\
             & 00    & 10    & -0.135984E+00 &              \\
             & 00    & 11c   &  0.385040E-01 &              \\
             & 10    & 00    & -0.135984E+00 &              \\
             & 11c   & 00    &  0.385040E-01 &              \\
        \bottomrule
	\end{tabular}
	\caption{DIFF-L2pol $x=1.0$}
	\label{}
\end{table}	

\begin{table}[ht]
    \newcolumntype{d}{D{.}{.}{9}}
	\begin{tabular}{llldd}
        \toprule
		Pair & $l_a$ & $l_b$ &  \rho         &  \alpha      \\
        \midrule
		O O  & 00    & 00    &  0.585650E+01 & 0.184777E+01 \\
             & 00    & 10    &  0.168560E-01 &              \\
             & 10    & 00    &  0.168560E-01 &              \\
             & 00    & 20    &  0.109910E-01 &              \\
             & 00    & 22c   & -0.108392E+00 &              \\
             & 20    & 00    &  0.109910E-01 &              \\
             & 22c   & 00    & -0.108392E+00 &              \\
		O H  & 00    & 00    &  0.467033E+01 & 0.201391E+01 \\
             & 00    & 10    & -0.190363E+00 &              \\
             & 00    & 11c   &  0.127970E-01 &              \\
             & 10    & 00    & -0.126610E-01 &              \\
             & 20    & 00    &  0.400000E-04 &              \\
             & 22c   & 00    & -0.178575E+00 &              \\
		H H  & 00    & 00    &  0.370112E+01 & 0.195166E+01 \\
             & 00    & 10    & -0.166826E+00 &              \\
             & 00    & 11c   &  0.468670E-01 &              \\
             & 10    & 00    & -0.166826E+00 &              \\
             & 11c   & 00    &  0.468670E-01 &              \\
        \bottomrule
	\end{tabular}
	\caption{L3pol $x=0.0$ (IP)}
	\label{}
\end{table}

\begin{table}[ht]
    \newcolumntype{d}{D{.}{.}{9}}
	\begin{tabular}{llldd}
        \toprule
		Pair & $l_a$ & $l_b$ &  \rho         &  \alpha      \\
        \midrule
		O O  & 00    & 00    &  0.578240E+01 & 0.186982E+01 \\
             & 00    & 10    & -0.879600E-02 &              \\
             & 10    & 00    & -0.879600E-02 &              \\
             & 00    & 20    &  0.851400E-02 &              \\
             & 00    & 22c   & -0.120035E+00 &              \\
             & 20    & 00    &  0.851400E-02 &              \\
             & 22c   & 00    & -0.120035E+00 &              \\
		O H  & 00    & 00    &  0.475443E+01 & 0.193819E+01 \\
             & 00    & 10    & -0.254994E+00 &              \\
             & 00    & 11c   &  0.177050E-01 &              \\
             & 10    & 00    & -0.421200E-02 &              \\
             & 20    & 00    &  0.604100E-02 &              \\
             & 22c   & 00    & -0.172302E+00 &              \\
		H H  & 00    & 00    &  0.367475E+01 & 0.202611E+01 \\
             & 00    & 10    & -0.157909E+00 &              \\
             & 00    & 11c   &  0.398460E-01 &              \\
             & 10    & 00    & -0.157909E+00 &              \\
             & 11c   & 00    &  0.398460E-01 &              \\
        \bottomrule
	\end{tabular}
	\caption{L3pol $x=0.5$}
	\label{}
\end{table}	

\begin{table}[ht]
    \newcolumntype{d}{D{.}{.}{9}}
	\begin{tabular}{llldd}
        \toprule
		Pair & $l_a$ & $l_b$ &  \rho         &  \alpha      \\
        \midrule
		O O  & 00    & 00    &  0.573132E+01 & 0.188375E+01 \\
             & 00    & 10    & -0.170380E-01 &              \\
             & 10    & 00    & -0.170380E-01 &              \\
             & 00    & 20    &  0.959900E-02 &              \\
             & 00    & 22c   & -0.122018E+00 &              \\
             & 20    & 00    &  0.959900E-02 &              \\
             & 22c   & 00    & -0.122018E+00 &              \\
		O H  & 00    & 00    &  0.478851E+01 & 0.190116E+01 \\
             & 00    & 10    & -0.308985E+00 &              \\
             & 00    & 11c   &  0.193160E-01 &              \\
             & 10    & 00    & -0.472900E-02 &              \\
             & 20    & 00    &  0.899500E-02 &              \\
             & 22c   & 00    & -0.169783E+00 &              \\
		H H  & 00    & 00    &  0.369505E+01 & 0.203243E+01 \\
             & 00    & 10    & -0.151968E+00 &              \\
             & 00    & 11c   &  0.375710E-01 &              \\
             & 10    & 00    & -0.151968E+00 &              \\
             & 11c   & 00    &  0.375710E-01 &              \\
        \bottomrule
	\end{tabular}
	\caption{DIFF-L3pol $x=1.0$}
	\label{}
\end{table}		  

\begin{table}[ht]
    \newcolumntype{d}{D{.}{.}{9}}
	\begin{tabular}{llldd}
        \toprule
		Pair & $l_a$ & $l_b$ &  \rho         &  \alpha      \\
        \midrule
		O O  & 00    & 00    &  0.566135E+01 & 0.192579E+01 \\
             & 00    & 10    & -0.163670E-01 &              \\
             & 10    & 00    & -0.163670E-01 &              \\
             & 00    & 20    &  0.108010E-01 &              \\
             & 00    & 22c   & -0.116228E+00 &              \\
             & 20    & 00    &  0.108010E-01 &              \\
             & 22c   & 00    & -0.116228E+00 &              \\
		O H  & 00    & 00    &  0.481617E+01 & 0.187771E+01 \\
             & 00    & 10    & -0.362695E+00 &              \\
             & 00    & 11c   &  0.232330E-01 &              \\
             & 10    & 00    & -0.641400E-02 &              \\
             & 20    & 00    &  0.965000E-02 &              \\
             & 22c   & 00    & -0.169048E+00 &              \\
		H H  & 00    & 00    &  0.374351E+01 & 0.197627E+01 \\
             & 00    & 10    & -0.135674E+00 &              \\
             & 00    & 11c   &  0.385730E-01 &              \\
             & 10    & 00    & -0.135674E+00 &              \\
             & 11c   & 00    &  0.385730E-01 &              \\
        \bottomrule
	\end{tabular}
	\caption{L3pol $x=1.5$}
	\label{}
\end{table}		  

\begin{table}[ht]
	\begin{tabular}{lll|l|l}
        \toprule
                     & IP    &         & DIFF  &         \\
		             & $x=0$ & $x=0.5$ & $x=1$ & $x=1.5$ \\
        \midrule
		L3:          &       &         &       &         \\
		$\beta_{OO}$ & 1.926 & 1.588   & 1.25  & 0.912   \\
		$\beta_{OH}$ & 1.926 & 1.698   & 1.47  & 1.242   \\
		$\beta_{HH}$ & 1.926 & 1.963   & 2.00  & 2.037   \\ \hline
		L2:          &       &         &       &         \\
		$\beta_{OO}$ & --    & --      & 1.25  & --      \\
		$\beta_{OH}$ & --    & --      & 1.57  & --      \\
		$\beta_{HH}$ & --    & --      & 2.00  & --      \\ \hline
		L1:          &       &         &       &         \\
		$\beta_{OO}$ & 1.926 & 1.588   & 1.25  & 0.912   \\
		$\beta_{OH}$ & 1.926 & 1.803   & 1.68  & 1.557   \\
		$\beta_{HH}$ & 1.926 & 1.963   & 2.00  & 2.037   \\
        \bottomrule
	\end{tabular}
	\caption{
        Polarisation damping parameters used for each model used in this work.
        The column titled ``IP'' indicates the damping based on the ionization 
        potential of water (see text for details), and ``DIFF'' indicates the 
        optimized damping for the DIFF models. 
        This table is also presented in the main paper.
    }
	\label{tab:poldamping}
\end{table}

\begin{table}[ht]
    \newcolumntype{d}{D{.}{.}{6}}
	\begin{tabular}{lddddd}
		\toprule
		Pair ($ab$) & \C{ab}{6}    & \C{ab}{8}   & \C{ab}{10}  & \C{ab}{12}  & \BETAdisp{ab} \\
        \midrule
		O O         & 24.34089  & 489.9063 & 12519.45 & 238364.1 & 1.7794 \\
		O H         & 4.335086  & 55.94859 & 1174.193 & 13116.46 & 1.9011 \\
		H H         & 0.7833591 & 4.356823 & 90.61106 & 771.3764 & 2.0227 \\
        \bottomrule
	\end{tabular}
	\caption{
       Dispersion coefficients from the localized ISA-Pol model and site-site damping 
       parameters. All terms in atomic units.
    }
	\label{}
\end{table}

\begin{table}[hb]
    \newcolumntype{d}{D{.}{.}{9}}
	\begin{tabular}{ldld}
        \toprule
		O   &	          & &  \\
		$t$	&   \Q{O}{t}  & &  \\
        \midrule               
		00  &   -0.825458 & &  \\
		10  &   -0.170731 & &  \\
		20  &    0.013320 & &  \\
		22c &    0.446098 & &  \\
		30  &   -0.111202 & &  \\
		32c &   -0.116581 & &  \\
		40  &   -0.395115 & &  \\
		42c &    0.449626 & &  \\
		44c &    0.017959 & &  \\
        \midrule
		H1  &              & H2  &     \\
		$t$	&   \Q{H1}{t}  & $t$ & \Q{H2}{t}   \\
        \midrule                       
		00  &    0.413222  & 00	 &  0.413227  \\
		10  &    0.016268  & 10	 &  0.016266  \\
		11c &   -0.022715  & 11c & -0.022713  \\
		20  &    0.026170  & 20	 &  0.026166  \\
		21c &   -0.012333  & 21c & -0.012333  \\
		22c &    0.023062  & 22c &  0.023061  \\
		30  &    0.022590  & 30	 &  0.022592  \\
		31c &    0.009128  & 31c &  0.009123  \\
		32c &   -0.000725  & 32c & -0.000723  \\
		33c &   -0.001680  & 33c & -0.001680  \\
		40  &   -0.047819  & 40	 & -0.047805  \\
		41c &    0.039014  & 41c &  0.039005  \\
		42c &   -0.027919  & 42c & -0.027914  \\
		43c &   -0.000806  & 43c & -0.000803  \\
		44c &    0.005395  & 44c &  0.005393  \\
        \bottomrule
	\end{tabular}
	\caption{Non-zero components of the DF-ISA rank 4 multipole model in the local axes frame.
       Note that symmetry is not imposed so that there are small differences in the multipoles
       on the two hydrogen sites.}
	\label{table:waterisa}
\end{table}

\begin{table}
	\caption{$\A{OO}{tu}$ for the ISA-Pol L3pol model. Terms are expressed in the local-axis
    frame. Units are atomic units.}
    \newcolumntype{d}{D{.}{.}{12}}
	\begin{tabular}{ldld}
        \toprule
		$t,u$   &  \A{OO}{tu}     & $t,u$   &  \A{OO}{tu}      \\
        \midrule
		10,10   & 6.938561067132  & 21s,31s & -5.695850943476  \\
		10,20   & 0.283014503203  & 21s,33s & -5.544559210436  \\
		10,22c  & 1.297065784153  & 22c,22c & 31.003555795636  \\
		10,30   & -1.924692628858 & 22c,30  & -2.134821966453  \\
		10,32c  & -5.065926516432 & 22c,32c & 6.387970997851   \\
		11c,11c & 6.758614571817  & 22s,22s & 28.975306251152  \\
		11c,21c & 2.609015559476  & 22s,32s & -5.423168252926  \\
		11c,31c & 5.869123330783  & 30,30   & 197.513252679545 \\
		11c,33c & -6.605301843408 & 30,31c  & -0.013502436204  \\
		11s,11s & 7.522295544935  & 30,21c  & 39.341637391107  \\
		11s,21s & 0.534498269999  & 30,33c  & -0.016042300561  \\
		11s,31s & -3.263639064343 & 31c,31c & 196.876749295609 \\
		11s,33s & -9.492140869903 & 31c,32c & 0.025580817673   \\
		20,20   & 27.634920511971 & 31c,33c & 22.885787318830  \\
		20,22s  & 6.178197154040  & 31s,31s & 194.533538449987 \\
		20,30   & -1.850509303928 & 31s,33s & 32.298750884673  \\
		20,32c  & -1.859913227916 & 32c,32c & 237.101696439329 \\
		21c,21c & 27.989582837200 & 32c,33c & 0.025727884497   \\
		21c,31c & 3.459918679126  & 32s,32s & 235.740220888427 \\
		21c,33c & 5.231198193522  & 33c,33c & 258.178588469019 \\
		21s,21s & 33.110619559210 & 33s,33s & 202.634991496274 \\
        \bottomrule
	\end{tabular}
\end{table}

\begin{table}
	\caption{$\A{HH}{tu}$ for the ISA-Pol L3pol model. Terms are expressed in the local-axis
    frame. Units are atomic units.}
    \newcolumntype{d}{D{.}{.}{12}}
	\begin{tabular}{ldld}
        \toprule
		$t,u$   &  \A{HH}{tu}     & $t,u$   &  \A{HH}{tu}      \\
        \midrule
		10,10   & 2.133530241229  & 21c,30  & 0.491372594083  \\
		10,11c  & 0.014806076479  & 21c,31c & 3.244889231824  \\
		10,21c  & -1.962195867167 & 21c,32c & -0.164192398715 \\
		10,21s  & 0.558394807976  & 21c,33c & 0.502961820740  \\
		10,22c  & -0.139007994819 & 21s,21s & 0.862161214828  \\
		10,30   & 2.624600448864  & 21s,22s & 0.067763055234  \\
		10,31c  & -0.387671591787 & 21s,31s & 4.658283502883  \\
		10,32c  & -0.285952156735 & 21s,32s & -0.299828269759 \\
		10,33c  & -0.039088323730 & 21s,33s & 0.123635495859  \\
		11c,10  & 0.014806076479  & 22c,22c & 2.258497604970  \\
		11c,11c & 0.768893987100  & 22c,30  & -0.014559136681 \\
		11c,20  & -0.060959104003 & 22c,31c & 0.073289640942  \\
		11c,21c & 0.204839449553  & 22c,32c & 1.932260882992  \\
		11c,22c & 0.146065219982  & 22c,33c & 0.846069547066  \\
		11c,30  & -0.084814571913 & 22s,22s & 2.358486780599  \\
		11c,31c & 0.025623412166  & 22s,31s & -0.083428643181 \\
		11c,32c & 0.374664191337  & 22s,32s & 2.019910136457  \\
		11c,33c & -0.399931921642 & 22s,33s & 0.526987026651  \\
		11s,11s & 0.787159280680  & 30,30   & -2.732173577046 \\
		11s,21s & 0.553458688663  & 30,31c  & -0.976616012299 \\
		11s,22s & 0.183476194197  & 30,32c  & 5.916676794459  \\
		11s,31s & -0.581574706631 & 30,33c  & 2.459537631952  \\
		11s,32s & 0.335756030493  & 31c,31c & 11.389988339701 \\
		11s,33s & -0.332939366208 & 31c,32c & 4.464304862624  \\
		20,20   & 4.693088404029  & 31c,33c & 4.569874943117  \\
		20,21c  & -0.677855615155 & 31s,31s & 13.964067932662 \\
		20,22c  & 0.725066406875  & 31s,32s & 3.094566951749  \\
		20,30   & 1.210062157884  & 31s,33s & -0.301357361307 \\
		20,31c  & 1.041758490711  & 32c,32c & 0.281721397105  \\
		20,32c  & 0.105983816308  & 32c,33c & -2.903496260436 \\
		20,33c  & 0.018679739312  & 32s,32s & -4.032641998208 \\
		21c,21c & 1.423976650100  & 32s,33s & 0.783324615248  \\
		21c,22c & 0.121006332199  & 33s,33s & 11.308337868268 \\
        \bottomrule
	\end{tabular}
\end{table}
\clearpage

\begin{table}
    \newcolumntype{d}{D{.}{.}{12}}
	\begin{tabular}{ld}
        \toprule
		$t,u$   &  \A{OO}{tu}       \\
        \midrule
		10,10   & 6.978439619408  \\
		10,20   & 0.298266643849  \\
		10,22c  & 1.284142535503  \\
		11c,11c & 6.923882571524  \\
		11c,21c & 2.590522565151  \\
		11s,11s & 7.712578185325  \\
		11s,21s & 0.590880476972  \\
		20,20   & 27.998687464761 \\
		20,22c  & 6.115649837756  \\
		21c,21c & 27.431664736301 \\
		21s,21s & 33.400963369514 \\
		22c,22c & 31.418111682887 \\
		22s,22s & 30.005436201383 \\
        \bottomrule
	\end{tabular}
	\caption{$\A{OO}{tu}$ for the ISA-Pol L2pol model. Terms are expressed in the local-axis
    frame. Units are atomic units.}
\end{table}

\begin{table}
    \newcolumntype{d}{D{.}{.}{12}}
	\begin{tabular}{llll}
        \toprule
		$t,u$   &  \A{HH}{tu}     & $t,u$   &  \A{HH}{tu}      \\
        \midrule
		10,10   & 2.080105897425  & 11s,22s & 0.302067978718  \\
		10,11c  & 0.053116146435  & 20,20   & 4.814477514669  \\
		10,20   & -1.854642739739 & 20,21c  & -0.684744785441 \\
		10,21c  & 0.655677997532  & 20,22c  & 0.730912286810  \\
		10,22c  & -0.203666996438 & 21c,21c & 1.438600109316  \\
		11c,11c & 0.712203468085  & 21c,22c & 0.121807515267  \\
		11c,20  & -0.123521050531 & 21,21s  & 0.870977300150  \\
		11c,21c & 0.166372245640  & 21s,22s & 0.064467545108  \\
		11c,22c & 0.159427069532  & 22c,22c & 2.257351971348  \\
		11s,11s & 0.699536201785  & 22s,22s & 2.378902651023  \\
		11s,21s & 0.443538935284  &         &                 \\
        \bottomrule
	\end{tabular}
	\caption{$\A{HH}{tu}$ for the ISA-Pol L2pol model. Terms are expressed in the local-axis
    frame. Units are atomic units.}
\end{table}

\begin{table}
    \newcolumntype{d}{D{.}{.}{12}}
	\begin{tabular}{ld}
        \toprule
		$t,u$   &  \A{OO}{tu}       \\
        \midrule
		10,10    &   6.636912518476   \\
		11c,11c  &   6.397939492453   \\
		11s,11s  &   6.746166610856   \\
        \bottomrule
	\end{tabular}
	\caption{$\A{OO}{tu}$ for the ISA-Pol L1pol model. Terms are expressed in the local-axis
    frame. Units are atomic units.}
\end{table}

\begin{table}
    \newcolumntype{d}{D{.}{.}{12}}
	\begin{tabular}{ld}
        \toprule
		$t,u$   &  \A{HH}{tu}       \\
        \midrule
		10,10    &  2.221233947114    \\
		10,11c   &  -0.013585535815   \\
		11c,11c  &   1.021508106671   \\
		11s,11s  &   1.188309162710   \\
        \bottomrule
	\end{tabular}
	\caption{$\A{HH}{tu}$ for the ISA-Pol L1pol model. Terms are expressed in the local-axis
    frame. Units are atomic units.}
\end{table}

\clearpage

\section{Plots for the two-body interaction}

\begin{figure}
	\caption{
		The total interaction energy along the profile of the dimer global minimum for all DIFF models.
	}
	\begin{center}
		\includegraphics[scale=0.4]{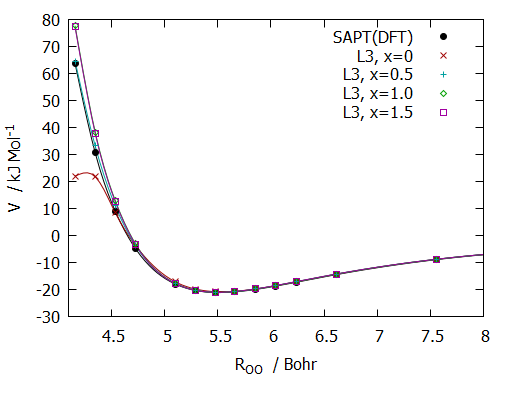}
		\includegraphics[scale=0.4]{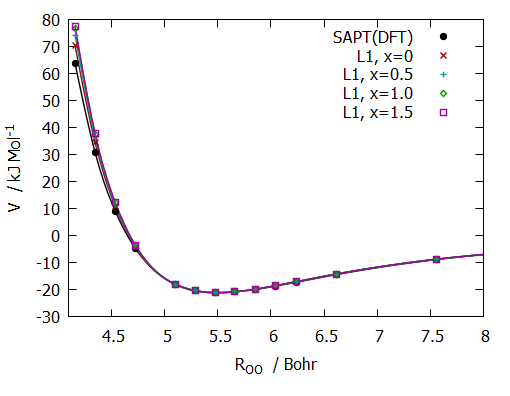}
		\includegraphics[scale=0.4]{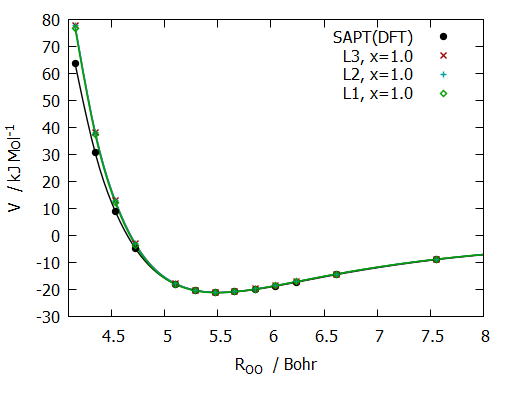}
	\end{center}
	\label{fig:profile}
\end{figure}

\begin{figure}
	\caption{
		Scatter plot of total interaction energy for all DIFF models $V$ versus SAPT(DFT) total interaction energy $E$.
	}
	\begin{center}
		\includegraphics[scale=0.4]{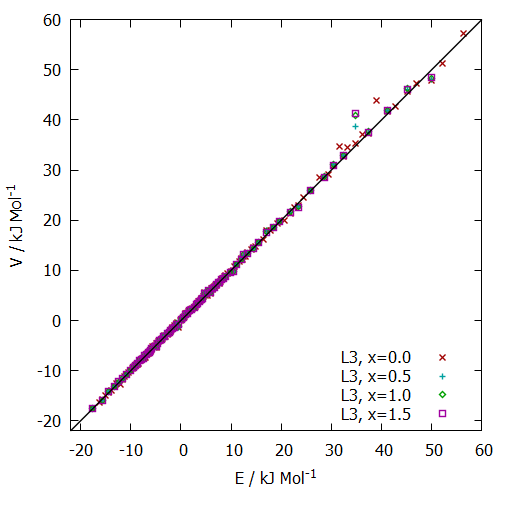}
		\includegraphics[scale=0.4]{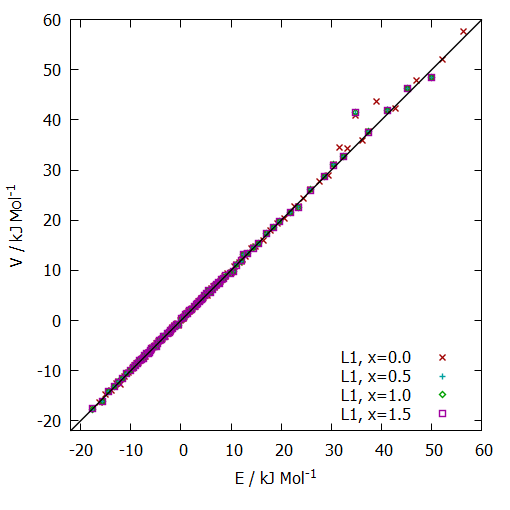}
		\includegraphics[scale=0.4]{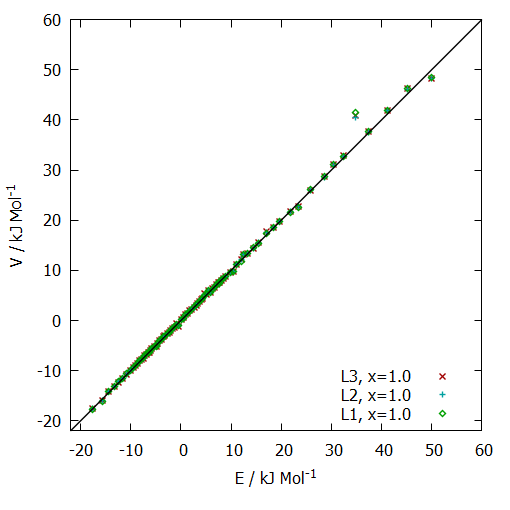}
	\end{center}
	\label{fig:scatter}
\end{figure}

\clearpage

\subsection{Second virial coefficient}

The second virial coefficient $B(T)$ is calculated using
\begin{equation}
B(T) = -\frac{1}{2} \int \int (e^{\EINT / kT} - 1) d \Omega dr^3 + \frac{\hbar^2}{24 (kT)^3}(\frac{<\mathbf{F}^2>_0}{M}+\sum_{\alpha}\frac{<\mathbf{T_{\alpha}}^2>_0}{I_{\alpha \alpha}})
\end{equation}
where the first term above is the classical result $B(T)_{\mathrm{Cl}}$ from integrating the Mayer function (the integration here is over separations and orientations) and the second term gives the quantum correction. Here $<\mathbf{F}^2>_0$ and $<\mathbf{T_{\alpha}}^2>_0$ are the mean square force and components of mean square torque on the molecule respectively and $I_{\alpha \alpha}$ are the molecule's moments of inertia.

\begin{figure}[h]
	\caption{\small Second virial coefficient for water for all L3 models and for the $x=1.0$ L2 and L1 models  plotted against temperature. Experimental data taken from Mas et al. (2000) \cite{MasBSG00}. }
	\includegraphics[scale=0.4]{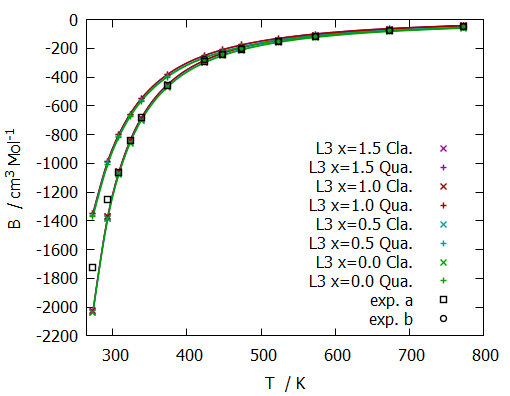}
	\includegraphics[scale=0.4]{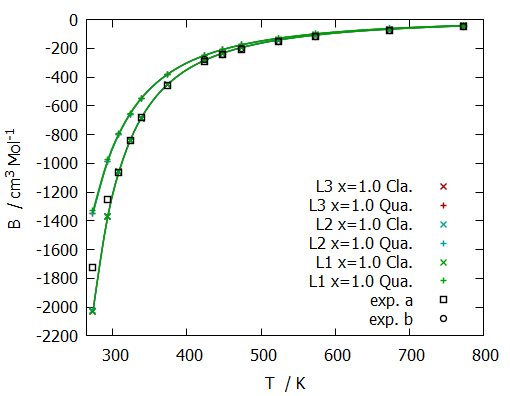}
	\label{fig:H2Ovirial}
\end{figure}

\clearpage

\section{Water cluster structures}
Sources for the water cluster structures:
\begin{itemize}
  \item \waterN{3}: Two sets were used. One from Liu \etal \cite{liu2017capturing} and the 
    600 trimer set from Akin-Ojo and Szalewicz \cite{akin-ojo_how_2013}.
  \item \waterN{6}: These structures were taken from Bates \& Tschumper \cite{bates2009ccsd}.
  \item \waterN{16}: These were from Yoo and Xantheas \cite{yoo2017structures} and are also 
    provided by G{\'o}ra \etal \cite{gora_predictions_2014} in their SI.
    Note that for the boat-a structure given in these references the O and H sites are not 
    ordered. The structure with the correct ordering is provided below.
  \item \waterN{24}: As for \waterN{16}. 
\end{itemize}

\subsection{\waterN{16} boat-a isomer}
This is the structure with the atoms ordered according to the water molecule they belong to:
\begin{verbatim}
48
boat-a: MP2/aug-cc-pVTZ (opt) E = -1221.536142 a.u.
O  0.12742587  1.14684972 -1.58027855
H -0.71725804  1.53849679 -1.87225414
H -0.00427839  0.17400371 -1.58650759
O  4.23247769 -0.26271901 -1.24153281
H  4.97071820 -0.25118561 -1.85737926
H  3.72092881  0.57286259 -1.40972697
O -2.42866591  2.23165734 -1.85770461
H -2.68572875  3.03888825 -2.31286191
H -2.49329627  2.42735616 -0.88330238
O -4.23429048  0.26350591  1.23919725
H -3.72176515 -0.57149271  1.40711309
H -4.97199711  0.25127986  1.85568760
O  2.76159464  2.38202662  1.41229013
H  3.27367235  1.54634517  1.57145967
H  3.16989060  3.05563656  1.96351943
O  2.52685413 -2.46453621 -0.77925849
H  1.63876110 -2.24899979 -1.11640748
H  3.10657890 -1.74710268 -1.09353796
O -2.52476503  2.46384789  0.77992184
H -1.63736906  2.24746385  1.11841417
H -3.10585832  1.74726359  1.09349706
O -2.76186721 -2.38048819 -1.41506704
H -3.27465488 -1.54522054 -1.57436672
H -3.16870741 -3.05438894 -1.96703452
O -2.79264028 -1.95698424  1.44172471
H -1.85687894 -1.75682833  1.63115432
H -2.78903010 -2.31202936  0.53726329
O  4.08043374  0.07727714  1.57432064
H  3.56805875 -0.70756435  1.83742253
H  4.32624834 -0.10262958  0.64945285
O -0.11657254 -1.55258996 -1.14833997
H -0.97416083 -1.94951222 -1.39379643
H -0.16336722 -1.46208551 -0.17107250
O  2.79243517  1.95862738 -1.44410922
H  1.85680583  1.75649358 -1.63217101
H  2.78916008  2.31417338 -0.53981244
O -4.08049979 -0.07673860 -1.57722005
H -3.56777805  0.70839215 -1.83878183
H -4.32866081  0.10305124 -0.65302683
O -0.12623544 -1.14911860  1.58450896
H  0.71887096 -1.54099591  1.87520218
H  0.00519896 -0.17621894  1.59153759
O  2.43054159 -2.23102984  1.85838070
H  2.49447002 -2.42706168  0.88398964
H  2.68572680 -3.03885326  2.31351095
O  0.11727910  1.55092462   1.15243575
H  0.97484648  1.94864826   1.39674109
H  0.16355228  1.45911335   0.17525371
\end{verbatim}

\section{Hexamer data}

\begin{table}[ht]
 \newcolumntype{d}{D{.}{.}{3}}
 \begin{tabular}{lddddddd}
   \toprule
      Model    &\multicolumn{1}{c}{ \text{2B-6B} }
                            &\multicolumn{1}{c}{ \text{2B} }
                                       &\multicolumn{1}{c}{ \text{3B} }
                                                   &\multicolumn{1}{c}{ \text{4B} }
                                                            &\multicolumn{1}{c}{ \text{5B} }
                                                                     &\multicolumn{1}{c}{ \text{6B} }
                                                                              &\multicolumn{1}{c}{ \text{>2B} } \\
   \midrule
     \multicolumn{8}{l}{Prism} \\
   \midrule
     CCSD(T)-F12 & -200.957 &  -161.711 &  -36.735 &  -2.761 &   0.251 &      0.000 &  -39.245  \\ 
     MB-pol      & -201.543 &  -163.008 &  -36.568 &  -2.175 &   0.209 &      0.000 &  -38.534  \\
     DIFF-L3pol  & -202.522 &  -158.044 &  -40.359 &  -4.429 &   0.299 &      0.012 &  -44.478  \\ 
     DIFF-L2pol  & -201.371 &  -158.025 &  -39.561 &  -4.154 &   0.357 &      0.012 &  -43.345  \\ 
     DIFF-L1pol  & -201.491 &  -158.930 &  -38.532 &  -4.227 &   0.188 &      0.010 &  -42.560  \\  
   \bottomrule
 \end{tabular}
 \caption{
    Water hexamer isomer intermolecular energies and many-body decomposition. 
    Reference CCSD(T)-F12 and MB-pol model energies are from Medders \etal 
    \cite{medders_representation_2015}.
    The columns $n$B show the $n$-body non-additive interaction energies, and the total interaction 
    energy is given in column ``2B-6B''. The sum of the terms of 3B to 6B is given in column ``>2B''.
    All energies are in \kJmol.
 }
 \label{tab:hexamer_prism}
\end{table}

\begin{table}[ht]
 \newcolumntype{d}{D{.}{.}{3}}
 \begin{tabular}{lddddddd}
   \toprule
      Model    &\multicolumn{1}{c}{ \text{2B-6B} }
                            &\multicolumn{1}{c}{ \text{2B} }
                                       &\multicolumn{1}{c}{ \text{3B} }
                                                   &\multicolumn{1}{c}{ \text{4B} }
                                                            &\multicolumn{1}{c}{ \text{5B} }
                                                                     &\multicolumn{1}{c}{ \text{6B} }
                                                                              &\multicolumn{1}{c}{ \text{>2B} } \\

   \midrule
     \multicolumn{8}{l}{Cage} \\
   \midrule
     CCSD(T)-F12 & -199.869 &  -159.828 &  -37.823 &  -2.217 &   0.041 &      0.000 &  -39.999  \\  
     MB-pol      & -200.204 &  -161.000 &  -37.363 &  -1.966 &   0.125 &      0.000 &  -39.204  \\
     DIFF-L3pol  & -199.435 &  -154.967 &  -40.915 &  -3.686 &   0.160 &     -0.025 &  -44.467  \\ 
     DIFF-L2pol  & -197.694 &  -154.743 &  -39.758 &  -3.356 &   0.188 &     -0.024 &  -42.951  \\ 
     DIFF-L1pol  & -198.644 &  -156.088 &  -39.077 &  -3.528 &   0.072 &     -0.021 &  -42.555  \\  
   \bottomrule
 \end{tabular}
 \caption{
    Water hexamer isomer intermolecular energies and many-body decomposition. 
    See the caption to \tabrff{tab:hexamer_prism} for an explanation of the columns.
 }
\end{table}

\begin{table}[ht]
 \newcolumntype{d}{D{.}{.}{3}}
 \begin{tabular}{lddddddd}
   \toprule
      Model    &\multicolumn{1}{c}{ \text{2B-6B} }
                            &\multicolumn{1}{c}{ \text{2B} }
                                       &\multicolumn{1}{c}{ \text{3B} }
                                                   &\multicolumn{1}{c}{ \text{4B} }
                                                            &\multicolumn{1}{c}{ \text{5B} }
                                                                     &\multicolumn{1}{c}{ \text{6B} }
                                                                              &\multicolumn{1}{c}{ \text{>2B} } \\
   \midrule
     \multicolumn{8}{l}{Book-1} \\
   \midrule
     CCSD(T)-F12 & -198.070 &  -149.619 &  -43.764 &  -4.518 &  -0.167 &      0.000 &  -48.450  \\
     MB-pol      & -196.648 &  -149.787 &  -42.927 &  -3.849 &  -0.041 &      0.000 &  -46.818  \\
     DIFF-L3pol  & -197.651 &  -146.005 &  -44.903 &  -6.337 &  -0.355 &     -0.049 &  -51.645  \\ 
     DIFF-L2pol  & -196.382 &  -145.936 &  -44.022 &  -6.068 &  -0.308 &     -0.044 &  -50.445  \\ 
     DIFF-L1pol  & -198.735 &  -147.359 &  -44.480 &  -6.434 &  -0.409 &     -0.051 &  -51.376  \\  
   \bottomrule
 \end{tabular}
 \caption{
    Water hexamer isomer intermolecular energies and many-body decomposition. 
    See the caption to \tabrff{tab:hexamer_prism} for an explanation of the columns.
 }
\end{table}

\begin{table}[ht]
 \newcolumntype{d}{D{.}{.}{3}}
 \begin{tabular}{lddddddd}
   \toprule
      Model    &\multicolumn{1}{c}{ \text{2B-6B} }
                            &\multicolumn{1}{c}{ \text{2B} }
                                       &\multicolumn{1}{c}{ \text{3B} }
                                                   &\multicolumn{1}{c}{ \text{4B} }
                                                            &\multicolumn{1}{c}{ \text{5B} }
                                                                     &\multicolumn{1}{c}{ \text{6B} }
                                                                              &\multicolumn{1}{c}{ \text{>2B} } \\
   \midrule
     \multicolumn{8}{l}{Book-2} \\
   \midrule
     CCSD(T)-F12 & -196.899 &  -150.038 &  -42.593 &  -4.184 &  -0.083 &      0.000 &  -46.860  \\
     MB-pol      & -195.853 &  -150.331 &  -41.965 &  -3.556 &   0.000 &      0.000 &  -45.521  \\
     DIFF-L3pol  & -196.080 &  -145.769 &  -43.950 &  -6.043 &  -0.277 &     -0.039 &  -50.310  \\ 
     DIFF-L2pol  & -194.686 &  -145.725 &  -42.989 &  -5.715 &  -0.221 &     -0.034 &  -48.961  \\ 
     DIFF-L1pol  & -196.131 &  -147.212 &  -42.741 &  -5.860 &  -0.280 &     -0.036 &  -48.919  \\  
   \bottomrule
 \end{tabular}
 \caption{
    Water hexamer isomer intermolecular energies and many-body decomposition. 
    See the caption to \tabrff{tab:hexamer_prism} for an explanation of the columns.
 }
\end{table}

\begin{table}[ht]
 \newcolumntype{d}{D{.}{.}{3}}
 \begin{tabular}{lddddddd}
   \toprule
      Model    &\multicolumn{1}{c}{ \text{2B-6B} }
                            &\multicolumn{1}{c}{ \text{2B} }
                                       &\multicolumn{1}{c}{ \text{3B} }
                                                   &\multicolumn{1}{c}{ \text{4B} }
                                                            &\multicolumn{1}{c}{ \text{5B} }
                                                                     &\multicolumn{1}{c}{ \text{6B} }
                                                                              &\multicolumn{1}{c}{ \text{>2B} } \\
   \midrule
     \multicolumn{8}{l}{Bag} \\
   \midrule
     CCSD(T)-F12 & -195.016 &  -146.481 &  -43.639 &  -4.853 &  -0.083 &      0.041 &  -48.576  \\
     MB-pol      & -193.719 &  -147.444 &  -42.467 &  -3.765 &  -0.041 &      0.000 &  -46.275  \\
     DIFF-L3pol  & -195.552 &  -143.327 &  -44.970 &  -6.769 &  -0.531 &      0.047 &  -52.224  \\ 
     DIFF-L2pol  & -193.862 &  -143.275 &  -43.878 &  -6.313 &  -0.441 &      0.047 &  -50.586  \\ 
     DIFF-L1pol  & -195.347 &  -144.690 &  -43.828 &  -6.393 &  -0.481 &      0.044 &  -50.657  \\  
   \bottomrule
 \end{tabular}
 \caption{
    Water hexamer isomer intermolecular energies and many-body decomposition. 
    See the caption to \tabrff{tab:hexamer_prism} for an explanation of the columns.
 }
\end{table}

\begin{table}[ht]
 \newcolumntype{d}{D{.}{.}{3}}
 \begin{tabular}{lddddddd}
   \toprule
      Model    &\multicolumn{1}{c}{ \text{2B-6B} }
                            &\multicolumn{1}{c}{ \text{2B} }
                                       &\multicolumn{1}{c}{ \text{3B} }
                                                   &\multicolumn{1}{c}{ \text{4B} }
                                                            &\multicolumn{1}{c}{ \text{5B} }
                                                                     &\multicolumn{1}{c}{ \text{6B} }
                                                                              &\multicolumn{1}{c}{ \text{>2B} } \\
   \midrule
     \multicolumn{8}{l}{Ring} \\
   \midrule
     CCSD(T)-F12 & -193.593 &  -135.645 &  -49.664 &  -7.447 &  -0.794 &     -0.041 &  -57.906  \\
     MB-pol      & -190.915 &  -135.728 &  -48.701 &  -6.066 &  -0.418 &      0.000 &  -55.186  \\
     DIFF-L3pol  & -192.801 &  -134.565 &  -47.909 &  -8.854 &  -1.340 &     -0.131 &  -58.236  \\ 
     DIFF-L2pol  & -192.230 &  -134.785 &  -47.508 &  -8.563 &  -1.255 &     -0.118 &  -57.445  \\ 
     DIFF-L1pol  & -197.511 &  -136.184 &  -49.872 &  -9.725 &  -1.557 &     -0.171 &  -61.327  \\  
   \bottomrule
 \end{tabular}
 \caption{
    Water hexamer isomer intermolecular energies and many-body decomposition. 
    See the caption to \tabrff{tab:hexamer_prism} for an explanation of the columns.
 }
\end{table}

\begin{table}[ht]
 \newcolumntype{d}{D{.}{.}{3}}
 \begin{tabular}{lddddddd}
   \toprule
      Model    &\multicolumn{1}{c}{ \text{2B-6B} }
                            &\multicolumn{1}{c}{ \text{2B} }
                                       &\multicolumn{1}{c}{ \text{3B} }
                                                   &\multicolumn{1}{c}{ \text{4B} }
                                                            &\multicolumn{1}{c}{ \text{5B} }
                                                                     &\multicolumn{1}{c}{ \text{6B} }
                                                                              &\multicolumn{1}{c}{ \text{>2B} } \\
   \midrule
     \multicolumn{8}{l}{Cyclic-boat-1} \\
   \midrule
     CCSD(T)-F12 & -189.367 &  -133.971 &  -47.864 &  -6.819 &  -0.669 &     -0.041 &  -55.354  \\
     MB-pol      & -187.275 &  -133.971 &  -47.279 &  -5.648 &  -0.376 &      0.000 &  -53.304  \\
     DIFF-L3pol  & -187.391 &  -131.902 &  -46.071 &  -8.154 &  -1.149 &     -0.112 &  -55.488  \\ 
     DIFF-L2pol  & -187.222 &  -132.176 &  -45.916 &  -7.938 &  -1.087 &     -0.102 &  -55.046  \\ 
     DIFF-L1pol  & -191.391 &  -133.581 &  -47.585 &  -8.783 &  -1.301 &     -0.139 &  -57.809  \\  
   \bottomrule
 \end{tabular}
 \caption{
    Water hexamer isomer intermolecular energies and many-body decomposition. 
    See the caption to \tabrff{tab:hexamer_prism} for an explanation of the columns.
 }
\end{table}

\begin{table}[ht]
 \newcolumntype{d}{D{.}{.}{3}}
 \begin{tabular}{lddddddd}
   \toprule
      Model    &\multicolumn{1}{c}{ \text{2B-6B} }
                            &\multicolumn{1}{c}{ \text{2B} }
                                       &\multicolumn{1}{c}{ \text{3B} }
                                                   &\multicolumn{1}{c}{ \text{4B} }
                                                            &\multicolumn{1}{c}{ \text{5B} }
                                                                     &\multicolumn{1}{c}{ \text{6B} }
                                                                              &\multicolumn{1}{c}{ \text{>2B} } \\
   \midrule
     \multicolumn{8}{l}{Cyclic-boat-2} \\
   \midrule
     CCSD(T)-F12 & -188.949 &  -133.720 &  -47.823 &  -6.736 &  -0.669 &     -0.041 &  -55.228  \\
     MB-pol      & -187.275 &  -133.971 &  -47.237 &  -5.648 &  -0.376 &      0.000 &  -53.262  \\
     DIFF-L3pol  & -187.961 &  -132.491 &  -46.165 &  -8.050 &  -1.141 &     -0.112 &  -55.469  \\ 
     DIFF-L2pol  & -187.691 &  -132.736 &  -45.895 &  -7.865 &  -1.089 &     -0.103 &  -54.954  \\ 
     DIFF-L1pol  & -192.553 &  -134.151 &  -48.133 &  -8.805 &  -1.320 &     -0.141 &  -58.401  \\  
   \bottomrule
 \end{tabular}
 \caption{
    Water hexamer isomer intermolecular energies and many-body decomposition. 
    See the caption to \tabrff{tab:hexamer_prism} for an explanation of the columns.
 }
\end{table}

\clearpage

\section{Energies for 16-mers and 24-mers}

\begin{table}[ht]
{\footnotesize
    \newcolumntype{d}{D{.}{.}{3}}
	\begin{tabular}{ldddddddd}
		\toprule
		Model      &\multicolumn{1}{c}{ \text{2B} }
                              &\multicolumn{1}{c}{ \text{3B} }
                                         &\multicolumn{1}{c}{ \text{4B} }
                                                   &\multicolumn{1}{c}{ \text{5B} }
                                                            &\multicolumn{1}{c}{ \text{>5B} }
                                                                     &\multicolumn{1}{c}{ \text{>4B} }
                                                                              &\multicolumn{1}{c}{ \text{2B-4B} }
                                                                                         &\multicolumn{1}{c}{ \text{3B(opt.geom.)} } \\
        \midrule
		4444-a     &          &          &         &        &        &        &          &            \\ 
        \midrule
		SAMBA      & -572.898 & -135.662 & -3.837  &\text{---}&\text{---}&\text{---}& -712.397 &\text{---}  \\
		DIFF-L3pol & -547.526 & -155.976 & -9.685  & 3.364  & -0.245 & 3.119  & -713.187 & -151.447   \\
		DIFF-L2pol & -546.715 & -148.381 & -8.565  & 3.175  & -0.301 & 2.875  & -703.661 & -144.902   \\
		DIFF-L1pol & -550.645 & -157.518 & -14.594 & 1.521  & -0.320 & 1.201  & -722.757 & -153.392   \\ 
        \midrule
		4444-b     &          &          &         &        &        &        &          &            \\ 
        \midrule
		SAMBA      & -566.288 & -141.013 & -3.933  &\text{---}&\text{---}&\text{---}& -711.234 &\text{---}  \\
		DIFF-L3pol & -541.546 & -160.606 & -11.721 & 2.586  & -0.161 & 2.424  & -713.873 & -155.949   \\
		DIFF-L2pol & -540.797 & -152.345 & -9.809  & 2.756  & -0.159 & 2.596  & -702.951 & -148.789   \\
		DIFF-L1pol & -545.097 & -162.315 & -15.100 & 1.199  & -0.181 & 1.018  & -722.512 & -158.058   \\ 
        \midrule
		boat-a     &          &          &         &        &        &        &          &            \\ 
        \midrule
		SAMBA      &\text{---}&\text{---}&\text{---}&\text{---}&\text{---}&\text{---}&\text{---}&\text{---} \\
		DIFF-L3pol & -531.955 & -167.257 & -17.023 & 1.523  & 0.097  & 1.620  & -716.235 & -162.406  \\
		DIFF-L2pol & -531.351 & -159.495 & -15.345 & 1.670  & 0.077  & 1.746  & -706.191 & -155.735  \\
		DIFF-L1pol & -536.249 & -166.501 & -18.932 & 0.478  & 0.03   & 0.508  & -721.682 & -162.093  \\ 
        \midrule
		boat-b     &          &          &         &        &        &        &          &           \\ 
        \midrule
		SAMBA      & -556.493 & -152.206 & -8.481  &\text{---}&\text{---}&\text{---}& -717.18  &\text{---}  \\
		DIFF-L3pol & -530.607 & -168.049 & -17.790 & 1.220  & 0.164  & 1.384  & -716.446 & -163.180  \\
		DIFF-L2pol & -530.093 & -160.155 & -15.766 & 1.528  & 0.16   & 1.688  & -706.014 & -156.384  \\
		DIFF-L1pol & -535.048 & -167.883 & -18.851 & 0.561  & 0.179  & 0.740  & -721.782 & -163.425  \\
        \midrule
		anti-boat  &          &          &         &        &        &        &          &           \\
        \midrule
		SAMBA      & -553.288 & -150.498 & -11.979 &\text{---}&\text{---}&\text{---}& -715.765 &\text{---}  \\
		DIFF-L3pol & -530.969 & -166.678 & -18.956 & 1.207  & 0.288  & 1.495  & -716.603 & -161.912  \\
		DIFF-L2pol & -530.696 & -159.556 & -17.364 & 1.292  & 0.204  & 1.496  & -707.616 & -155.802  \\
		DIFF-L1pol & -534.791 & -170.783 & -23.615 & -0.677 & 0.039  & -0.638 & -729.189 & -166.228  \\ 
        \bottomrule
	\end{tabular}
	\caption{Decomposition of many-body energies up to four-body contributions, for each model using 
    DIFF (i.e. $x=1$) damping compared with the SAMBA energies from G{\'o}ra \etal 
    \cite{gora_predictions_2014}. The final column gives the 3B non-additive energies where the
    molecular properties have been replaced by those from the water monomer in a conformation
    optimized using CCSD(T)/cc-pVTZ using the Psi4 program.}
	\label{table:n16-energy-decomposition}
} 
\end{table}

\begin{table}[ht]
{\footnotesize
  \newcolumntype{d}{D{.}{.}{3}}
  \begin{tabular}{lddddddd}
    \toprule
      Model        &\multicolumn{1}{c}{ \text{2B} }
                              &\multicolumn{1}{c}{ \text{3B} }
                                         &\multicolumn{1}{c}{ \text{4B} }
                                                   &\multicolumn{1}{c}{ \text{5B} }
                                                            &\multicolumn{1}{c}{ \text{>5B} }
                                                                     &\multicolumn{1}{c}{ \text{2B-4B} }
                                                                                 &\multicolumn{1}{c}{ \text{2B-$\infty$B} } \\
    \midrule
    \multicolumn{8}{l}{ISOMER 316}                                                         \\
    \midrule
      SAMBA        & -801.370 & -246.864 & -29.292 &        &        & -1077.526 &           \\
      DIFF-L3pol   & -769.749 & -257.040 & -43.971 & -2.921 & 0.628  & -1070.760 & -1073.053 \\
      DIFF-L3pol(opt)&        & -249.883 & -42.548 & -2.809 &        &           &           \\
      DIFF-L2pol   & -769.741 & -244.892 & -39.843 & -1.996 & 0.676  & -1054.476 & -1055.796 \\
      DIFF-L1pol   & -778.151 & -259.324 & -45.453 & -3.629 & 0.528  & -1082.927 & -1086.028 \\
      CC-pol-8s-NB & -769.254 & -191.104 & -27.882 & -1.460 & 0.360  & -988.240  &           \\
      CCpol23+     & -769.387 & -214.434 & -27.188 & -1.339 & 0.314  & -1011.009 &           \\
    \midrule
      \multicolumn{8}{l}{ISOMER 308}                                                         \\
    \midrule
      SAMBA        & -808.048 & -241.798 & -28.305 &        &        & -1078.151 &           \\
      DIFF-L3pol   & -772.943 & -255.382 & -43.003 & -2.473 & 0.728  & -1071.328 & -1073.074 \\
      DIFF-L3pol(opt)&         & -248.242 & -41.597 & -2.361 &        &           &           \\
      DIFF-L2pol   & -772.855 & -243.089 & -38.998 & -1.667 & 0.707  & -1054.941 & -1055.902 \\
      DIFF-L1pol   & -781.169 & -255.401 & -45.545 & -3.772 & 0.439  & -1082.115 & -1085.448 \\
      CCpol-8s-NB  & -773.203 & -185.623 & -30.882 & -2.556 & 0.079  & -989.708  &           \\
      CCpol23+     & -773.576 & -210.497 & -28.125 & -1.929 & 0.126  & -1012.197 &           \\
    \midrule
      \multicolumn{8}{l}{Difference: 316 - 308}                                     \\
    \midrule
      SAMBA        & 6.678    & -5.067   & -0.987  &        &        & 0.625     &           \\
      DIFF-L3pol   & 3.194    & -1.658   & -0.968  & -0.448 & -0.100 & 0.568     & 0.021     \\
      DIFF-L3pol(opt)&        & -1.641   & -0.951  & -0.448 &        & 0.463     &           \\
      DIFF-L2pol   & 3.114    & -1.803   & -0.845  & -0.329 & -0.031 & 0.465     & 0.106     \\
      DIFF-L1pol   & 3.018    & -3.923   & 0.092   & 0.143  & 0.089  & -0.812    & -0.580    \\
      CCpol-8s-NB  & 3.954    & -5.481   & 2.996   & 1.092  & 0.280  & 1.469     &           \\
      CCpol23+     & 4.188    & -3.941   & 0.937   & 0.590  & 0.188  & 1.188     &           \\
    \bottomrule
	\end{tabular}
	\caption{
       Decomposition of interaction energies for two variants of the \waterN{24} tetradecahedron
       isomers and the differences between them, using the DIFF models. 
       We also present the CCpol23+, CC-pol-8s+NB and reference SAMBA results from 
       G{\'o}ra \etal \cite{gora_predictions_2014}
       In the DIFF-L3pol(opt) rows we present energies obtained using the DIFF-L3pol models
       with molecular properties (multipoles and polarizabilities) replaced with those from 
       the water monomer in a conformation optimized using CCSD(T)/cc-pVTZ using the Psi4 program.
    }
	\label{table:n24}
 } 
\end{table}

\clearpage

\providecommand*{\mcitethebibliography}{\thebibliography}
\csname @ifundefined\endcsname{endmcitethebibliography}
{\let\endmcitethebibliography\endthebibliography}{}